%% file: main.tex
\pdfoutput=1
\documentclass[aps, superscriptaddress, reprint, amsmath, amssymb, nofootinbib, prl]{revtex4-2}
\usepackage{graphicx}
\usepackage{multirow}
\usepackage{dcolumn}
\usepackage[usenames]{color}
\usepackage{booktabs}
\usepackage{makecell}
\usepackage[version=4]{mhchem}
\usepackage[utf8]{inputenc}
\usepackage{longtable}
\usepackage{caption}
\usepackage{bibunits}
\usepackage[colorlinks=true, allcolors=blue]{hyperref}
\usepackage{siunitx}
\DeclareSIUnit\angstrom{\text{Å}}
\DeclareSIUnit{\debye}{D}
\DeclareSIUnit{\calorie}{cal}
\DeclareSIUnit{\kcal}{\kilo\calorie}
\DeclareSIUnit{\step}{step}

\usepackage{pifont}
\newcommand{\cmark}{\ding{51}}
\newcommand{\xmark}{\ding{55}}
\makeatletter
\expandafter\let\csname longtable*\endcsname\relax
\expandafter\let\csname endlongtable*\endcsname\relax
\makeatother

\begin{document}

\title{SO3LR Tauri: Generally Applicable Force Fields for Reactive (Bio)Molecular Simulations Covering the Periodic Table}

\author{Adil Kabylda}
\email{kabylda@gmail.com}
\affiliation{Department of Physics and Materials Science, University of Luxembourg, L-1511 Luxembourg City, Luxembourg}

\author{Florian N. Brünig}
\affiliation{Department of Physics and Materials Science, University of Luxembourg, L-1511 Luxembourg City, Luxembourg}

\author{Miguel Gallegos}
\affiliation{Department of Physics and Materials Science, University of Luxembourg, L-1511 Luxembourg City, Luxembourg}

\author{J. Thorben Frank}
\affiliation{Machine Learning Group, Technische Universit\"at Berlin, 10587 Berlin, Germany}
\affiliation{Berlin Institute for the Foundations of Learning and Data -- BIFOLD, 10587 Berlin, Germany}

\author{Evgeny Moerman}
\affiliation{Department of Physics and Materials Science, University of Luxembourg, L-1511 Luxembourg City, Luxembourg}

\author{Oliver T.\ Unke}
\affiliation{Google DeepMind, 10117 Berlin, Germany}

\author{Stefan Chmiela}
\affiliation{Machine Learning Group, Technische Universit\"at Berlin, 10587 Berlin, Germany}
\affiliation{Berlin Institute for the Foundations of Learning and Data -- BIFOLD, 10587 Berlin, Germany}

\author{Klaus-Robert M\"{u}ller}
\email{klaus-robert.mueller@tu-berlin.de}
\affiliation{Machine Learning Group, Technische Universit\"at Berlin, 10587 Berlin, Germany}
\affiliation{Berlin Institute for the Foundations of Learning and Data -- BIFOLD, 10587 Berlin, Germany}
\affiliation{Google Research, 10117 Berlin, Germany}
\affiliation{Max Planck Institute for Informatics, Stuhlsatzenhausweg, 66123 Saarbr{\"u}cken, Germany}
\affiliation{Department of Artificial Intelligence, Korea University, Anam-dong, Seongbuk-gu, 02841 Seoul, Korea}

\author{Alexandre Tkatchenko}
\email{alexandre.tkatchenko@uni.lu}
\affiliation{Department of Physics and Materials Science, University of Luxembourg, L-1511 Luxembourg City, Luxembourg}

\begin{abstract}
\noindent 
Machine-learned force fields (MLFFs) aim to create a unified model that describes all major classes of (bio)molecules at quantum-mechanical accuracy. We present SO3LR Tauri, a family of MLFFs developed toward this goal, coupling a fast SO3krates network to explicit, physics-based interaction modules: pair-specific short-range repulsion, environment-dependent dispersion, and electrostatics, each improved beyond its original formulation. The models are trained on 148 million structures spanning $85$ elements, drawn from several quantum-chemical datasets. Against coupled-cluster references, the models accurately reproduce nucleobase-pair and charged protein--ligand interactions, carbohydrate conformer ordering, and molecular dipoles. They also resolve the transition pathways of organic and enzymatic reactions. Dynamical simulations recover experimental densities of water and organic solvents, the hydration structure of mono- and divalent ions, the p$K_\mathrm{a}$ values of amino-acid side chains, and the structure and dynamics of proteins, membranes, nucleic acids, glycoproteins, and metal-containing biomolecules, all without system-specific fitting. Released in three sizes, the models reach $0.7$--$3.2$~\si{\micro\second}/atom and scale to $27$--$84$ million atoms on a single GPU. Much work remains to be done: achieving predictive accuracy for static and dynamic liquid properties, reaction barriers, coverage of chemical and conformational space, and the sampling and inference speed needed to describe biomolecular processes.
\end{abstract}

\maketitle

\section{Introduction}

Molecular dynamics (MD) simulations have become a central tool for studying biomolecular structure and function~\cite{hollingsworth2018molecular}, and their reliability rests on the accuracy of the underlying force field. For decades this role has been filled by classical, empirical force fields, whose careful parametrization has made them both efficient and remarkably successful across a vast range of biological problems~\cite{tian2020ff19sb, huang2017charmm36m, jorgensen1996development}. Given their simple functional forms, accuracy is achieved by fitting separate parameter sets for each class of biomolecules, with dedicated models for water, ions and metal centers~\cite{jorgensen1983comparison, joung2008determination, li2017metal}. These force fields are well established for proteins and lipids, while nucleic acids and carbohydrates remain more challenging~\cite{foley2012carbohydrate, sponer2018rna}. Simulating several classes of biomolecules together therefore requires combining independently developed parameters~\cite{yoo2018new}, which were not designed for use together.

Machine-learned force fields (MLFFs) hold a different prospect~\cite{unke2021machine, unke2024biomolecular, kovacs2025mace, anstine2024aimnet2}. With far greater descriptive power, a single model could in principle represent all of these classes consistently and with close to quantum-mechanical accuracy. Biomolecular applications further require that such a model scale to large assemblies and include the long-range electrostatic and dispersion interactions that shape their structure and dynamics~\cite{honig1995classical, wennberg2013lennard}. SO3LR follows a recent line of models that tackle these requirements by augmenting local message-passing with explicit physical terms~\cite{unke2021spookynet, anstine2024aimnet2}, coupling a fast and accurate SO3krates message-passing backbone~\cite{frank2024euclidean} to terms for short-range repulsion, electrostatics, and dispersion. Its first version, trained on 4M PBE0+MBD structures over eight elements and scaling to 200k atoms on a single GPU, demonstrated the promise of this approach for biomolecular simulation~\cite{kabylda2025molecular}.

Here we present SO3LR Tauri\footnote{\scriptsize Named after the T Tauri phase, an early stage of stellar evolution.\par}, a new generation of SO3LR models built toward this generality. We retain the fast SO3krates backbone and extend the model with a much broader training set and refined interaction physics terms that transfer across the periodic table. Since no single consistent quantum-chemical dataset covers the biologically relevant elements and their bonding, we assemble a corpus of roughly 148 million structures across 85 elements, a 40-fold increase over SO3LR, drawing on several recent datasets including QCML~\cite{ganscha2025qcml}, QCell~\cite{kabylda2026qcell}, OMol25~\cite{levine2025open}, and OPoly26~\cite{levine2026open}. 

These datasets mix two levels of theory. The empirically optimized $\omega$B97M-V functional excels at main-group thermochemistry and reaction barriers~\cite{mardirossian2016omegab97m, goerigk2017look}, whereas PBE0+MBD(-NL) is built with minimal empiricism and treats long-range many-body dispersion interactions explicitly and accurately~\cite{adamo1999pbe0, ambrosetti2014mbd, hermann2020mbdnl}. Unlike recent general-purpose models, which condition each prediction on one level of theory~\cite{batatia2025cross, wood2026family}, we first train separate heads on a shared representation and then combine both levels in a single production head during late-stage fine-tuning. This approach aims to combine the strengths of both levels of theory.

The short-range repulsion, electrostatics, and dispersion terms of SO3LR are each revised so that they remain valid for 85 elements and for condensed phases. Close interatomic contacts are rare in reference datasets, which leaves a learned repulsion poorly constrained. The learned ZBL-type repulsion is thus replaced by pair-specific Nordlund--Lehtola--Hobler (NLH) potentials~\cite{nordlund2025repulsive}, pre-tabulated from dedicated DFT calculations, which keep the short-range repulsion physically correct even at high temperature and pressure. SO3LR rescaled free-atom $C_6$ coefficients by a single learned Hirshfeld ratio, which is too restrictive across the periodic table, so the network now predicts per-atom polarizability and $C_6$ ratios. Since electrostatics and dispersion decay too slowly to be truncated in large or periodic systems, both are evaluated with particle-mesh Ewald summation~\cite{darden1993particle, loche2025fast}.

Training and running models at this scale also required changes to the data pipeline, backbone, training recipe, and inference code. The resulting models are released in three sizes, SO3LR-T-S/M/L, spanning a range of accuracy and cost. In what follows, we describe the architecture and training, benchmark the models on static properties, reactivity, and dynamics, including their limitations, and evaluate their speed and scaling.

\begin{figure*}
\centering
\includegraphics[width=\linewidth]{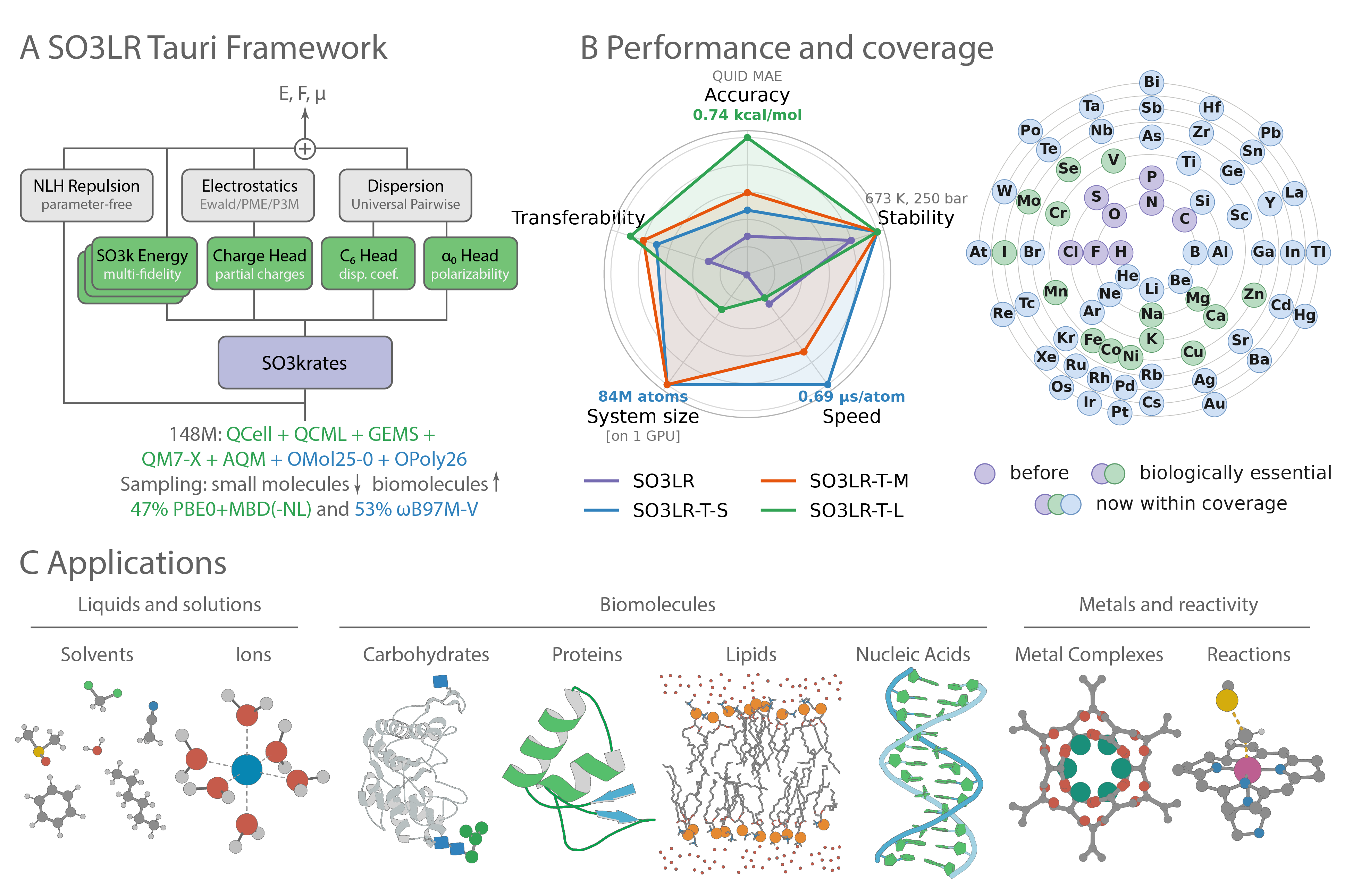}
\caption{
\textbf{Overview of SO3LR Tauri. (A)} A SO3krates neural network ingests reference quantum-mechanical data spanning $\sim$148M molecular fragments drawn from the QCell, QCML, GEMS, QM7-X, AQM, OMol25, and OPoly26 datasets. Per-atom heads predict semi-local energies, partial charges $q_i$, atomic polarizabilities $\alpha_0$, and dispersion coefficients $C_6$, which parametrize long-range electrostatics and dispersion; NLH repulsion contains no trainable parameters. The terms sum to the total energy $E$, with forces $F = -\nabla E$ and the dipole $\mu$ from the predicted charges. \textbf{(B)}~Capabilities of SO3LR-T-S/M/L compared to SO3LR. Radar chart: accuracy is the interaction-energy MAE on the QUID benchmark (kcal/mol), speed the measured cost in \si{\micro\second}/atom, and system size is the largest system that fits on one GPU. Stability (assessed up to \SI{673}{K} and \SI{250}{bar}) and transferability are qualitative. Periodic spiral (Ernst Huth, 1884): elemental coverage grows from 8 elements (purple) to elements $Z = 1$--$85$ (purple, green and blue), including the biologically essential set (purple and green). The lanthanides Ce--Lu are covered but not shown. \textbf{(C)} SO3LR Tauri enables simulations of solvents and ions, all major classes of biomolecules (carbohydrates, proteins, lipids, and nucleic acids), and metal complexes and reactions}
\label{fig:overview}
\end{figure*}

\section{Model Architecture and Training}

We describe the SO3LR Tauri model family: its modular architecture and physically grounded interaction terms, the expanded training corpus and the multi-head scheme that combines two levels of theory, and the three released model sizes.  Each design choice targets biomolecular simulation. Full architectural, dataset, and training details are provided in the Supplementary Information (Sec.~\ref{sec:architecture} and Sec.~\ref{sec:training}).

\subsection{Model architecture}

SO3LR has a modular structure, in which a neural-network backbone supplies per-atom features and physical quantities to a set of explicit interaction terms. The total potential energy is
\begin{align}
    E_\text{total} = \underbrace{E_\text{rep}}_{\text{short-range}}
    + \underbrace{E_\text{SO3k}}_{\text{semi-local}}
    + \underbrace{E_\text{elec} + E_\text{disp}}_{\text{long-range}},
    \label{eq:mlff-potential-energy}
\end{align}
with contributions from short-range repulsion, the SO3krates semi-local interaction, electrostatics, and dispersion. Every term is differentiable, and all modules are trained jointly on quantum-chemical reference data so that they remain mutually consistent. The SO3krates backbone is carried over from SO3LR with minimal changes (described in Sec.~\ref{sec:architecture}), while the three physical modules have each been refined.

The neural network takes the atomic positions $R$, atomic numbers $Z$, total charge $Q$, and spin multiplicity $S$, and predicts the per-atom quantities that enter the physical terms: the atomic energy $E_i$, the partial charge $q_i$, the polarizability ratio $h_i$, and the dispersion-coefficient ratio $r^{C_6}_i$
\begin{equation}
    E_i, q_i, \, h_i,\, r^{C_6}_i = \text{SO3krates}(R, Z, Q, S).
    \label{eq:predicted_quantities}
\end{equation}
The polarizability and $C_6$ coefficients follow from these ratios via Eq.~\eqref{eq:cp-ratios}. Each message-passing layer couples an atom only to neighbors within the radial cutoff $r_\text{c}$, so after $N$ layers $E_\text{SO3k}$ depends on an environment of at most $N r_\text{c}$.

We recover long-range physics through the explicit modules rather than by stacking message-passing layers to enlarge the receptive field, since this has been shown to be insufficient for learning pairwise long-range effects~\cite{frank2026machine}. Deep message-passing neural networks also do not obey the correct asymptotic decay~\cite{esders2025analyzing} and cannot transmit interactions across a vacuum. In addition, a larger receptive field would require reference data for mesoscopic configurations, which are absent from current datasets and prohibitively expensive to generate. Modules with the correct $1/r$ and $1/r^{6}$ asymptotics remove the need for the network to learn the long-range behavior from data.

\textbf{Short-range repulsion: ZBL $\to$ NLH.} 
Long trajectories, high-temperature or high-pressure sampling, and geometries from generative AI models often contain close contacts that are rare in reference datasets. In these configurations, a repulsion learned from data can become unphysical. In SO3LR this repulsion was a ZBL-inspired screening function fitted during training~\cite{ziegler1985stopping, unke2021spookynet}. We replace it with the pre-tabulated, pair-specific potentials of Nordlund, Lehtola, and Hobler (NLH)~\cite{nordlund2025repulsive}, fitted to dedicated self-consistent DFT calculations. The repulsion is a screened nuclear Coulomb interaction
\begin{equation}
E_\text{rep} = \sum_{i<j} k_e\,\frac{Z_i Z_j}{r_{ij}}\,\phi_{Z_iZ_j}(r_{ij})\,s(r_{ij}),
\label{eq:nlh}
\end{equation}
with the pair-specific screening function given by a sum of three exponentials,
\begin{equation}
\phi_{Z_iZ_j}(r) = \sum_{k=1}^{3} a^{Z_iZ_j}_k\,e^{-b^{Z_iZ_j}_k\, r},
\label{eq:nlh-screen}
\end{equation}
where the coefficients $\{a_k, b_k\}$ are tabulated for all combinations of $Z = 1$--$92$, and $s(r)$ is a switching function that removes the term beyond $\sim$1.5~\AA. NLH agrees with MP2 reference data to within $\sim$2\% above $30$~eV, compared to $\sim$5--10\% for ZBL~\cite{nordlund2025repulsive}. Keeping the NLH term fixed does not bias the model, because the semi-local SO3krates energy can still learn corrections to the short-range repulsion wherever the training data require them. The change from ZBL to NLH mainly improves robustness, which we tested under the extreme conditions of the water simulations (up to \SI{673}{\kelvin} and \SI{250}{\bar}; see \textit{Liquid Properties} and Fig.~\ref{fig:water_high_tp}). 

\textbf{Electrostatics.} The network predicts per-atom charges $q_i$ that are supervised indirectly through molecular dipoles, and which are used in a damped analytical pairwise Coulomb energy that models long-range electrostatics
\begin{equation}
E_\text{elec} = \sum_{i<j} k_e\, q_i q_j\,\frac{\operatorname{erf}(r_{ij}/\sigma)}{r_{ij}} \;%
\label{eq:elec}
\end{equation}
where $\sigma$ sets the short-range damping width. For periodic systems the Coulomb sum is evaluated by particle-mesh Ewald (JAX-PME) summation~\cite{darden1993particle, loche2025fast}. The real-space term is energy- and force-shifted so that it vanishes smoothly at the real-space cutoff, which reduces integration drift.

\textbf{Dispersion: $h_i$ $\to$ $(C_6,\alpha_0)$.}
We describe dispersion with universal pairwise interatomic van der Waals potentials derived from quantum Drude oscillators (QDO)~\cite{Khabibrakhmanov2023}.  SO3LR derived all atomic dispersion parameters from one learned Hirshfeld volume ratio, $h_i = V_{\text{eff}}/V_{\text{free}}$~\cite{hirshfeld1977bonded}. This ratio scaled the free-atom polarizability by $h_i$ and the free-atom $C_6$ coefficient by $h_i^2$. Such a single scalar rescaling is not flexible enough to describe dispersion across the broad chemical space that the new models cover. SO3LR Tauri therefore predicts two environment-dependent ratios per atom, a polarizability ratio $r^{\alpha}_i$ and a $C_6$ ratio $r^{C_6}_{i}$. Both are trained on per-atom labels from the QCML~\cite{ganscha2025qcml} and QCell~\cite{kabylda2026qcell} datasets and scale the free-atom references
\begin{equation}
\alpha_i = \alpha^\text{free}_{z_i}\, r^{\alpha}_i, \qquad C_{6,i} = C^\text{free}_{6,z_i}\, r^{C_6}_i.
\label{eq:cp-ratios}
\end{equation}
Pairwise coefficients follow from the Casimir--Polder mixing rules
\begin{equation}
\alpha_{ij} = \tfrac{1}{2}\,(\alpha_i+\alpha_j), \qquad
C_{6,ij} = \frac{2\,C_{6,i}\,C_{6,j}\,\alpha_i\,\alpha_j}{\alpha_i^2\, C_{6,j} + \alpha_j^2\, C_{6,i}},
\label{eq:cp-mixing}
\end{equation}
with higher-order terms from the QDO relations $C_{8,ij} = 5\,C_{6,ij}/\gamma_{ij}$ and $C_{10,ij} = 245\,C_{6,ij}/(8\gamma_{ij}^2)$, where $\gamma_{ij}$ is determined from $\alpha_{ij}$~\cite{Khabibrakhmanov2023}. 
The dispersion energy sums the $C_6$, $C_8$, and $C_{10}$ contributions, 
\begin{equation}
E_\text{disp} = -\sum_{i<j}\;\sum_{n=3}^{5} \frac{C_{2n,ij}}{\,r_{ij}^{2n} + R_{ij}^{2n}\,} \;
\label{eq:disp}
\end{equation}
where the damping radius $R_{ij}\propto \alpha_{ij}^{1/7}$ prevents the short-range divergence as $r_{ij}\!\to\!0$ for the leading term. For periodic systems, a reciprocal-space term extends the leading $C_6$ interaction through the same PME implementation as the electrostatics, so the two long-range interactions share a unified treatment (Eq.~\ref{eq:disp-pme}).

\begin{figure*}
\centering
\includegraphics[width=1.0\linewidth]{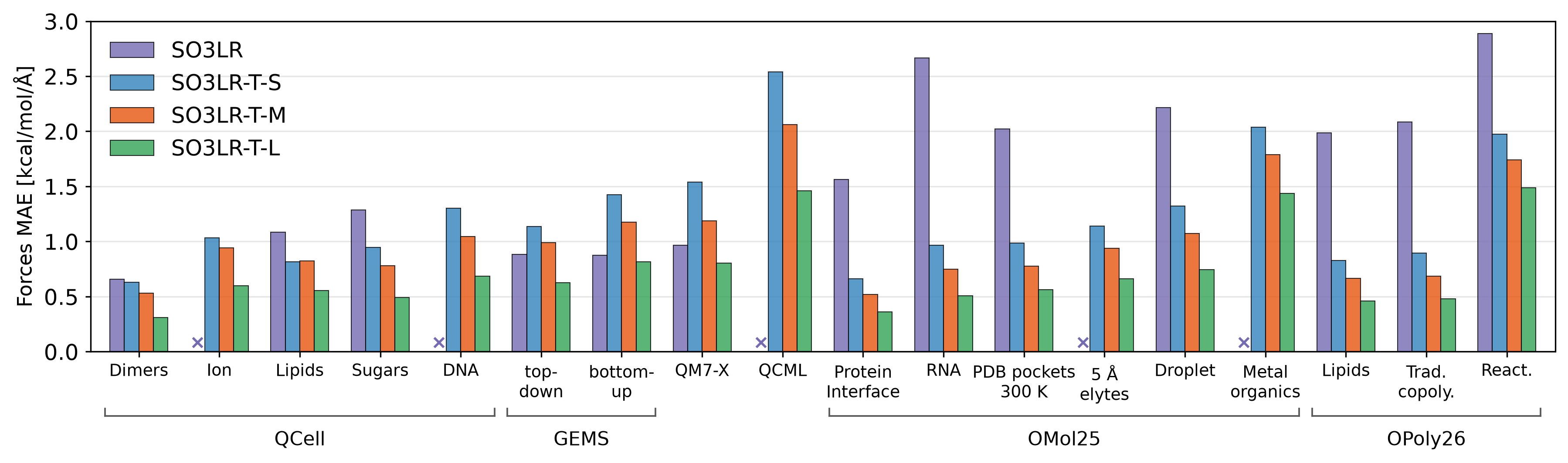}
\caption{\textbf{Test set errors.} Force mean absolute errors (MAEs, kcal/mol/\AA{}) of SO3LR and SO3LR-T-S, -M, and -L across test sets: QCell (with subsets), GEMS, QM7-X, QCML, OMol25 (protein interfaces and pockets, nucleic acids, electrolytes, metal-organic systems), and OPoly26 (lipids, copolymers, reactive systems). SO3LR is the original model, which supports eight elements (H, C, N, O, F, P, S, Cl); structures containing any other element are excluded in full, leaving $55\%$ of the test set. Its bar is therefore shown only for subsets where at least half of the structures pass this filter. Full per-subset tables are given in the SI (Tables~\ref{tab:pbe0-forces}--\ref{tab:opoly26-forces})}%
\label{fig:test-set-errors}
\end{figure*}

\subsection{Training}

\textbf{Training data.} A general-purpose biomolecular model needs training data broader than any single quantum-chemical dataset offers, so the corpus is assembled from seven complementary sources (Table~\ref{tab:Datasets}). It holds $\sim$148 million structures over $85$ elements (${\sim}6.9$B atoms), a 40-fold increase on the $\sim$4 million structures and 8 elements behind the predecessor. It combines the SO3LR data (GEMS bottom-up fragments~\cite{unke2024biomolecular}, QM7-X~\cite{hoja2021qm7}, AQM~\cite{medrano2024dataset}, SPICE dipeptides~\cite{eastman2023spice}, and DES dimers~\cite{donchev2021quantum}) with four recently released datasets, QCML~\cite{ganscha2025qcml}, QCell~\cite{kabylda2026qcell}, OMol25~\cite{levine2025open}, and OPoly26~\cite{levine2026open}. 

If sampled in proportion to their size, these datasets would be dominated by near-redundant conformers of small molecules, biasing the model toward systems for which existing MLFFs already perform well. We therefore downsample the largest small molecule subsets and upsample less abundant biomolecular motifs, so that the training distribution reflects target chemical space coverage rather than the number of available structures (Tables~\ref{tab:Datasets} and~\ref{tab:sampling}; Fig.~\ref{fig:dataset}).

\textbf{Theory-level heads.} The constituent datasets are computed at two levels of theory: PBE0+MBD(-NL) for the SO3LR data, QCML, and QCell, and $\omega$B97M-V for OMol25 and OPoly26. Rather than mixing them under a single target, the final layer predicts a per-atom energy for each reference level $t$
\begin{equation}
    \varepsilon_{i,t} = s_{Z_i,t}\,\mathrm{MLP}_t(x_i) + b_{Z_i,t},
    \label{eq:multihead}
\end{equation}
where $s_{Z,t}$ and $b_{Z,t}$ are a per-element scale and offset for each head $t$. A one-hot mask selects the head that matches the level of theory of each structure, so that each head is trained only on its own data. The head-specific scales and offsets absorb the systematic differences between the levels of theory. This allows the backbone to learn a general, method-independent representation~\cite{gardner2026understanding}.

\textbf{Training.} Training proceeds in five stages (Table~\ref{tab:si_stages}). Stages~1 and~2, about ${\sim}80$\% of the optimizer steps, train both heads with a \qty{10}{\angstrom} long-range cutoff, which keeps them inexpensive. From stage~3 onward, the long-range cutoff is increased to \qty{1000}{\angstrom} which includes all atom pairs, and a single production head, initialized from the PBE0+MBD(-NL) head, is trained on data from both levels of theory. Stages~4 and~5 then rebalance the water content of the training set to correct the over-structured hydrogen bonding inherited from PBE0+MBD(-NL). Each model has been trained for ${\sim}400$ GPU-hours on a single NVIDIA A100 40~GB. Full details, including dataset composition, sampling weights, loss, schedule, and training curves, are given in  Sec.~\ref{sec:training}.

\begin{table}
\centering
\caption{\textbf{SO3LR Tauri model hyperparameters}}
\label{tab:model_sizes}
\begin{tabular}{lccc}
\toprule
\textbf{Parameter} & \textbf{Small} & \textbf{Medium} & \textbf{Large} \\
\midrule\midrule
Cutoff radius (\unit{\angstrom}) & 4.5 & 4.5 & 5.0 \\ 
Feature dimension & 128 & 128 & 256 \\
Message-passing layers & 2 & 3 & 3 \\
Attention heads & 4 & 4 & 8 \\
Maximal degree & 4 & 4 & 4 \\
Radial basis functions & 32 & 32 & 64 \\
Parameters (M) & 0.41 & 0.53 & 1.79 \\
\bottomrule
\end{tabular}%
\end{table}

\textbf{Model sizes.} We release the model in three sizes (Table~\ref{tab:model_sizes}): SO3LR-T-S/M/L, spanning a range of accuracy and cost that we characterize below. SO3LR-T-M shares its backbone configuration with SO3LR, making it a natural choice for analysing the benefits of our improved components and data. We therefore focus on \mbox{SO3LR-T-M} throughout the article and use it for all biomolecular dynamics simulations below, with the exception of the reactivity study, which uses \mbox{SO3LR-T-L}. Overall accuracy improves with model size, although SO3LR-T-M is competitive with, and on some benchmarks exceeds, the larger SO3LR-T-L (Table~\ref{tab:so3lr-performance}).

\section{Benchmark Accuracy}\label{sec:bench}

Before turning to dynamics, we evaluated the accuracy of SO3LR and the SO3LR Tauri models at fixed geometries. We use five families of benchmarks: held-out test splits, metal--organic frameworks, intermolecular interaction energies, conformational energetics, and molecular dipoles. To avoid a bias toward a particular density functional, we primarily selected external benchmarks computed at the coupled-cluster level (Table~\ref{tab:so3lr-performance}).

\textbf{Test set error.} We first evaluate the models on the held-out test splits, measuring interpolation across the training data (Fig.~\ref{fig:test-set-errors}, Sec.~\ref{sec:static-benchmarks}). SO3LR-T-M performs well across the four biomolecular classes, with no chemistry standing out as systematically harder. Force MAEs for proteins fall in the range $0.52$--$1.17$~kcal/mol/\AA{} across the OMol25 interface and core, PDB fragment and pocket, and GEMS peptide subsets; carbohydrates (QCell sugars) reach $0.78$~kcal/mol/\AA{}; lipids span $0.67$--$0.82$~kcal/mol/\AA{} across the OPoly26 and QCell lipids subsets; and nucleic acids span $0.70$--$1.04$~kcal/mol/\AA{} across the OMol25 DNA/RNA and QCell DNA subsets. SO3LR is slightly more accurate than SO3LR-T-M on several subsets from its own training sources (such as QM7-X and both GEMS sets). These sources formed most of the training set of SO3LR. In SO3LR Tauri, they share the training budget with four new datasets, and QM7-X is further downsampled (Table~\ref{tab:sampling}).

The subsets with higher errors are concentrated in chemistry that is either intrinsically harder or deliberately undersampled. For example, on the OPoly26 reactivity and charged-fragment subsets, SO3LR-T-M reaches force MAEs of $1.74$ and $1.61$~kcal/mol/\AA{}, as bond breaking and strong electrostatics are more demanding than conformational sampling (and generally have higher values). Small organic molecules remain accurate despite their reduced sampling weight (QM7-X $1.19$, AQM $1.49$~kcal/mol/\AA), whereas QCML is the most challenging small-molecule source ($2.06$~kcal/mol/\AA{}), reflecting its broad and unique chemical coverage. The transition-metal and metal-organic subsets carry the largest errors ($1.79$--$5.00$~kcal/mol/\AA{}); these mostly lie outside the target biomolecular core and were downsampled accordingly. We examine metal-containing chemistry in more detail below.

\textbf{Periodic solids.} The training corpus contains no periodic structures. We therefore tested transferability of the models on metal--organic frameworks (MOFs). On $50$ thermal snapshots of MIL-53(Al) and UiO-66(Zr)~\cite{vandenhaute2023mof} recomputed here at PBE0+MBD-NL, SO3LR-T-M reaches a force MAE of $3.55$~kcal/mol/\AA{}, against $6.80$ for PBE-D3 itself (Fig.~\ref{fig:mof}). On MOFSimBench, a curated set of $100$ nanoporous frameworks~\cite{krass2025mofsimbench}, SO3LR-T-M reproduces the PBE-D3(BJ)-optimized cell volumes with a median error of $2.3\%$ and the atomic positions with a median RMSD of $0.26$~\AA{}. Since our models are trained at a higher level of theory than these references, part of this residual may reflect the functional difference instead of model error (Sec.~\ref{sec:mof}).

\begin{table*}
\caption{\textbf{Performance of SO3LR models on interaction-energy, conformational, and dipole benchmarks}. Energies are reported in kcal/mol, dipole magnitudes in Debye, and angles in degrees. MAE\% for PLA15 denotes the mean absolute relative error. Sets probe intermolecular interaction energies (SAPT10k~\cite{villot2024ab}; JSCH-2005 nucleobase pairs~\cite{jurevcka2006benchmark}; QUID~\cite{puleva2025extending} and PLA15 protein--ligand active sites~\cite{kriz2020benchmarking}), conformational energetics (TorsionNet500 torsion profiles~\cite{rai2022torsionnet, kabylda2025molecular}; GlcMal glucose/$\alpha$-maltose conformers~\cite{marianski2016assessing}), and dipoles (QM7b, AlphaML)~\cite{yang2019quantum}}
\centering
\setlength{\tabcolsep}{3pt}
\begin{tabular}{l cc cc ccc cc cc cc cc cc}
\toprule
 & \multicolumn{2}{c}{SAPT10k}
 & \multicolumn{2}{c}{QUID}
 & \multicolumn{2}{c}{JSCH-2005}
 & \multicolumn{3}{c}{PLA15}
 & \multicolumn{2}{c}{TorsionNet500}
 & \multicolumn{2}{c}{GlcMal}
 & \multicolumn{2}{c}{QM7b}
 & \multicolumn{2}{c}{AlphaML}\\
 & & & & & & & & & & & & &
   & \multicolumn{1}{c}{mag.} & \multicolumn{1}{c}{ang.}
   & \multicolumn{1}{c}{mag.} & \multicolumn{1}{c}{ang.} \\
\cmidrule(lr){2-3} \cmidrule(lr){4-5} \cmidrule(lr){6-7} \cmidrule(lr){8-10}
\cmidrule(lr){11-12} \cmidrule(lr){13-14} \cmidrule(lr){15-16} \cmidrule(lr){17-18}
\scriptsize & \scriptsize MAE & \scriptsize RMSE
            & \scriptsize MAE & \scriptsize RMSE
            & \scriptsize MAE & \scriptsize RMSE
            & \scriptsize MAE\% & \scriptsize MAE & \scriptsize RMSE
            & \scriptsize MAE & \scriptsize RMSE
            & \scriptsize MAE & \scriptsize RMSE
            & \scriptsize MAE & \scriptsize MAE
            & \scriptsize MAE & \scriptsize MAE \\
\midrule
SO3LR    & 0.90 & 1.88 & 2.67 & 3.07 & 1.75 & 2.24 & 5.8 & 8.75 & 9.85 & 1.01 & 1.47 & 2.28 & 3.05 & 0.13 & 5.1 & 0.14 & 15.8 \\
SO3LR-T-S  & 0.60 & 1.30 & 1.58 & 2.17 & 1.01 & 1.27 & 2.5 & 3.74 & 4.50 & 0.81 & 1.20 & 1.37 & 1.72 & 0.11 & 4.1 & 0.13 & 12.6 \\
SO3LR-T-M  & 0.45 & 0.83 & 1.24 & 1.59 & 0.78 & 1.04 & 2.1 & 3.81 & 5.48 & 0.62 & 0.93 & 1.36 & 1.68 & 0.10 & 3.3 & 0.13 & 6.3  \\
SO3LR-T-L  & 0.49 & 1.22 & 0.74 & 1.00 & 0.58 & 0.76 & 4.3 & 6.42 & 7.36 & 0.51 & 0.78 & 0.93 & 1.16 & 0.10 & 2.9 & 0.10 & 5.2  \\
\bottomrule
\end{tabular}
\label{tab:so3lr-performance}%
\end{table*}

\textbf{Interaction energies (SAPT10k, QUID, JSCH-2005, PLA15).} 
Non-covalent interactions shape biomolecular structure through a balance of electrostatics, polarization, and dispersion, which we probe using four external benchmarks (Table~\ref{tab:so3lr-performance}). On SAPT10k~\cite{villot2024ab}, a chemically diverse set of intermolecular interactions with SAPT2+(3)(CCD)/aug-cc-pVTZ references, SO3LR-T-M reaches an MAE of $0.45$~kcal/mol, half the $0.90$~kcal/mol of SO3LR. For protein--ligand dimers, we use QUID with LNO-CCSD(T)/CBS reference energies~\cite{puleva2025extending}, for which SO3LR-T-M yields $1.24$~kcal/mol. For nucleic-acid interactions, we use the 102-complex subset of JSCH-2005 with estimated CCSD(T)/CBS reference energies, including hydrogen-bonded and stacked nucleobase pairs~\cite{jurevcka2006benchmark}, where SO3LR-T-M achieves $0.78$~kcal/mol (Table~\ref{tab:jsch-reference-subsets} gives the full set including the MP2/CBS references).

Finally, PLA15 comprises fifteen protein--ligand active sites, providing a demanding test of large, charged non-covalent complexes~\cite{kriz2020benchmarking}. Its reference interaction energies are based on MP2-F12 and DLPNO-CCSD(T) calculations, and SO3LR-T-M reaches an MAE of $3.8$~kcal/mol ($2.1\%$ relative error), outperforming most heavier pretrained MLFFs and semi-empirical methods, whose MAEs span $7.7$--$112$~kcal/mol (Table~\ref{tab:pla15-extended}). We note that PLA15 is the only benchmark where SO3LR-T-L considerably increases deviation compared to SO3LR-T-M. However, the relative error is 4.3\%, and the reference numbers could well be in error of a few percent from CCSD(T)/CBS level. 

\textbf{Conformational energies (TorsionNet500, GlcMal).} Accurate intramolecular conformational energetics determine the populations sampled in dynamics. On TorsionNet500~\cite{rai2022torsionnet}, druglike torsion profiles recomputed at PBE0+MBD~\cite{kabylda2025molecular}, SO3LR-T-M reaches a profile MAE of $0.62$~kcal/mol, improving on SO3LR's $1.01$~kcal/mol. Carbohydrates have long been difficult for additive force fields, which often mis-rank conformers that differ in their hydroxyl hydrogen-bonding networks. On the GlcMal set~\cite{marianski2016assessing} of glucose and $\alpha$-maltose conformers (CCSD(T)/CBS and DLPNO-CCSD(T)/CBS references), SO3LR-T-M reaches an MAE of $1.36$~kcal/mol and reproduces the conformer ordering with Spearman $\rho = 0.89$. This is a marked improvement over the dedicated classical carbohydrate force fields and semi-empirical methods (MAE $2.96$--$5.55$~kcal/mol, $\rho = 0.38$--$0.77$; Table~\ref{tab:marianski-methods}).

\textbf{Dipole moments (QM7b, AlphaML).} Molecular dipoles probe the charge distribution that underlies the electrostatics module. Against LR-CCSD/d-aug-cc-pVDZ references, SO3LR-T-M reaches an MAE of $0.10$~D on QM7b and $0.13$~D on a more diverse AlphaML set with larger molecules. At the same time, the dipole orientation improves considerably, with the AlphaML angular error decreasing from $15.8^\circ$ for SO3LR to $6.3^\circ$. For reference, PBE0+MBD itself differs from the LR-CCSD reference by $0.10$~D on this set, so models reach hybrid-DFT accuracy for the magnitude. The improvements over SO3LR, most visible in the orientation, follow from the QCML small-molecule dipoles, which improve the inferred partial charges.

\section{Reactive Chemistry}\label{sec:reactions}

\begin{figure*}[!t]
\centering
\includegraphics[width=\linewidth]{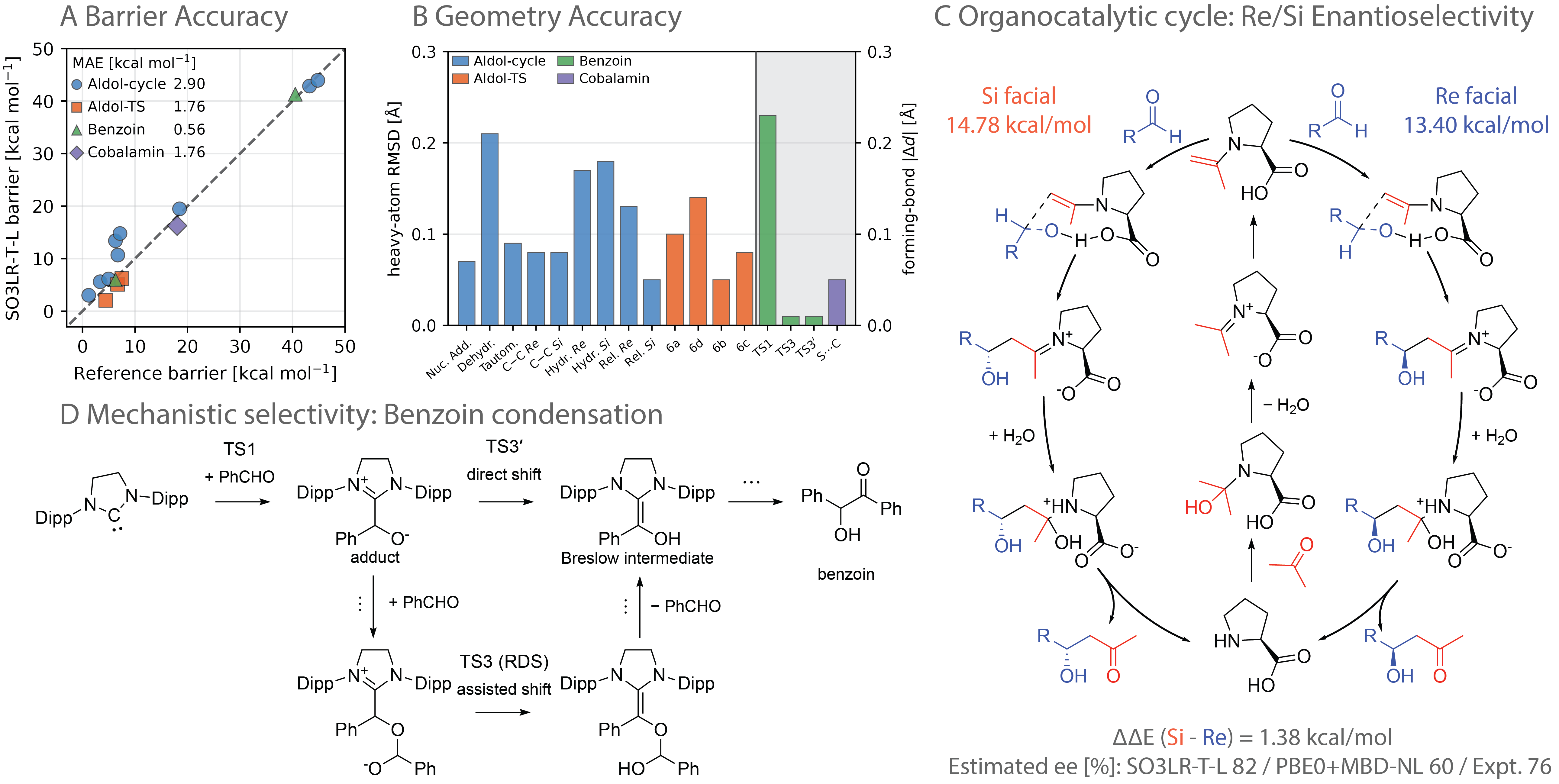}
\caption{\textbf{Transition-state reactivity of SO3LR-T-L.} 
\textbf{(A)} Barrier accuracy: PBE0+MBD-NL single points at the SO3LR-T-L geometries for the aldol cycle, the cobalamin methyl transfer with the 63-atom model of Ref.~\citenum{jensen2003meth} recalculated here at PBE0+MBD-NL,
and the free-energy barriers for the benzoin condensation~\cite{huang2022benzoin}; the legend gives the mean absolute error of each series. %
\textbf{(B)} Saddle-geometry errors: heavy-atom RMSD of the aldol saddles to their reference structures and absolute deviations in key forming-bond distances for benzoin and cobalamin (shaded region, right axis). \textbf{(C)} Proline-catalysed aldol cycle, showing the competing \emph{Re}- and \emph{Si}-facial transition states, their SO3LR-T-L barriers, and the estimated enantiomeric excess ($R=\mathrm{CH}_3$). \textbf{(D)} SIPr-catalysed benzoin condensation via direct (TS3$'$) or aldehyde-assisted (TS3) proton transfer to the Breslow intermediate; TS1 denotes the initial carbene addition, and dotted arrows mark steps omitted for clarity (RDS, rate-determining step; Ph, phenyl; Dipp, 2,6-diisopropylphenyl)} %
\label{fig:reactivity}
\end{figure*}

The benchmarks above test SO3LR models on structures where no bonds change. MLFFs can also describe bond formation and breaking, which lie outside a fixed bonding topology employed in classical force fields. The absence of a topology does not by itself guarantee that a model resolves chemical reactions. Locating a transition state, a first-order saddle point, requires a model to reproduce the curvature of the potential-energy surface as well as energies and forces. Such structures are sparsely represented in the reference data, DFT barrier heights vary by several kcal/mol between functionals~\cite{goerigk2017look}, and the barrier differences that determine selectivity are often below $1$~kcal/mol. We therefore employed the largest model, SO3LR-T-L, to locate transition states (TS) with the automated \textsc{mars-rover} workflow~\cite{gallegos2026marsrover} (detailed protocol and SO3LR-T-M results are described in Sec.~\ref{sec:si-reactions}). Three reactions were chosen to probe distinct aspects of reactivity (Fig.~\ref{fig:reactivity}): stereoselectivity in an organocatalytic cycle, mechanistic selectivity between competing pathways, and enzymatic reactivity at a transition-metal cofactor in a solvated cluster.

\textbf{Organocatalytic cycle.} Proline-catalyzed aldol addition~\cite{list2000proline} is a classic example of enamine organocatalysis (Fig.~\ref{fig:reactivity}C). Its enantioselectivity is decided at the \ce{C-C} bond-forming step, where the enamine can approach either face of the aldehyde (\emph{Re} or \emph{Si}). The two barriers differ by only about $1$~kcal/mol. The full cycle involves several types of bond changes, including the formation and cleavage of \ce{C-N} and \ce{C-O} bonds, proton transfers and the formation of a \ce{C-C} bond. SO3LR-T-L locates all nine first-order saddles, from carbinolamine formation to hydrolytic release, each connected to its neighboring minima by an intrinsic reaction coordinate (IRC; Sec.~\ref{sec:si-reactions}). At the SO3LR-T-L saddle geometries, the barriers agree with PBE0+MBD-NL single points~\cite{adamo1999pbe0, hermann2020mbdnl} with a mean absolute error of $2.9$~kcal/mol (largest deviation $7.6$~kcal/mol, at the \emph{Si} \ce{C-C} step; Fig.~\ref{fig:reactivity}A). The saddle geometries lie within a mean heavy-atom RMSD of $0.12$~\AA{} of the PBE0-D3(BJ)-optimized structures (Fig.~\ref{fig:reactivity}B and Table~\ref{tab:si_react_cycle}). 

At the stereodetermining step, SO3LR-T-L favors the \emph{Re} face by $1.38$~kcal/mol, which corresponds to an enantiomeric excess of $82\%$ at $298$~K. For comparison, PBE0+MBD-NL at the DFT-optimized saddles gives $0.82$~kcal/mol ($60\%$, Table~\ref{tab:si_react_mbd}), and the measured value for the analogous reaction in DMSO is $76\%$~\cite{list2000proline, allemann2010stereoselectivities}. Overall, SO3LR-T-L describes the complete catalytic cycle and predicts the experimentally observed facial selectivity.

\textbf{Mechanistic selectivity.} The benzoin condensation joins two aldehydes through an acyl-anion equivalent (Fig.~\ref{fig:reactivity}D)~\cite{breslow1958}. Under aprotic conditions, no external acid or base is available. Therefore, the rate-determining proton migration cannot proceed by an unassisted \mbox{1,2-shift}, and a second aldehyde has to relay the proton. SO3LR-T-L reproduces both saddle points and their energy difference. The relayed migration has a barrier of $5.9$~kcal/mol and the direct shift a barrier of $41.3$~kcal/mol, close to the reference values of $6.3$ and $40.5$~kcal/mol~\cite{huang2022benzoin} (Fig.~\ref{fig:reactivity}A). The model therefore correctly identifies the relayed pathway as the preferred mechanism. IRC calculations confirm that each saddle point connects the expected intermediates (Table~\ref{tab:si_react_benzoin}).

\begin{figure*}[t]
\centering
\includegraphics[width=0.99\linewidth]{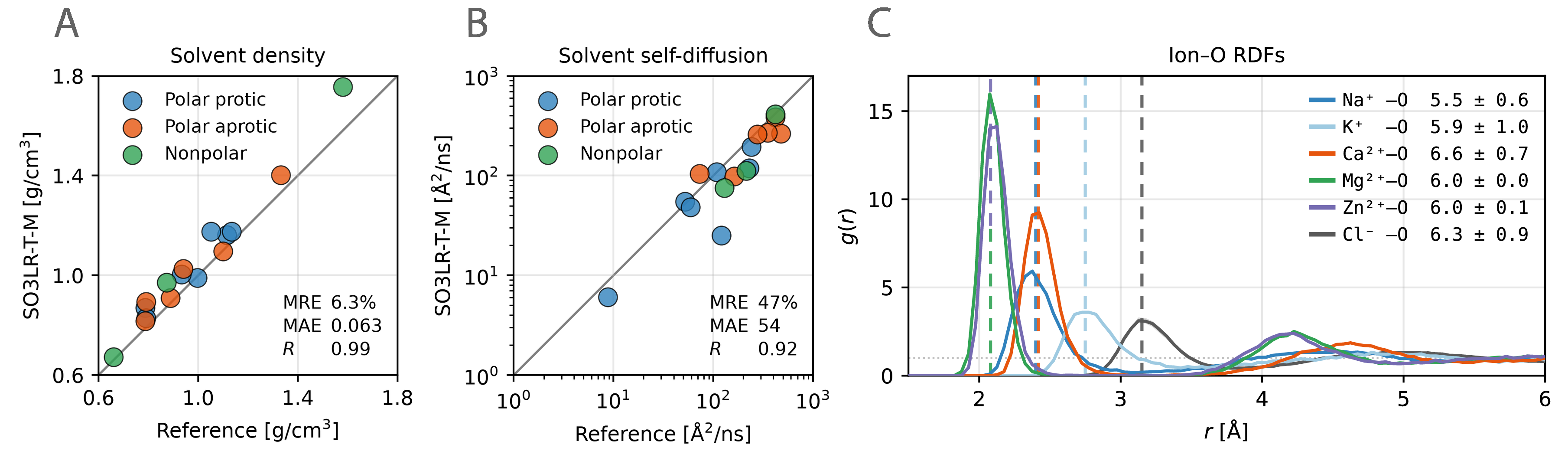}
\caption{\textbf{Condensed-phase properties of SO3LR-T-M versus experiment.} \textbf{(A,\,B)}~Density and self-diffusion for 16 neat solvents, coloured by solvent class; the diagonal marks perfect agreement. Self-diffusion includes the Yeh--Hummer finite-size correction evaluated with the experimental shear viscosity~\cite{yeh2004system}. Mean relative errors and the uncertainties are defined in Tables~\ref{tab:bulk_density} and~\ref{tab:bulk_diffusion}. \textbf{(C)}~Single-ion hydration from five net-neutral salt boxes: ion--O radial distribution functions, dashed lines at the experimental first-shell distances, and the mean first-shell coordination number $\pm$ standard deviation in the legend. Per-ion values and references are in Table~\ref{tab:ion_hydration}}%
\label{fig:solvents}
\end{figure*}

\textbf{Enzymatic reactivity.} The third reaction involves a transition-metal cofactor in a solvated environment. Methionine synthase transfers the cobalt-bound methyl of vitamin B12 to a homocysteine thiolate (modeled here as methanethiolate), one of the few biological reactions involving a metal--carbon bond (shown in Fig.~\ref{fig:overview}C)~\cite{matthews2001methyltransferases}. SO3LR-T-L converges to a first-order saddle whose single imaginary mode ($-498$~cm$^{-1}$) is the backside methyl transfer along the \mbox{Co$\cdots$C$\cdots$S} axis, and an IRC connects it to methylcobalamin and the product (Sec.~\ref{sec:si-reactions}). The role of the axial base is captured as well: removing the imidazole raises the barrier by $6.2$~kcal/mol, consistent with the trans-base assistance ascribed to the enzyme. The barrier of $16.3$~kcal/mol compares with $18.0$~kcal/mol at the PBE0+MBD-NL level for the bare 63-atom complex, for which the original B3LYP/6-31G(d) study reported $4.9$~kcal/mol, likely due to an unconverged basis set~\cite{jensen2003meth} (Tab.~\ref{tab:si_react_cobalamin_recalc}). The forming \mbox{S$\cdots$C} distance at the saddle is $2.35$~\AA{}, within $0.05$~\AA{} of the implicit-solvent reference of $2.40$~\AA. 

Across all three reactions, SO3LR-T-L reproduces saddle geometries, barriers, and the correct stereochemical and mechanistic selectivity, with IRC-confirmed connectivity. However, only the aldol saddles were found blindly; benzoin and cobalamin were refined from published structures~\cite{huang2022benzoin, jensen2003meth}. This remains a stringent test, since the model must yield a first-order saddle with the correct imaginary mode.

\section{Molecular Dynamics}\label{sec:biomol}

Accuracy on fixed geometries does not necessarily translate to reliable dynamics, as errors can accumulate in long simulations of large systems~\cite{fu2022forces}. We therefore evaluate SO3LR-T-M under simulation conditions, on the densities and diffusion of organic liquids, the hydration of ions, protonation free energies in peptides, the structure and dynamics of biomolecules, and finally its computational performance.

\subsection{Liquid Properties}\label{sec:liquid-properties}

\textbf{Densities.} Liquid densities probe the joint effect of the broader training data and the PME treatment of electrostatics and dispersion. We simulated water and fifteen organic solvents of varied polarity and hydrogen-bonding character in four independent NPT replicas of $1$--$11$~ns at $300$~K and $1$~atm (Sec.~\ref{sec:si-neat}). SO3LR-T-M reproduces the experimental densities with a mean relative error (MRE) of $6.3\%$ (Fig.~\ref{fig:solvents}A), against $2.4\%$ for OPLS4 (Table~\ref{tab:bulk_density}). The MRE is computed from log ratios, so over- and underestimation count equally (Eq.~\ref{eq:si-mre}). We remark that a predictive MLFF should perform better than 6\% in liquid densities. Part of this overestimation may stem from nuclear quantum effects (NQE), which lower the densities of organic liquids by up to ${\sim}5\%$~\cite{ugur2025nuclear}. Future studies should analyze non-covalent interaction curves in detail and probe the role of NQEs in anharmonic MLFFs. 

\textbf{Diffusion coefficients.} Density is a static property that depends mainly on the average packing of the liquid and can be reproduced by a model with approximately correct average forces. Self-diffusion, a dynamical property, is more demanding, since it depends on the barriers between local configurations and on hydrogen-bond lifetimes, which are sensitive to small errors in the forces.

The diffusion coefficients are accordingly reproduced less tightly than the densities (Fig.~\ref{fig:solvents}B), with an MRE of $47\%$ for SO3LR-T-M (i.e. a deviation from experiment by a factor of $1.47$) against $28\%$ for OPLS4 and $50\%$ for GAFF (Table~\ref{tab:bulk_diffusion}). SO3LR-T-M and OPLS4 values include the Yeh--Hummer finite-size correction, GAFF values do not. %
In principle, one could finetune a model directly on the experimental densities and diffusion coefficients and reproduce both by construction. We take the view, however, that these properties should emerge from a correct description of the intermolecular interactions rather than being imposed as fitting targets.

\textbf{Water properties.} Water is present in nearly every biomolecular simulation and is also an instructive case, since our two reference functionals deviate from experiment in opposite directions. We observe that PBE0 over-structures the hydrogen-bond network into an ice-like liquid near $0.94$~g/cm$^3$, whereas $\omega$B97M-V under-structures it toward $1.1$--$1.2$~g/cm$^3$~\cite{batatia2026mace}; trained on both through the multi-head scheme, and with additional $\omega$B97M-V water added in the late fine-tuning stages to offset the residual PBE0 over-structuring, SO3LR-T-M gives a density of $0.990 \pm 0.006$~g/cm$^3$ (standard error over four replicas; Table~\ref{tab:bulk_density}). Some over-structuring remains, with the first oxygen--oxygen peak of the radial distribution function reaching $3.05$ against the experimental $2.58$ (Fig.~\ref{fig:water_rdf_levels}), and it also affects the dynamics, with water diffusing at $119$~\AA$^2$/ns against the experimental $230$, about half the measured value. We remark that nuclear quantum effects may affect the self-diffusion coefficient strongly by up to a factor of two across previous simulations~\cite{Berta2020}.

\textbf{Stability of water.} To test the stability of the model beyond ambient conditions, we ran $2$~ns simulations of water at six state points between $300$ and $673$~K, at $1$~bar up to $330$~K and at $25$~MPa above, covering the liquid, compressed and supercritical regimes (Fig.~\ref{fig:water_high_tp}). High temperatures and pressures can push MLFFs outside their training data. All simulations remained stable, and both temperature and density stabilized within the first $0.5$~ns. The structural response follows the expected qualitative trend, with the first $g_\text{OO}$ peak falling monotonically from $3.05$ to $1.85$ as the hydrogen-bond network is progressively weakened. The density agrees with the IAPWS-95 reference equation of state to within $1\%$ up to $400$~K, and deviates by $+3.1\%$ at $500$~K and $+12.4\%$ at $600$~K~\cite{wagner2002iapws}. Above the critical point, at $673$~K, the model retains a liquid-like density of $0.665$~g/cm$^3$ (IAPWS-95: $0.167$~g/cm$^3$) and does not reproduce the supercritical density drop.

The long-range contribution to the energy is much smaller than the local covalent and repulsive terms. Omitting it therefore has little effect on test set errors or observables derived from short MD trajectories. Even for bulk properties, a local model may give a reasonable density, because the short-range part of the potential can partly absorb the missing long-range interactions through compensating errors~\cite{esders2025analyzing}. With electrostatics and dispersion treated explicitly by PME, correct bulk properties are less likely to result from error cancellation.

\subsection{Ion Solution Properties}

Most biomolecular simulations contain ions. Nucleic acids, metalloproteins, and membrane surfaces all require the model to describe each ion and its hydration shell together with the surrounding water. We therefore simulated five net-neutral salt solutions, \ce{NaCl}, \ce{KCl}, \ce{CaCl2}, \ce{MgCl2}, and \ce{ZnCl2}, each a single cation with its charge-balancing \ce{Cl-} in roughly $900$ water molecules, for $2$~ns under NPT at $300$~K and $1$~atm (Sec.~\ref{sec:si-ions}).

First-shell geometry is reproduced for all ions (Fig.~\ref{fig:solvents}C). The cation--oxygen peak positions lie within the experimental range for all ions except \ce{Mg^2+} and \ce{Zn^2+}, which deviate by $0.02$--$0.04$~\AA{}. The mean coordination numbers agree with experiment for all ions (Tab.~\ref{tab:ion_hydration}). The \ce{Cl-}--O distributions from the five simulations with different cations overlap, indicating that the hydration of \ce{Cl-} is not affected by the counter-cation.

The distribution of coordination numbers provides further insight (Fig.~\ref{fig:solvents}C, legend). \ce{Mg^2+} and \ce{Zn^2+} retain a rigid six-fold shell, with essentially no water exchange during the simulations. In contrast, \ce{Ca^2+} and \ce{Na+} sample several coordination states, and \ce{K+} shows the widest distribution. The resulting order of shell rigidity, $\ce{Mg^2+} \approx \ce{Zn^2+} > \ce{Ca^2+} \approx \ce{Na+} > \ce{K+}$, agrees with the measured water-exchange rates, which correspond to residence times ranging from microseconds for \ce{Mg^2+} to tens of picoseconds for \ce{K+}~\cite{helm2005}.

\subsection{Proton transfer in aqueous solution}

Proton transfer between biomolecules and water is ubiquitous and shapes protein folding, enzymatic catalysis and protein--ligand binding. Its thermodynamics is quantified by the acidity constant p$K_\mathrm{a}$. To compute it, an MLFF must describe both the transfer of the proton to a neighbouring water molecule and the solvation of the released proton in bulk water. We computed the side-chain p$K_\mathrm{a}$ values of alanine-based pentapeptides with one titratable residue each (Asp, Glu, His, Cys, Tyr, Lys or Arg) and of the C- and N-terminal groups of penta-alanine, using SO3LR-T-M with adaptive biasing force sampling in explicit water via PySAGES~\cite{zubieta2024pysages, darve2008adaptive, xu2026quantitative}, in four independent 8--12~ns replicas per group (Sec.~\ref{sec:si-pka}).

\begin{figure}[t]
  \centering
  \includegraphics[width=0.36\textwidth]{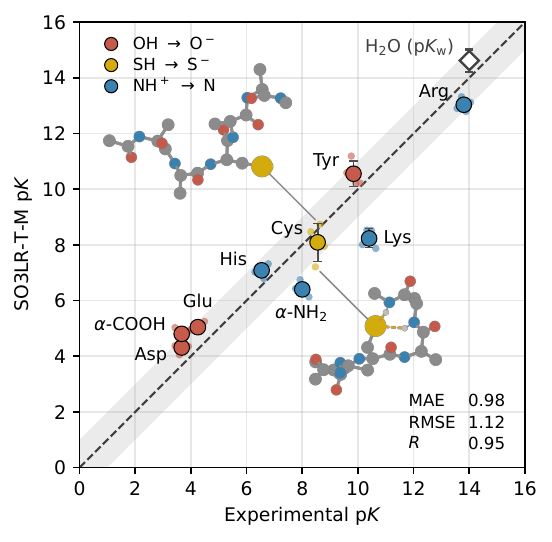}
  \caption{\textbf{Side-chain p$K_\mathrm{a}$ values from SO3LR-T-M.} Comparison with experiment~\cite{thurlkill2006pk, Fitch2015}: means and standard deviations over replicas, with individual replicas as small dots (Table~\ref{tab:pk_replicas}). The shaded band marks $\pm 1$ p$K$ unit; the open diamond is p$K_\mathrm{w}$. Insets show representative frames of the two Cys thiolate conformations, with S$^-$ H-bonded by water only (upper; waters and hydrogens are omitted) or by the backbone N--H (lower)} %
  \label{fig:pka}
\end{figure} 

\textbf{Acidity of peptides.} Over the nine titratable groups, SO3LR-T-M reproduces the experimental p$K_\mathrm{a}$ values with an MAE of 0.98 p$K$ units, or 1.35~kcal/mol in deprotonation free energy (RMSE 1.12, $R = 0.95$; Fig.~\ref{fig:pka})~\cite{thurlkill2006pk, Fitch2015}. For the four groups that release the proton from an oxygen atom (the carboxylic acids of Asp, Glu and the C-terminus, and the phenol of Tyr), the p$K_\mathrm{a}$ is overestimated by 0.6--1.1 units, which suggests a systematic offset for O--H deprotonation. The largest outliers are Lys and the N-terminus, both primary amines, whose p$K_\mathrm{a}$ values are underestimated by 2.2 and 1.6 units (8.2 vs.\ 10.4 and 6.4 vs.\ 8.0).

\textbf{Autoionization of water.} To find the origin of these deviations, we also computed p$K_\mathrm{w}$ (based on p$K_\mathrm{a}$ and p$K_\mathrm{b}$ of Lys), which depends on how hydronium (\ce{H3O+}) and hydroxide (\ce{OH-}) are described in bulk water. SO3LR-T-M gives p$K_\mathrm{w} = 14.6 \pm 0.4$, close to the experimental 14.0. The upward shift of 0.6 units is similar to the offset of the O--H acids and suggests a slight under-stabilization of the solvated hydronium. For Lys and the N-terminus, this means that the deviation comes from the balance between the amine and the ammonium group.

\textbf{Conformation-dependent acidity of Cys.} The independent runs of most residues agree to within 0.4--0.7 units (Table~\ref{tab:pk_replicas}). However, the Cys runs differ by up to 1.5 units. We found that this difference arises from the local environment of the thiolate. In parts of the simulations, the C-terminus folds into a helical turn and two backbone N--H groups donate hydrogen bonds to the thiolate (Fig.~\ref{fig:pka}, insets), which lowers the p$K_\mathrm{a}$ by $\sim$2.6 units compared to the thiolate hydrated by water only (6.7 vs.\ $9.3 \pm 0.4$; Sec.~\ref{sec:si-pka}). This agrees with earlier experiments and simulations, where hydrogen bonds from backbone amides lowered the Cys p$K_\mathrm{a}$~\cite{kortemme1995ionisation, roos2013understanding}.

\subsection{Biomolecular dynamics}

\textbf{Reproducing the SO3LR systems.} We re-ran the systems used to characterize SO3LR, a polyalanine helix, solvated crambin, a POPC bilayer, and a glycoprotein, to confirm that the larger corpus and refined modules maintain or improve performance in the production regime (details are given in Sec.~\ref{sec:md-biomol}). In short, a gas-phase Ala$_{15}$ helix at $300$~K samples a $68{:}32$ mixture of $\alpha$- and $3_{10}$-helix among its helical residues over three $30$-ns trajectories (Fig.~\ref{fig:ala15-ss}). Solvated crambin maintains its native fold across four replicas, with a heavy-atom RMSD of $1.10 \pm 0.22$~\AA{} and a radius of gyration stable to $1\%$, and backbone flexibility indistinguishable from SO3LR (Fig.~\ref{fig:si-crambin}). The 1K7C glycoprotein retains its fold while its glycans stay mobile, giving the same result under SO3LR and SO3LR-T-M (protein RMSD ${\sim}1.3$~\AA, carbohydrate ${\sim}2.4$~\AA; Fig.~\ref{fig:glyco-so3lrm}).

The POPC bilayer is a particularly demanding test, since its structure depends on both the dispersion-driven packing of the lipid tails and the electrostatic and hydrogen-bonding interactions of the headgroups with water. In isotropic NPT at $303$~K, SO3LR-T-M reproduces the head-to-head thickness ($D_{HH} = 37.5 \pm 0.8$~\AA{} against an experimental $36.5$~\AA) and the acyl-chain deuterium order parameters along both chains (combined RMSD $0.022$), including the characteristic dip at the \textit{sn}-2 double bond (Fig.~\ref{fig:popc-so3lrm}). Predicting the tensionless area per lipid in a semi-isotropic ensemble remains challenging, as it is highly sensitive to the delicate balance of lateral pressures across the bilayer.

\begin{figure*}[t]
\centering
\includegraphics[width=\linewidth]{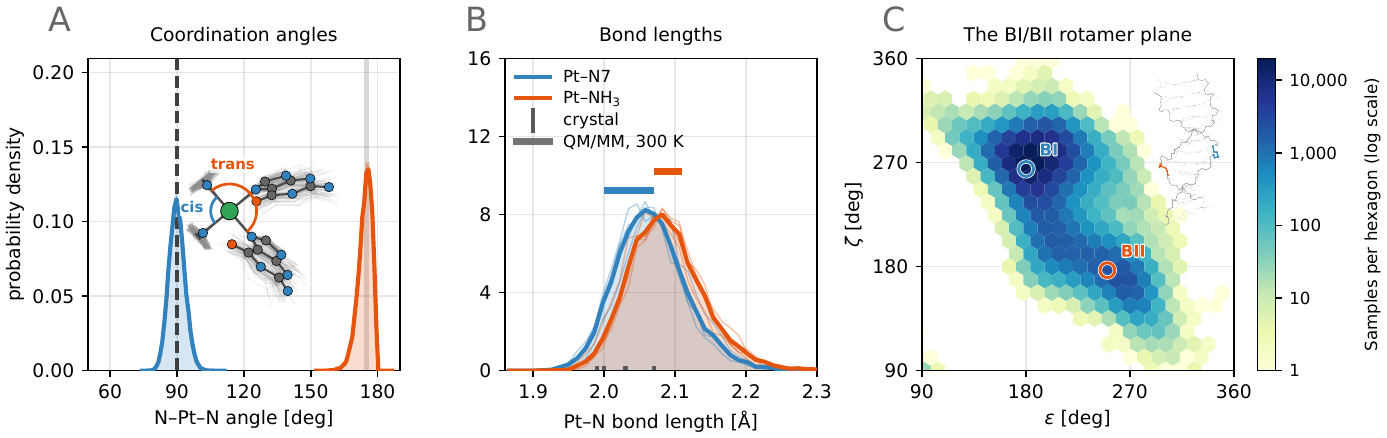}
\caption{\textbf{Platinum--DNA crosslink and B-DNA.} \textbf{(A)}~Cis and trans N--Pt--N angles at the cis-platin 1,2-d(GpG) crosslink (PDB: 3LPV) compared with the deposited crystal values (dashed line, shaded band). Inset, 60 frames superposed on the PtN$_4$ core, with one cis and one trans angle marked. 
\textbf{(B)}~Pt--N bond lengths by donor type compared with the deposited crystal values (ticks) and with the range of QM/MM Car--Parrinello MD averages over three models at 300~K (bars)~\cite{spiegel2004cisplatin}.
\textbf{(C)}~Backbone rotamer plane of the Drew--Dickerson dodecamer as a log-density hexbin. Open circles mark representative BI and BII conformations. Inset, a trajectory snapshot with BI (blue) and BII (orange) backbone segments highlighted on opposite strands of the duplex (grey). More details are given in Figs.~\ref{fig:si_cis} and \ref{fig:si_ddd}}\vspace{-2mm}
\label{fig:main3}
\end{figure*}

\textbf{Cis-platin crosslink and B-DNA duplex.} Metal-containing biomolecules remain challenging for classical force fields, which often need separate parameters for each coordination environment. For example, platinum parameters are derived one complex at a time~\cite{mcpb2016}, and the bonded model fixes the coordination geometry with improper torsions whose force constants require dedicated quantum-chemical calculations~\cite{li2017metal}. Hence, neither the coordination number nor the charge distribution can respond to the environment. We therefore ran the cis-platin 1,2-d(GpG) crosslink~\cite{pt_dna_3lpv}, a Pt(II) centre bridging two guanines inside a solvated polyelectrolyte with its own condensed counterion atmosphere, alongside an unmodified Drew--Dickerson dodecamer~\cite{drew1981}, with one pretrained model. Both were run in explicit solvent under NPT at $300$~K, in four replicas each ($10$~ns in total for the crosslink and $15$~ns for the dodecamer).

The coordination sphere is preserved throughout the simulation. The Pt(II) site stays four-coordinated in every frame of every replica, with no ligand exchange and a count independent of the cutoff, at angles consistent with the deposited geometry ($90.0^\circ$ and $174.4^\circ$; Fig.~\ref{fig:main3}A) and Pt--N bond lengths within $0.1$~\AA{} of crystallographic averages over cis-platin--d(GpG) structures~\cite{gkionis2013} (Fig.~\ref{fig:main3}B). We also benchmarked the site against PBE0+MBD-NL, where the forces correlate at $r = 0.99$ with a mean absolute error of $0.10$~eV~\AA$^{-1}$, and $0.27$~eV~\AA$^{-1}$ on platinum (Fig.~\ref{fig:si_cis}). No restraints or improper torsions were imposed in our simulations, so square-planar coordination emerges from the MLFF model.

The Drew--Dickerson dodecamer is a canonical B-DNA motif and a long-standing benchmark for the development and testing of nucleic-acid force fields~\cite{perez2007dynamics}. We find that SO3LR-T-M maintains the overall B-form structure of the duplex over the simulated timescale. The mean helical twist is $33.4^\circ$, close to the crystallographic reference of $33.6^\circ$. The listed backbone and glycosidic torsions differ from the crystallographic values by less than $8^\circ$, with overall agreement comparable to that of the dedicated DNA force fields (Table~\ref{tab:ddd_table3}). Notably, the backbone samples both rotameric substates, BI and BII, as two distinct basins in $\varepsilon$--$\zeta$ (Fig.~\ref{fig:main3}C). Determining their exact populations and interconversion kinetics would require longer sampling than the few nanoseconds run here. The sugar pucker, however, is sampled well enough to show a clear deficiency. Its distribution has an excess population near O4$'$-endo, where reference free-energy profiles indicate a barrier. 

\begin{figure}[h]
  \centering
  \includegraphics[width=0.49\textwidth]{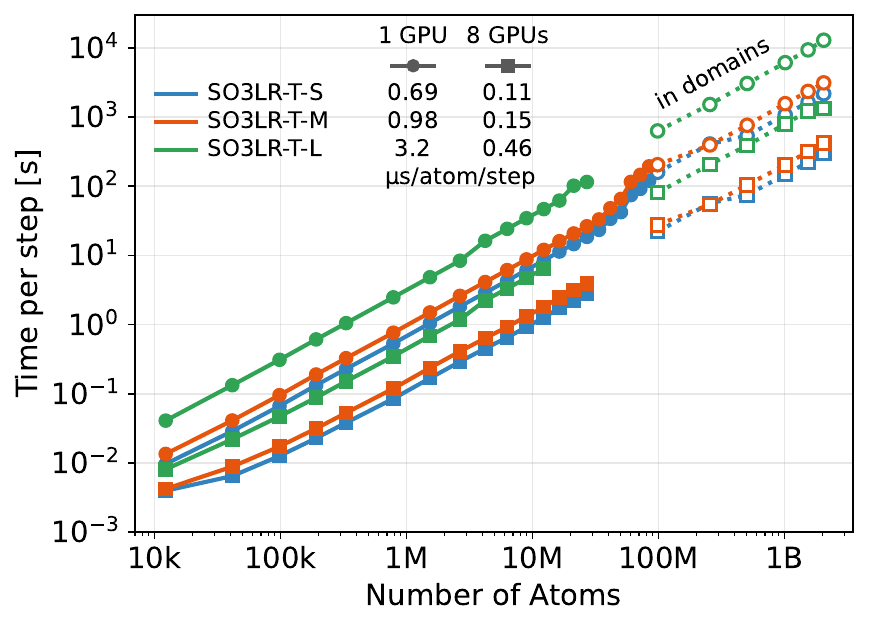}
  \caption{\textbf{Computational performance.} Time per MD step versus system size for SO3LR-T-S/M/L on periodic liquid water, measured on AMD MI355X GPUs (\qty{288}{GB}), on one (circles) or eight (squares) GPUs. Dotted lines denote the forces evaluated one domain at a time (one force call per point, with only real-space long-range interactions)}\vspace{-5mm}
  \label{fig:so3lr-single-gpu-scaling}
\end{figure} 

\subsection{Computational Performance} 

Biomolecular systems range from a solvated protein with $\sim$$10$k atoms, a membrane protein with $\sim$$100$k atoms up to a whole minimal cell, which would correspond to more than $6$B atoms in an all-atom model~\cite{stevens2023molecular}. We measured the cost of SO3LR Tauri models on periodic liquid water on AMD MI355X GPUs (\qty{288}{GB}) in NVT with PME for electrostatics and dispersion (Fig.~\ref{fig:so3lr-single-gpu-scaling}).

The cost scales linearly with the number of atoms, at $0.69$, $0.98$, and $3.2$~\si{\micro\second}/atom/step on one GPU for SO3LR-T-S, -M, and -L (legacy SO3LR matches -M). For a $50$k-atom system this corresponds to $2.5$, $1.8$, and $0.54$~ns/day at a $1$~fs time step, and one GPU holds up to $84$M atoms with SO3LR-T-S/M and $27$M with -L. The evaluation can also be split over the GPUs of a node: from $100$k atoms on, eight GPUs are $6$--$7$ times faster than one, at $0.11$, $0.15$, and $0.46$~\si{\micro\second}/atom/step.

Larger assemblies require domain decomposition, which has been demonstrated across GPUs for JAX-based MLFFs at the million-atom scale~\cite{fuchs2025chemtrain}. Within our implementation, a single GPU computed the forces on $2$B atoms at $1.1$, $1.5$, and $6.1$~\si{\micro\second}/atom, or \qty{52}{min} per force call with SO3LR-T-M; eight GPUs are $7$--$8$ times faster. The domain evaluation is so far limited to force evaluations with real-space long-range interactions. These sizes approach that of a minimal cell. Nevertheless, systems beyond $\sim$$10$--$100$M atoms remain out of reach of brute-force MD sampling; they are, however, within reach of geometry optimization and of generative models that could propose configurations to be evaluated~\cite{roney2022state, lewis2025scalable, diez2026transferable}.

This performance comes from improvements to the inference code. The forces are computed in fused GPU kernels that take the atomic positions and accumulate the interactions of each atom directly, with dedicated backward kernels instead of automatic differentiation, so that no per-edge arrays are stored; the neighbors, including the long-range pairs, are found on a cell grid at every call instead of being kept in a neighbor list, and the learned radial functions are tabulated~\cite{lysogorskiy2021performant, fu2026train}. Compared with the neighbor-list implementation, the new one needs $140$--$430$ times less memory per atom for SO3LR-T-M ($9.4$~KB in the fastest mode and $3.1$~KB in the most economical one, against $1.3$~MB) and runs $3.7$--$4.0$ times faster. For the largest systems, memory modes recompute intermediate arrays instead of storing them. Implementation details, results on NVIDIA A100 GPUs, and numerical precision checks are given in Sec.~\ref{sec:si-perf}.

\vspace{-2mm}
\section{Conclusion}
\vspace{-1mm}

SO3LR Tauri is a step toward a single force field for biomolecular matter at near-quantum-mechanical accuracy. It couples a fast SO3krates backbone to explicit repulsion, electrostatics, and dispersion, and is trained on $148$ million structures over $85$ elements. The applications cover proteins, nucleic acids, lipids, carbohydrates, water and organic liquids, ions, protonation equilibria, metal-containing biomolecules and frameworks, and reactive pathways, without a separate parametrization for each system. Released in three sizes with open code and data, the models simulate tens of millions of atoms on a single GPU at $0.7$--$3.2$~\si{\micro\second}/atom.

We probed the models on a broad range of systems to expose limitations that may guide future work. The discrepancies in liquid transport require an understanding of how learned interactions determine properties that are not fitted during training, such as diffusion coefficients, conformational populations, and free energies. We currently treat long-range interactions as pairwise electrostatics and dispersion, ignoring anisotropic many-body effects beyond the receptive field~\cite{kabylda2026atoms}. Closing this gap will require architectures that represent delocalized collective modes, or explicit non-local many-body physics coupled to learned local terms. Nuclear quantum effects remain to be assessed, particularly for water, and quantitative reaction kinetics will require barriers closer to chemical accuracy and training data richer in transition-state regions. More broadly, long simulations with uncertainty estimates can reveal the unfilled gaps in (bio)molecular chemical space and guide additional reference calculations, including beyond DFT.

Reaching the timescales of protein folding, nucleic acid, membrane dynamics, and other slow conformational processes will require faster inference and more efficient sampling. Multiscale descriptions of interactions~\cite{thurlemann2026amp} and scalable, hardware-aware implementations offer routes to faster inference. Coupling SO3LR Tauri with generative models is a particularly promising route to more efficient sampling~\cite{roney2022state, lewis2025scalable, diez2026transferable}, combining the ability of generative models to sample conformational space without femtosecond time integration with the physical accuracy of a quantum-mechanically trained force field. 

\vspace{-2mm}
\section{Data, model and code availability}
\vspace{-1mm}
The three model sizes and the full inference and training code are freely available at \url{https://github.com/general-molecular-simulations/so3lr}. The repository contains a convenient command line interface, documentation, and tutorials. The quantum mechanical datasets used for training and testing SO3LR Tauri are available at \url{https://doi.org/10.5281/zenodo.22920714}. 

\vspace{-2mm}
\section{Supporting Information}
\vspace{-1mm}
Details of the SO3krates architecture changes, physical interaction modules, and model configurations; training datasets, sampling weights, loss functions, stage-by-stage training configurations, learning curves, and training cost; test-set errors, external benchmark comparisons, and metal--organic framework benchmarks; transition-state searches and validation, reference calculations, and a comparison of SO3LR-T-L with SO3LR-T-M; simulation protocols and additional analyses for neat liquids, water, and aqueous ions; free-energy calculations of peptide p$K_\mathrm{a}$ values; protocols and analyses for proteins, a glycoprotein, a lipid membrane, and nucleic acids; and computational performance on one and several AMD MI355X and NVIDIA A100 GPUs.

\vspace{-2mm}
\section{Acknowledgements}
\vspace{-1mm}
We thank Winfried Ripken (TU Berlin) for his help with the parallel TFDS dataloader, Mit Kotak (MIT) for sharing the legacy SO3LR minimal inference code, and Pranathi Vasireddy (AMD) for running the scaling benchmarks on MI355X GPUs. Claude Opus 5.5 assisted with profiling and optimizing the GPU inference code, with preparing the code and dataset releases, and with preparing the Supporting Information. The training and simulations were performed on the Luxembourg national supercomputer MeluXina.

A.K. acknowledges financial support from the Luxembourg National Research Fund (FNR AFR Ph.D. Grant 15720828). A.K., E.M. and A.T. acknowledge the European Research Council under ERC-AdG grant FITMOL (101054629). A.T. acknowledges the Luxembourg National Research Fund under grant FNR-CORE MBD-in-BMD (18093472). M.G. acknowledges support from the European Union’s Horizon Europe Marie Skłodowska-Curie Actions (MSCA) Postdoctoral Fellowship under Grant Agreement No. 101202630. K.R.M.\ was partly funded by the German Ministry for Education and Research (BMBF) as BIFOLD – Berlin Institute for the Foundations of Learning and Data - under Grants 01IS14013A-E, 01GQ1115, 01GQ0850, 01IS18025A, 031L0207D, and 01IS18037A. K.R.M.\ was also supported by the Institute of Information \& communications Technology Planning \& Evaluation (IITP) grants funded by the Korea government (MSIT) (No.\ RS-2019-II190079, Artificial Intelligence Graduate School Program of Korea University and No.\ RS-2024-00457882, AI Research Hub Project), and also by DFG and HFA. K.R.M.  was furthermore supported in part by the Korea University Grant.

\section*{Contributions}
A.K. and A.T. conceived the project. A.K. assembled the training data, implemented and trained the models, developed the GPU inference code and the PME treatment of dispersion, and designed, performed, and analyzed the benchmarks and simulations. F.N.B. implemented PME electrostatics and performed the neat-liquid simulations. M.G. carried out the reaction studies. E.M. implemented the interface to PySAGES and conducted the p$K_\mathrm{a}$ calculations. J.T.F. implemented the multi-head approach. O.T.U. and S.C. contributed to scientific discussions. A.T. and K.R.M. supervised the project. A.K. wrote the manuscript with contributions from all authors. Correspondence should be addressed to A.K., K.R.M., and A.T.

\twocolumngrid
\bibliography{references}

\clearpage
\onecolumngrid
\setcounter{secnumdepth}{3}
\setcounter{section}{0}
\setcounter{subsection}{0}
\setcounter{figure}{0}
\setcounter{table}{0}
\setcounter{equation}{0}
\renewcommand{\thesection}{S\arabic{section}}
\renewcommand{\thesubsection}{\thesection.\arabic{subsection}}
\renewcommand{\thefigure}{S\arabic{figure}}
\renewcommand{\thetable}{S\arabic{table}}
\renewcommand{\theequation}{S\arabic{equation}}
\renewcommand{\theHequation}{S\arabic{equation}}
\makeatletter
\renewcommand{\addcontentsline}[3]{%
  \def\si@lvl{#2}\def\si@sec{section}\def\si@sub{subsection}%
  \ifx\si@lvl\si@sec\si@tocwrite{#2}{#3}\else\ifx\si@lvl\si@sub\si@tocwrite{#2}{#3}\fi\fi}
\newcommand{\si@tocwrite}[2]{%
  \addtocontents{sitoc}{\protect\contentsline{#1}{#2}{\thepage}{\@currentHref}\protected@file@percent}}
\renewcommand{\tableofcontents}{\@starttoc{sitoc}}
\renewcommand{\p@subsection}{}
\renewcommand{\p@subsubsection}{}
\makeatother
\setcounter{tocdepth}{2}

\input{si}

\end{document}

%% file: si.tex
\section*{Supplementary Information}

This Supplementary Information provides the details underlying the model, its training, and the simulations reported in the main text. Section~\ref{sec:architecture} describes the modifications to the SO3krates backbone and the physical interaction modules. Section~\ref{sec:training} covers the training corpus, sampling weights, loss function, five-stage schedule, and computational cost. Section~\ref{sec:static-benchmarks} reports the per-subset test-set errors, along with the protocols and full results tables for the external static benchmarks. Section~\ref{sec:si-reactions} presents the transition-state searches, Section~\ref{sec:liquids} the liquid and ion simulations, Section~\ref{sec:si-pka} the pK$_a$ calculations, Section~\ref{sec:md-biomol} the biomolecular dynamics results, and Section~\ref{sec:si-perf} the GPU timing protocol.

{\small\tableofcontents}

\clearpage

\section{Model architecture}\label{sec:architecture}

\textbf{Model family.} SO3LR-Tauri-S/M/L keeps the SO3krates backbone of SO3LR: the equivariant message-passing trunk with invariant and equivariant feature channels, the Bernstein radial basis, and the polynomial cutoff function (see Refs.~\citenum{frank2024euclidean, kabylda2025molecular} for more details).  Main hyperparameters of the released models are listed in Tab.~\ref{tab:si_model_configs}. As in SO3LR, this backbone is coupled to erf-damped, shifted-force real-space electrostatics driven by predicted partial charges and to quantum-Drude-oscillator dispersion with $C_6/C_8/C_{10}$ damping~\cite{Khabibrakhmanov2023}. Beyond the changes described in the main text, three modifications to the interaction layers were introduced to improve training stability:

(i) \emph{Normalization.} SO3LR applied two learnable LayerNorms per interaction layer, one after the attention residual and one at the layer output. SO3LR-T keeps a single normalization, placed after the attention residual, and replaces LayerNorm by a parameter-free root-mean-square normalization, $\mathrm{RMS}(\mathbf{x}) = \mathbf{x}\,/\,\sqrt{\langle x_c^2\rangle_c + \epsilon}$ with $\epsilon = 10^{-6}$. Dropping the mean subtraction avoids coupling the invariant channels through their mean, and dropping the learnable per-channel gain and bias removes $2H$ parameters per normalization that the following dense layer can absorb in any case.

(ii) \emph{QK normalization.} The per-head queries and keys are RMS-normalized immediately after their linear projections and before the bilinear attention contraction,
\begin{equation}
  \alpha^{(h)}_{ij} = f_\mathrm{cut}(r_{ij})\,\frac{1}{\bar{n}} \sum_{c} \mathrm{RMS}\!\left(\mathbf{q}^{(h)}_i\right)_c\, w^{(h)}_{ij,c}\, \mathrm{RMS}\!\left(\mathbf{k}^{(h)}_j\right)_c ,
  \label{eq:si-qknorm}
\end{equation}
and identically for the equivariant branch. The attention logits are then bounded by the radial--spherical filter $w_{ij}$ alone, independently of the magnitude of the node features.

(iii) \emph{Residual scalars.} Each layer $l$ rescales the residual stream and re-injects the initial embedding $\mathbf{x}^{(0)}$ through two learnable scalars,
\begin{equation}
  \mathbf{x} \;\leftarrow\; \lambda^{(l)}_\mathrm{res}\, \mathbf{x} \;+\; \lambda^{(l)}_{0}\, \mathbf{x}^{(0)},
  \label{eq:si-residual}
\end{equation}
initialized to $\lambda_\mathrm{res} = 1$ and $\lambda_0 = 0.1$, following the per-layer residual scaling of nanochat~\cite{nanochat}. The attention weights of Eq.~\eqref{eq:si-qknorm} are invariant under $\mathbf{x} \to c\,\mathbf{x}$, so the layer normalizes its input again, only the ratio $\lambda^{(l)}_0/\lambda^{(l)}_\mathrm{res}$ is meaningful, not the overall scale of the two scalars. After training, this ratio has shifted the same way in all three models: every layer after the first places $77$--$95\,\%$ of its input weight on $\mathbf{x}^{(0)}$. This weighting suggests that the trained networks rely mainly on a direct connection from the atomic embedding to each layer. The actual contribution of each path also depends on the norms of $\mathbf{x}$ and $\mathbf{x}^{(0)}$.

\textbf{Output heads.} The energy readout provides several theory-level channels, and a one-hot mask selects the channel $t$ assigned to each structure,
\begin{equation}
    \varepsilon_{i,t} = s_{Z_i,t}\,\mathrm{MLP}_t(x_i) + b_{Z_i,t},
  \label{eq:si-multihead}
\end{equation}
with a per-element scale $s_{Z,t}$ and offset $b_{Z,t}$ for each head, so that in stages~1 and~2 each head is trained only on data from its own level of theory, which is what allows data of different provenance to be combined. From stage~3 on, a production head, initialized from the PBE0+MBD(-NL) head, is trained and used at inference (Sec.~\ref{sec:training-schedule}).

\textbf{Short-range repulsion.} The $10$-parameter learnable ZBL-type screening function of SO3LR is replaced by the fixed, element-pair-resolved NLH screened-nuclear-repulsion tables~\cite{nordlund2025repulsive},
\begin{equation}
  E^\mathrm{rep}_{ij} = \tfrac{1}{2}\, w(r_{ij})\, f_\mathrm{cut}(r_{ij})\, \frac{k_\mathrm{e} Z_i Z_j}{r_{ij}} \sum_{k=1}^{3} a^{(ij)}_{k}\, e^{-b^{(ij)}_{k} r_{ij}},
  \label{eq:si-nlh}
\end{equation}
with $a^{(ij)}_k, b^{(ij)}_k$ tabulated for all element pairs up to $Z = 92$ (SO3LR Tauri models cover $Z = 1$--$85$) and $w$ a smooth switching function that turns the term off beyond $1.5$~\AA. The repulsion therefore carries no trainable parameters and reproduces the correct nuclear-repulsion limit for every element pair.

\textbf{Behaviour at close contact.} At distances shorter than any chemical bond, the per-atom heads operate outside their training regime and can produce unphysical contributions that exceed the repulsion. To keep the model well behaved, the interatomic distance passed to the radial basis is floored for each element pair,
\begin{equation}
  \tilde d_{ij} =
  \begin{cases}
    d_{ij}, & d_{ij} \ge r_\mathrm{min} + w,\\[2pt]
    (r_\mathrm{min}+w) - w\,\tanh\!\dfrac{r_\mathrm{min}+w-d_{ij}}{w}, & d_{ij} < r_\mathrm{min} + w,
  \end{cases}
  \label{eq:si-floor}
\end{equation}
The floored distance $\tilde d_{ij}$ equals $d_{ij}$ above $r_\mathrm{min}+w$ and approaches $r_\mathrm{min}$ smoothly below it; the transition is continuously differentiable and has width $w = 0.05$~\AA. The onset $r_\mathrm{min}$ is set by each pair's own repulsion, as the distance at which the tabulated NLH term reaches $5$~eV; it therefore needs no per-pair tuning and moves outward with atomic number. The floor acts only on the descriptor. The repulsion and electrostatics use the true distance, so the NLH repulsion of Eq.~\eqref{eq:si-nlh} determines the energy at contact, while the learned quantities are kept out of their divergent regime.

\textbf{Electrostatics.} The unconstrained per-atom charges $\tilde q_i$ predicted by the network are corrected to the total charge $Q$ by distributing the excess equally over all $N$ atoms~\cite{unke2019physnet},
\begin{equation}
  q_i = \tilde q_i + \frac{1}{N}\Big(Q - \sum_{j=1}^{N} \tilde q_j\Big).
  \label{eq:charge-cons}
\end{equation}
The charges enter an erf-damped Coulomb sum, evaluated by particle-mesh Ewald for periodic systems,
\begin{equation}
  E_\mathrm{elec} = \sum_{i<j} k_e\, q_i q_j\, \frac{\operatorname{erf}(r_{ij}/\sigma)}{r_{ij}} \;+\; E^\mathrm{rec}_\mathrm{elec},
  \label{eq:si-elec}
\end{equation}
with smearing $\sigma$ and, for periodic systems, reciprocal-space term $E^\mathrm{rec}_\mathrm{elec}$. The real-space term is energy- and force-shifted so that it vanishes at the cutoff, as in the non-periodic case; the shift is not applied to the reciprocal-space part.

\textbf{Dispersion.} SO3LR predicted a single Hirshfeld volume ratio $v_i$ and set $\alpha_i = \alpha^\mathrm{free}_{z_i} v_i$ and $C_{6,i} = C^\mathrm{free}_{6,z_i} v_i^2$. SO3LR-T instead predicts a polarizability ratio $h_i$ and a $C_6$ ratio $r^{C_6}_i$ with two independent heads,
\begin{equation}
  \alpha_i = \alpha^\mathrm{free}_{z_i}\, h_i, \qquad C_{6,i} = C^\mathrm{free}_{6,z_i}\, r^{C_6}_i,
  \label{eq:si-cpratios}
\end{equation}
which removes the assumption that both scale with the same learned quantity. Pairwise coefficients follow from the Casimir--Polder mixing rules
\begin{equation}
  \alpha_{ij} = \tfrac{1}{2}\,(\alpha_i+\alpha_j), \qquad
  C_{6,ij} = \frac{2\,C_{6,i}\,C_{6,j}\,\alpha_i\,\alpha_j}{\alpha_i^2\, C_{6,j} + \alpha_j^2\, C_{6,i}}.
  \label{eq:si-cpmixing}
\end{equation}

\begin{table*}[t]
    \centering
    \caption{\textbf{Hyperparameters of the released models.} SO3LR (the original model) and SO3LR-T-S/M/L differ in the feature width $H$, the number of message-passing layers $T$, the number of attention heads $h$, the maximal degree $L$, the size of the Bernstein radial basis $k$, and the short-range cutoff $r_\mathrm{c}$, with $\bar{n}_\mathrm{nb}$ the average neighbor count used to normalize message passing (a configuration constant) and parameter counts measured from the released checkpoints. All remaining architectural choices are shared among SO3LR-T-S/M/L; the modifications relative to SO3LR are described in the text} 
    \label{tab:si_model_configs}
    \begin{tabular}{lcccccccc}
        \toprule
        \textbf{Model} & \textbf{Features ($H$)} & \textbf{Layers ($T$)} & \textbf{Heads ($h$)} & \textbf{Degrees ($L$)} & \textbf{Radial basis ($k$)} & \textbf{$r_\mathrm{c}$ (\unit{\angstrom})} & \textbf{$\bar{n}_\mathrm{nb}$} & \textbf{\#Params} \\
        \midrule
        SO3LR & $128$ & $3$ & $4$ & $1$--$4$            & $32$ & $4.5$ & $13.2$ & $530$k \\
        \midrule
        SO3LR-T-S & $128$ & $2$ & $4$ & $1$--$4$            & $32$ & $4.5$ & $16$ & $411$k \\
        SO3LR-T-M & $128$ & $3$ & $4$ & $1$--$4$            & $32$ & $4.5$ & $16$ & $532$k \\
        SO3LR-T-L & $256$ & $3$ & $8$ & $1$--$4$ ($\times2$) & $64$ & $5.0$ & $22$ & $1.786$M \\
        \bottomrule
    \end{tabular}
\end{table*}

The higher-order coefficients follow from the quantum-Drude-oscillator (QDO) relations $C_{8,ij} = 5\,C_{6,ij}/\gamma_{ij}$ and $C_{10,ij} = 245\,C_{6,ij}/(8\gamma_{ij}^2)$~\cite{Khabibrakhmanov2023}. The QDO parameter $\gamma_{ij}$ follows from $\alpha_{ij}$ (in atomic units) as
\begin{equation}
  \gamma_{ij} = \frac{1}{2\ell_{ij}^2}, \qquad
  \ell_{ij} = \sum_{k=0}^{3} b_k \big(R^\mathrm{vdW}_{ij}\big)^k, \qquad
  R^\mathrm{vdW}_{ij} = \alpha_\mathrm{fsc}^{-4/21}\,\alpha_{ij}^{1/7},
  \label{eq:si-gamma}
\end{equation}
where $\ell_{ij}$ is the width of the oscillator ground state, $R^\mathrm{vdW}_{ij}$ is the van der Waals radius of Ref.~\citenum{fedorov2018quantum}, and $\alpha_\mathrm{fsc}$ is the fine-structure constant. The cubic-fit coefficients are $b_0 = -0.00433008$, $b_1 = 0.24428889$, $b_2 = 0.04125273$, and $b_3 = -0.00078893$. The dispersion energy sums the three terms, each damped with a per-pair radius $R_{ij}\propto\alpha_{ij}^{1/7}$,
\begin{equation}
  E_\mathrm{disp} = -\sum_{i<j}\;\sum_{n=3}^{5} \frac{C_{2n,ij}}{\,r_{ij}^{2n} + R_{ij}^{2n}\,}.
  \label{eq:si-disp}
\end{equation}
For periodic systems, the leading $C_6$ term is additionally summed in reciprocal space by PME, using effective dispersion ``charges''
\begin{equation}
  q_i^{\text{disp}} = \sqrt{C_{6,\text{free}}[z_i]\,r_i^{C_6}},
  \label{eq:disp-pme}
\end{equation}
that map the $r^{-6}$ interaction onto the electrostatics implementation. This mapping is exact only for the geometric-mean combination $C_{6,ij}=q_i^{\text{disp}}q_j^{\text{disp}}$; the difference between the Casimir--Polder and geometric-mean coefficients is kept in the real-space sum, together with the short-ranged $C_8$ and $C_{10}$ terms.

\textbf{Charge and spin embedding.} The backbone is conditioned on the total charge $Q$ and spin multiplicity $S$, following the input convention of SO3LR. The charge also enters directly through the constraint of Eq.~\eqref{eq:charge-cons}. Because $Q$ and $S$ are supplied at inference, a single model can treat different protonation and oxidation states without separate training.

\clearpage

\section{Training}\label{sec:training}

SO3LR Tauri is trained in five stages (Table~\ref{tab:si_stages}). The model uses two neighbor lists: a short-range list at $r_\mathrm{c}$ and a long-range list at $r_\mathrm{lr}$. The long-range list is the more expensive to build, and extending it to every pair in the batch reduces throughput by 25--30\% (Fig.~\ref{fig:si_training_hardware}). We therefore keep the long-range cutoff at \qty{10}{\angstrom} in stages~1 and~2, which contain most of the optimizer steps, and extend it to all pairs from stage~3 onward. For each model size, the released weights are the stage-5 checkpoint.

\subsection{Training data}\label{sec:training-data}

The corpus combines seven sources computed at two levels of theory (Table~\ref{tab:Datasets}), split into the \num{52} subsets that the sampler treats independently (Table~\ref{tab:sampling}). QCML, QM7-X, GEMS, AQM and QCell are computed at PBE0+MBD(-NL). They supply energies and forces, all but GEMS also supply dipoles, and QCML and QCell add per-atom dispersion labels ($\alpha_0$ and $C_6$ ratios). OMol25-0 and OPoly26 are computed at $\omega$B97M-V level of theory, and we use only their forces.

\begin{table*}[h]
    \centering
    \caption{\textbf{Properties taken from each source.} Energies at the $\omega$B97M-V level are not used, and the per-atom dispersion labels ($\alpha_0$ and $C_6$ ratios) come only from PBE0+MBD-NL data. Sizes are the structures available in each source, before the sampling weights of Table~\ref{tab:sampling}. Targets a source does not provide are masked in the loss}
    \label{tab:Datasets}
    \setlength{\tabcolsep}{6pt}
    \renewcommand{\arraystretch}{1.1}
    \begin{tabular}{lccccccc}
    \toprule
    \textbf{Dataset} & \textbf{Size} & \textbf{Level of theory} & \textbf{Forces} & \textbf{Energies} & \textbf{Dipoles} & $\boldsymbol{\alpha_0}$ & $\boldsymbol{C_6}$ \\
    \midrule
    QCML    & 33.5M & PBE0+MBD-NL    & \cmark & \cmark & \cmark & \cmark & \cmark \\
    QCell   & 526k  & PBE0+MBD(-NL)  & \cmark & \cmark & \cmark & \cmark & \cmark \\
    QM7-X   & 4.2M    & PBE0+MBD       & \cmark & \cmark & \cmark & \xmark & \xmark \\
    AQM     & 60k   & PBE0+MBD       & \cmark & \cmark & \cmark & \xmark & \xmark \\
    GEMS    & 2.7M  & PBE0+MBD       & \cmark & \cmark & \xmark & \xmark & \xmark \\
    OMol25-0  & 100M  & $\omega$B97M-V & \cmark & \xmark & \xmark & \xmark & \xmark \\
    OPoly26 & 6.6M  & $\omega$B97M-V & \cmark & \xmark & \xmark & \xmark & \xmark \\
    \bottomrule
    \end{tabular}
\end{table*}

If the subsets were sampled at their natural rates, near-redundant conformers of small molecules would dominate the training stream. Six OMol25 subsets alone hold \num{56.8} million structures, each a conformer-rich collection of small molecules, and training on that distribution would bias the model toward the regime where transferable force fields already perform well. We therefore draw the subsets at the weights of Table~\ref{tab:sampling}, quoted relative to OPoly26. The six conformer-rich OMol25 subsets enter at $0.10\times$ and QCML with QM7-X at $0.68\times$, while the less abundant biomolecular subsets are upsampled: sugars, ions, AQM and GEMS bottom-up at $6.8\times$, DES at $12.5\times$, RNA, DNA, dipeptides and lipids at $13.5\times$, and the \num{12}k structures of GEMS top-down at $33.7\times$.

\begin{table*}[h]
\centering
\caption{\textbf{Sampling weights.} Subsets are grouped by the weight at which they are drawn, relative to OPoly26 ($1\times$). The \num{52} subsets are those of the stage-5 mixture (Table~\ref{tab:si_stages}): stage~1 uses \num{50}, and OPoly26 and the neutral-water subset are added later. Train $N$ is the number of structures available in a group, not the number drawn from it, so the two columns must be read together. 
\textsuperscript{a}~Conformer-rich subsets: ANI-1xBB, ANI-2x, OMol electrolytes, OMol metal-organics, GEOM-ORCA6, OrbNet-Denali.
\textsuperscript{b}~The remaining OMol25 subsets.
\textsuperscript{c}~Neutral water clusters, added in the final stage (Table~\ref{tab:si_stages})}
\label{tab:sampling}
\setlength{\tabcolsep}{6pt}
\renewcommand{\arraystretch}{1.1}
\begin{tabular}{lrrr}
\toprule
\textbf{Dataset group} & \textbf{Subsets} & \textbf{Train $N$} & \textbf{Sampling weight} \\
\midrule
  OMol25, conformer-rich\textsuperscript{a}            &  6 & 56.8M & $\times 0.10$ \\
  QCML $+$ QM7-X                                       &  2 & 36.8M & $\times 0.68$ \\
  OPoly26 $+$ OMol25, remainder\textsuperscript{b}     & 33 & 50.3M & $\times 0.99$ \\
  OMol25, neutral water\textsuperscript{c}             &  1 &  690k & $\times 3.02$ \\
  GEMS bottom-up, AQM, QCell sugars, QCell ions        &  4 &  2.8M & $\times 6.76$ \\
  QCell DES                                            &  1 &  393k & $\times 12.51$ \\
  QCell RNA, DNA, SPICE dipeptides, lipids             &  4 &   83k & $\times 13.48$ \\
  GEMS top-down                                        &  1 &   12k & $\times 33.73$ \\
\bottomrule
\end{tabular}
\end{table*}

\begin{figure*}[t]
    \centering
    \includegraphics[width=0.95\linewidth]{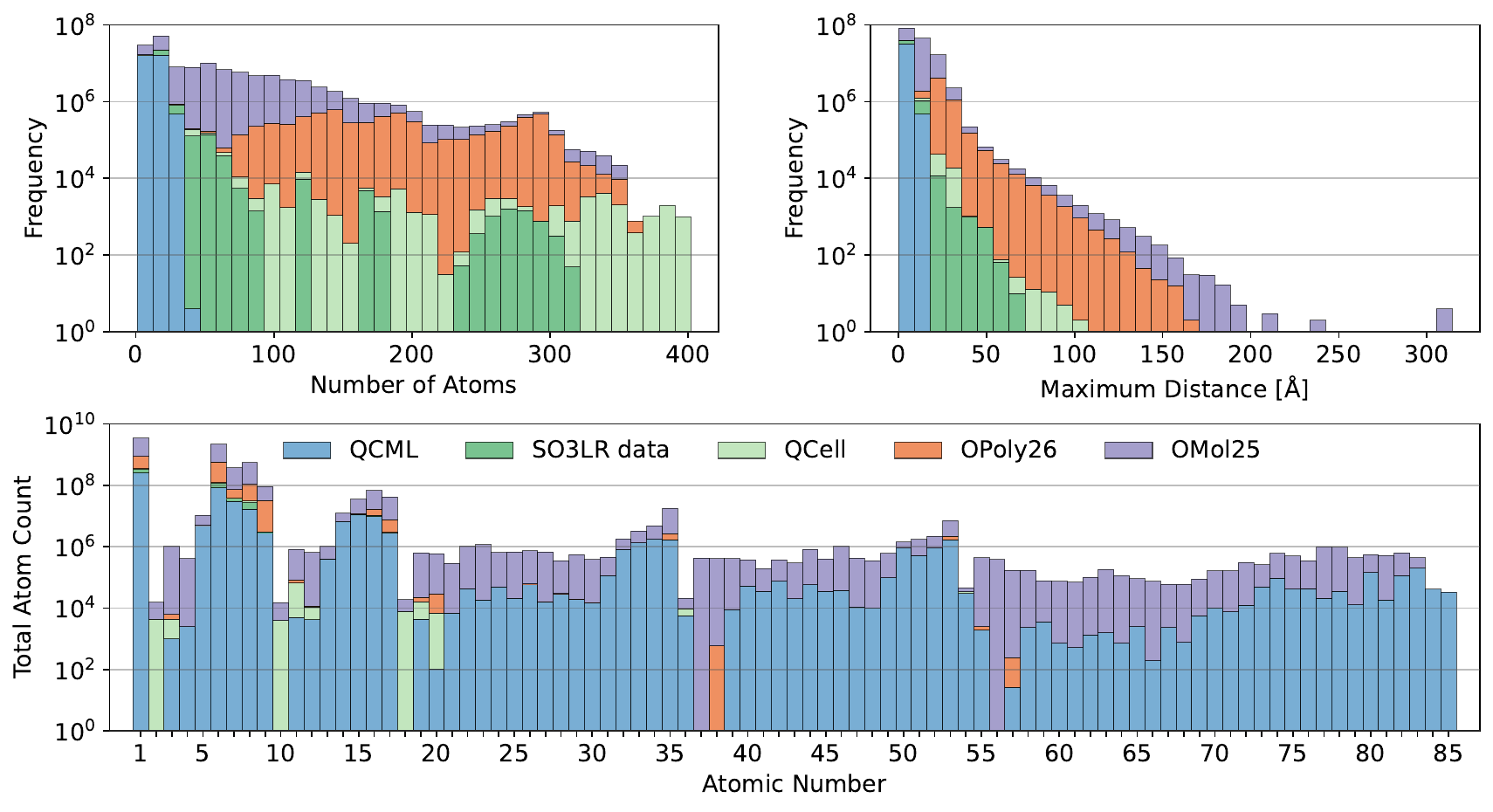}
    \caption{\textbf{Composition and structural diversity of the training corpus.} (top left) Molecule sizes, most structures holding fewer than \num{50} atoms with a tail reaching $\sim$\num{400}. (top right) Largest pairwise atomic distance within a structure, from small organic fragments below \qty{10}{\angstrom} to extended systems spanning several tens of \unit{\angstrom}. (bottom) Total atom count by atomic number, $Z = 1$ to $85$, coloured by source: QCML (blue), the SO3LR training data (dark green; QM7-X, GEMS, AQM, and SPICE dipeptides), QCell (light green; DES and the QCell sugar, lipid, nucleic-acid, and ion sets), OPoly26 (orange), and OMol25 (purple). Counts are the structures available in each source, before the sampling weights of Table~\ref{tab:sampling} are applied}
    \label{fig:dataset}
\end{figure*}

Coverage across the periodic table is uneven, and the sampling weights do not change this (Fig.~\ref{fig:dataset}). The light elements of biological matter dominate the atom counts, with a long sparse tail over heavier and metallic species. The sampling weights rebalance system sizes and interaction types, not elements.

\subsection{Streaming and batch composition}\label{sec:training-budgets}

During training the batches are assembled on the fly via the TFDS dataloader, which builds both neighbor lists, apply the sampling weights, and fill a small bounded queue. Since no preprocessing pass over the full corpus is needed, a training job starts producing gradients within minutes of launch. The loader also keeps up with the device: GPU utilization dips only during evaluation and checkpoint writing (Fig.~\ref{fig:si_training_hardware}).

Each batch is limited by four budgets: graphs, atoms, long-range pairs, and short-range edges (Table~\ref{tab:si_stages}). In practice only the edge budget is reached. Batches fill \qtyrange{96}{98}{\percent} of it and never approach the nominal \num{128} graphs; replaying \num{100} batches of the real training stream gives \numrange{85}{114} structures per batch. The effective batch size is therefore set by the data and the cutoffs rather than by a configuration parameter, and it can differ between the three models.

The difference is visible in the validation curves. The curve of SO3LR-T-L is noisy in stage~2 (Fig.~\ref{fig:si_metrics_core}) because its edge budget was smaller than that of S and M ($5\times10^{4}$ against $8\times10^{4}$ edges), so its validation batches held only \num{26.5} structures, against \numrange{57}{60}. Since every batch enters the reported average with equal weight, a single outlier structure shifted that average about twice as much. Doubling the budget at stage~3 raised the count to \num{57}, and reduced the noise.

\subsection{Loss}\label{sec:training-loss}

\textbf{Loss components.} Five properties are supervised: the energy $E$, the forces $F$, the molecular dipole $\mu$, and the per-atom polarizability ($\alpha_0$) and dispersion $C_6$ ratios,
\begin{equation}
    \mathcal{L} = \omega_F \mathcal{L}_F + \omega_E \mathcal{L}_E + \omega_\mu \mathcal{L}_\mu + \omega_{\alpha_0} \mathcal{L}_{\alpha_0} + \omega_{C_6} \mathcal{L}_{C_6},
    \qquad
    \mathcal{L}_p = \big\langle \rho\big(\hat{y}_p - y_p;\, \alpha_p, c_p\big) \big\rangle .
    \label{eq:loss_weighted}
\end{equation}
Forces carry the dominant weight, $\omega_F = 1.0$, and the energy enters weakly, with $\omega_E = 0.01$ ($0.001$ in stage~1; Table~\ref{tab:si_stages}). Dipoles, weighted with $\omega_\mu = 0.1$, supervise the partial charges indirectly, and the polarizability and $C_6$ ratios, weighted with $\omega_{\alpha_0} = \omega_{C_6} = 0.1$, supervise the dispersion module directly. Labels that a structure lacks are masked in the loss. This is what allows sources with different sets of labels to be mixed at all: only QCML and QCell supply dispersion labels, and the two $\omega$B97M-V sources supply forces alone.

\textbf{Adaptive Robust Loss.} A training corpus of this size, drawn from 52 subsets, cannot be inspected structure by structure, and some of its labels could be unconverged or incorrect. To limit the influence of such outliers, we employ the adaptive robust loss of Barron~\cite{barron2019general}, with a scale parameter $c_p$ and a shape parameter $\alpha_p \in [0,2]$, both learned independently for each target $p$. Residuals smaller than approximately $c_p$ are penalized quadratically, as under an $L_2$ loss. For residuals much larger than $c_p$, the loss grows as $|\hat y_p - y_p|^{\alpha_p}$, and logarithmically in the limit $\alpha_p \to 0$. For $\alpha_p < 1$, the gradient attains its maximum near $c_p$ and decreases for larger residuals, such that large outliers contribute only weakly to the parameter update.

\textbf{Learning curves.}
The training logs illustrate the effect of the robust loss. Individual batches exhibit errors up to two orders of magnitude above the running median (a batch MAE of up to $10.1$, compared with a median of $0.063$), yet neither the objective nor the gradient norm shows a corresponding increase. Under an $L_2$ loss, each of these batches would have produced a large update. The validation curves, which are averaged over many batches, are considerably smoother than the per-batch training signal, with the exception of the stage-2 noise of SO3LR-T-L discussed in Sec.~\ref{sec:training-budgets}.

\begin{figure*}[h]
    \centering
    \includegraphics[width=\linewidth]{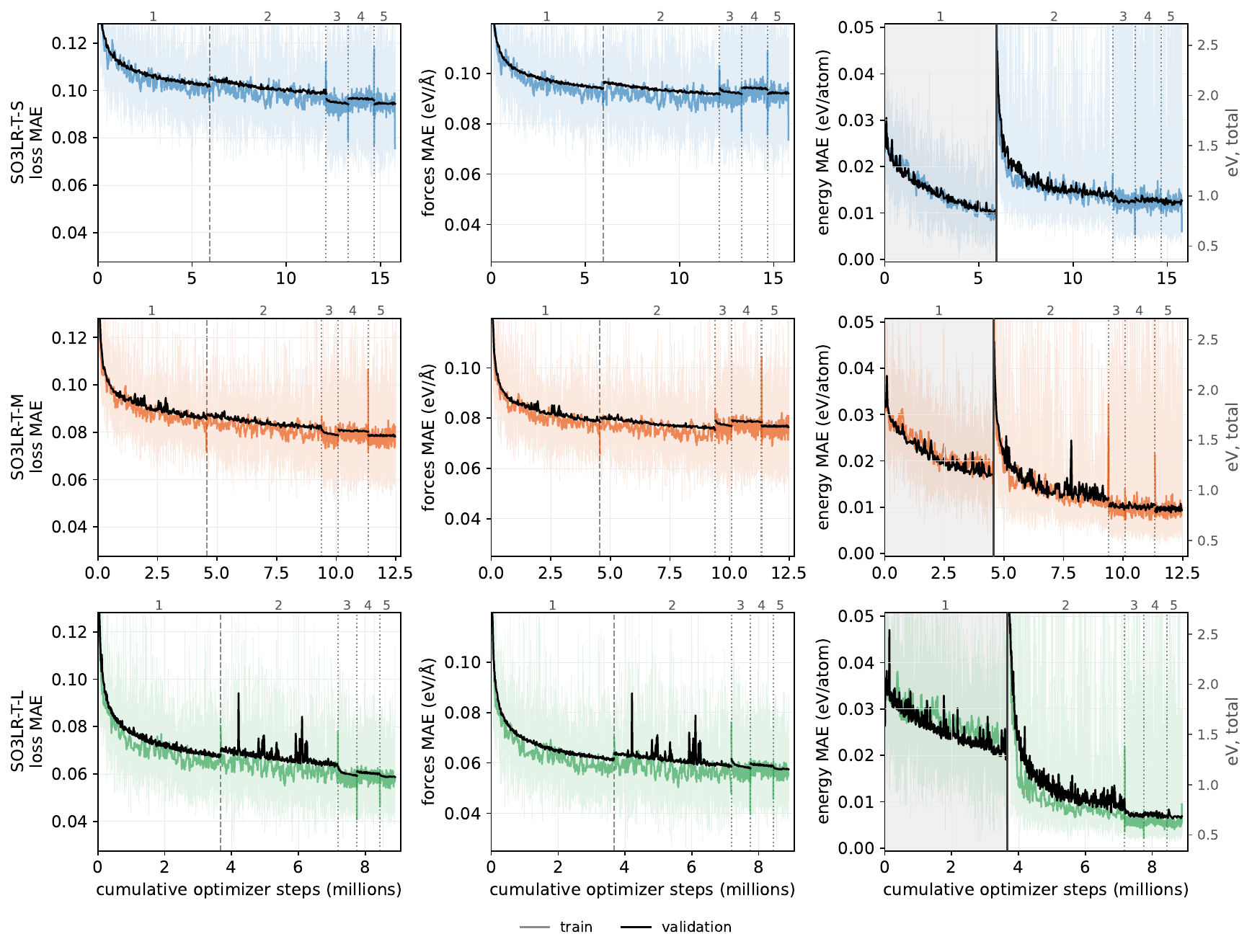}
    \caption{\textbf{Training and validation metrics over the five stages.} Loss MAE, force MAE and energy MAE of SO3LR-T-S, -M and -L, with the stages concatenated along the horizontal axis. The dashed and dotted vertical mark the stage changes, and the numbers along the top give the stage number. Training curves are running medians of the logged values, drawn over the raw values; validation is shown in black. The energy panel has two vertical axes because the energy loss was normalized per atom from stage~2 onward. Stage~1 (shaded background) refers to the right-hand axis, in eV per structure; stages~2--5 refer to the left-hand axis, in eV per atom. A solid vertical line marks the transition. Energies improve at stage~3, when the long-range cutoff opens from \qty{10}{\angstrom} to \qty{1000}{\angstrom}. The noisy validation curve of SO3LR-T-L in stage~2 is a consequence of its smaller edge budget, which left \num{26.5} structures in a validation batch against \numrange{57}{60} for S and M (Sec.~\ref{sec:training-budgets}); doubling the budget at stage~3 raised the count to \num{57}, and the noise decreased}
    \label{fig:si_metrics_core}
\end{figure*}

\begin{figure*}[h]
    \centering
    \includegraphics[width=\linewidth]{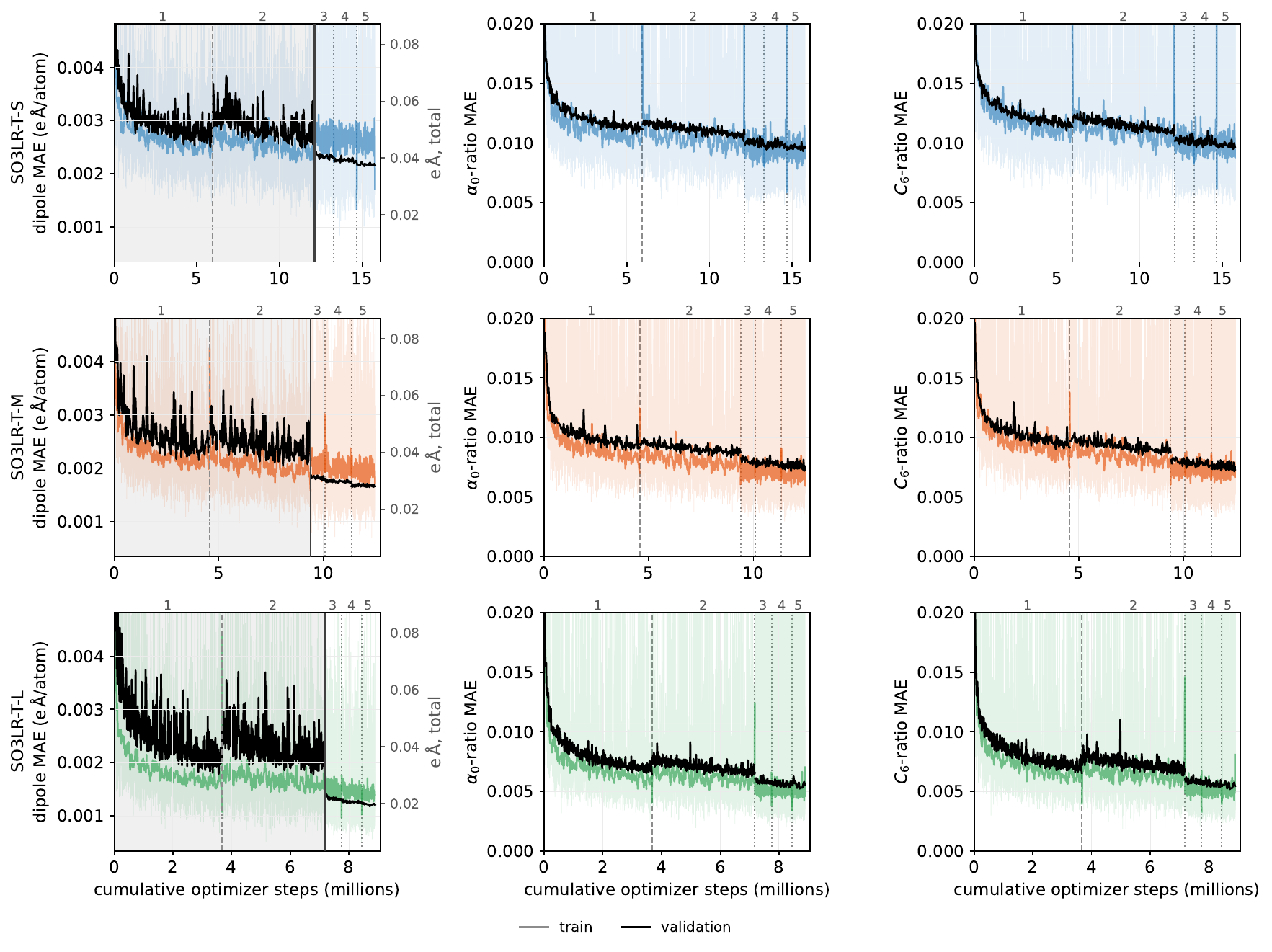}
    \caption{\textbf{Auxiliary training targets over the five stages.} MAE of the dipole vector and of the polarizability ($\alpha_0$) and $C_6$ ratios, drawn as in Fig.~\ref{fig:si_metrics_core}. These three targets enter the loss with weight $0.1$ each, against $1.0$ for forces and $0.01$ for energy ($0.001$ in stage~1; Table~\ref{tab:si_stages}). The dipole panel has two vertical axes because the dipole loss was normalized per atom from stage~3 onward. Stages~1--2 (shaded background) refer to the right-hand axis, in $e$\,\AA{} per structure; stages~3--5 refer to the left-hand axis, in $e$\,\AA{} per atom. A solid vertical line marks the transition}
    \label{fig:si_training_aux}
\end{figure*}

The energy and dipole loss terms were modified during training to use a per-atom normalization, which prevents large structures from dominating these terms. The change was applied at the stage-1/2 boundary for the energy and at the stage-2/3 boundary for the dipole. For the dipole, this choice is also physically motivated: in large systems, a single molecular dipole provides only a weak constraint on the per-atom charges, and the determination of partial charges from the dipole alone is ill-posed. Since the objective differs before and after these transitions, the corresponding panels of Figs.~\ref{fig:si_metrics_core} and~\ref{fig:si_training_aux} exhibit discontinuities at these points, which reflect the change in the loss definition rather than a change in model performance.

\subsection{Training Stages}\label{sec:training-schedule}

Training proceeds in five stages (Table~\ref{tab:si_stages}). Stage~1 uses \num{50} of the subsets. Stage~2 continues from stage~1 by adding OPoly26 to the training data. In this stage, the energy loss is switched to a per-atom normalization, and the energy weight is increased tenfold to $\omega_E = 0.01$ to compensate for the smaller magnitude of the normalized loss. Stage~3 extends the long-range cutoff to \qty{1000}{\angstrom}, such that all long-range interactions within a batch are treated explicitly. To accommodate the additional pairs, the per-batch pair budget is increased to $10^{6}$, and the learning rate is reduced to $10^{-4}$ with a slower decay.

Stages~4 and~5 were introduced to improve the description of water. PBE0 overestimates the structure of the hydrogen-bond network, and a model that inherits this bias exhibits suppressed diffusion and an ice-like density in molecular dynamics simulations. Both stages therefore modify the water content of the training mixture. Stage~4 downsamples the water clusters in GEMS bottom-up, and stage~5 adds the neutral water clusters of OMol25, computed at the $\omega$B97M-V level, as the $52^\mathrm{nd}$ subset, sampled at three times the rate of OPoly26. In addition, the dispersion damping parameter is reduced from $1.2$ to $1.0$ in stage~5 (in stage~4 for SO3LR-T-M). In the final stage, the learning rate is reduced to $10^{-5}$.

The energy readout provides up to \num{16} channels, each corresponding to a level of theory and selected by a one-hot mask, which allows additional levels of theory to be incorporated in the future. Two channels are used here: channel~0 for PBE0+MBD(-NL) and channel~1 for $\omega$B97M-V. Both are trained in stages~1 and~2. From stage~3 onward, only channel~0 is trained, together with all backbone weights, and this channel is used at inference. Channel~1 is not updated from stage~3 onward.

\begin{table*}[h]
    \centering
    \caption{\textbf{Training configuration, stage by stage.} Rows apply to all three sizes unless a size is named. The \textsc{muon} optimizer, gradient clipping at $\lVert g \rVert = 10$, the exponential form of the learning-rate schedule and the adaptive robust loss are the same in every stage and are not listed. Steps are reported in millions for each stage. Each stage ran on one NVIDIA A100-SXM4-40\,GB GPU per model. Energy heads are the columns of the $16$-way theory-level output that receive gradient. Dataset counts are subsets in the sense of Table~\ref{tab:sampling}}
    \label{tab:si_stages}
    \begin{tabular}{lcccccc}
        \toprule
        \textbf{Setting} &
        \textbf{\makecell{1\\from scratch}} &
        \textbf{\makecell{2\\resume}} &
        \textbf{\makecell{3\\long-range}} &
        \textbf{\makecell{4\\water}} &
        \textbf{\makecell{5\\water}} &
        \textbf{Total} \\
        \midrule
        Learning rate                       & $1\times10^{-3}$ & $1\times10^{-3}$ & $1\times10^{-4}$ & $1\times10^{-4}$ & $1\times10^{-5}$ & \\
        Schedule decay rate                 & $0.90$ & $0.90$ & $0.99$ & $0.99$ & $0.99$ & \\
        \midrule
        Dataset subsets                            & $50$ & $51$ & $51$ & $51$ & $52$ & \\
        \quad added                         & -- & OPoly26 & -- & -- & neutral water & \\
        GEMS bottom-up weight                   & $54\,291$ & $54\,291$ & $54\,291$ & $27\,139$ & $27\,139$ & \\
        \midrule
        Loss weight, energy                 & $0.001$ & $0.01$ & $0.01$ & $0.01$ & $0.01$ & \\
        Loss weight, forces                 & $1.0$ & $1.0$ & $1.0$ & $1.0$ & $1.0$ & \\
        Loss weight, dipole                 & $0.1$ & $0.1$ & $0.1$ & $0.1$ & $0.1$ & \\
        Loss weight, $\alpha_0$ ratio       & $0.1$ & $0.1$ & $0.1$ & $0.1$ & $0.1$ & \\
        Loss weight, $C_6$ ratio            & $0.1$ & $0.1$ & $0.1$ & $0.1$ & $0.1$ & \\
        \midrule
        $r_\mathrm{lr}$ (\unit{\angstrom})  & $10$ & $10$ & $1000$ & $1000$ & $1000$ & \\
        Graphs per batch                    & $128$ & $128$ & $128$ & $128$ & $128$ & \\
        Atoms per batch                     & $8000$ & $8000$ & $8000$ & $8000$ & $8000$ & \\
        Pairs per batch                     & $4\times10^{5}$ & $4\times10^{5}$ & $1\times10^{6}$ & $1\times10^{6}$ & $1\times10^{6}$ & \\
        Edges per batch, S / M              & $8\times10^{4}$ & $8\times10^{4}$ & $8\times10^{4}$ & $8\times10^{4}$ & $8\times10^{4}$ & \\
        Edges per batch, L                  & $5\times10^{4}$ & $5\times10^{4}$ & $1\times10^{5}$ & $1\times10^{5}$ & $1\times10^{5}$ & \\
        \midrule
        Electrostatics damping                    & $4.0$ & $4.0$ & $4.0$ & $4.0$ & $4.0$ & \\
        Dispersion damping                    & $1.2$ & $1.2$ & $1.2$ & $1.2$ & $1.0$ & \\
        Energy heads trained                & $2$ & $2$ & $1$ & $1$ & $1$ & \\
        \midrule
        Steps, S (M) & $5.94$ & $6.18$ & $1.17$ & $1.38$ & $1.11$ & $15.78$ \\
        Steps, M (M) & $4.56$ & $4.83$ & $0.69$ & $1.26$ & $1.13$ & $12.47$ \\
        Steps, L (M) & $3.67$ & $3.51$ & $0.57$ & $0.69$ & $0.45$ & $8.89$ \\
        Cost (GPUh, per model)            & 141 & 144 & $36$ & $48$ & $48$ & $417$ \\
        \bottomrule
    \end{tabular}
\end{table*}

\subsection{Training Cost}

\begin{figure*}[h]
    \centering
    \includegraphics[width=\linewidth]{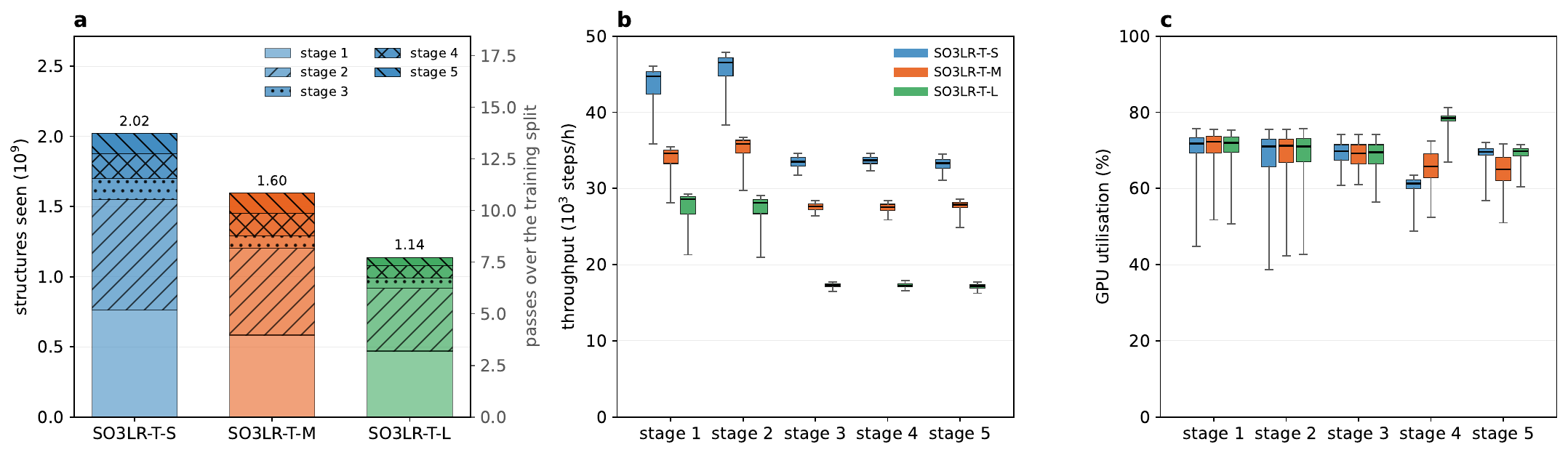}
    \caption{\textbf{Data seen, throughput and hardware use.} \textbf{(a)}~Number of structures seen, stacked by stage, with the corresponding number of passes over the training split on the right-hand axis. SO3LR-T-S, -M and -L see \num{2.02e9}, \num{1.60e9} and \num{1.14e9} structures, corresponding to \num{14}, \num{11} and \num{8} passes, respectively. Each batch is counted at its nominal size of \num{128} graphs, so these values are upper bounds. \textbf{(b)}~Throughput and \textbf{(c)}~mean GPU utilization for each stage. Boxes show the median and interquartile range over the logged samples of each stage, and whiskers indicate the 5th and 95th percentiles. Throughput is computed as the instantaneous rate between consecutive samples, excluding evaluation and checkpointing. Throughput decreases by \qtyrange{25}{30}{\percent} when the long-range cutoff is extended and the per-batch pair budget is increased. The GPU utilization drops to \qtyrange{40}{50}{\percent} coincide with evaluation and checkpoint writing}
    \label{fig:si_training_hardware}
\end{figure*}

The computational cost of training is modest. Each stage was run on a single NVIDIA A100-SXM4-\qty{40}{\giga\byte} GPU per model, amounting to \num{417} GPU-hours per model over all five stages, or approximately \num{1250} GPU-hours for the three released models combined. In total, SO3LR-T-S, -M and -L were trained for \num{15.8}, \num{12.5} and \num{8.89} million optimizer steps, respectively. Assuming the nominal batch size of \num{128} graphs, this corresponds to at most \num{2.02e9}, \num{1.60e9} and \num{1.14e9} structures, or \num{14}, \num{11} and \num{8} passes over the training split (Fig.~\ref{fig:si_training_hardware}).

The explicit treatment of all long-range pairs incurs only a moderate additional cost. Within a stage, throughput varies by a few percent. When the long-range cutoff is extended and the pair budget is increased by \qty{150}{\percent}, throughput decreases by only \qtyrange{25}{30}{\percent}. The drops of GPU utilization to \qtyrange{40}{50}{\percent} are attributable to evaluation and checkpoint writing.

\clearpage
\section{Static benchmarks}\label{sec:static-benchmarks}

\subsection{Test set errors}

Tables~\ref{tab:pbe0-forces}--\ref{tab:opoly26-forces} report the force errors of SO3LR and SO3LR-T-S, -M and -L on all held-out test splits (some shown in Fig.~\ref{fig:test-set-errors}), grouped by level of theory. Unless stated otherwise, all values below are force MAEs in kcal/mol/\AA. Since SO3LR supports only eight elements, structures containing other elements are excluded; the table captions give the resulting coverage, and subsets without any supported structure are marked with a dash. With the exception of a few small subsets, errors decrease with increasing model size.

On the PBE0+MBD(-NL) splits, SO3LR remains more accurate than SO3LR-T-S and -M on QM7-X, GEMS, AQM and the SPICE dipeptides. These datasets formed most of the SO3LR training data but represent a much smaller fraction of the SO3LR-T training corpus (Sec.~\ref{sec:training-data}). SO3LR-T-L outperforms SO3LR on all of these datasets except AQM, where its MAE is \num{1.36} compared with \num{0.95} for SO3LR. For chemistry that is newly covered by the expanded corpus, the improvement is substantial: on QCML, the MAE decreases from \num{10.23} for SO3LR (evaluated on the \qty{47}{\percent} of structures it supports) to \num{2.06} for SO3LR-T-M, and the errors on lipids and sugars decrease by \qty{24}{\percent} and \qty{40}{\percent}, respectively.

On OMol25, SO3LR-T-M achieves MAEs below \num{1} on most protein, nucleic acid and PDB fragment subsets, as well as on two electrolyte subsets (ml\_elytes and 5A\_elytes). The MAE increases to \numrange{1.5}{3} on the metal-organic, radical and reactive subsets (rpmd, pmechdb, orbnet\_denali, low\_spin\_23, ml\_mo and tm\_react), to \numrange{3.8}{3.9} on the high-energy ANI samples, to \numrange{4}{7} on rmechdb, mo\_hydrides and the reactive and stretched electrolytes, and to \num{11.7} on the \num{44} noble-gas compounds. On the subsets where SO3LR can be evaluated, its MAE is typically two to three times higher than that of SO3LR-T-M, with ratios ranging from \num{1.1} to \num{3.9}. SO3LR-T-S shows an outlier on mo\_hydrides, where its RMSE of \num{3167} greatly exceeds its MAE of \num{26.4}, indicating that a small number of structures with extreme force errors dominate the RMSE. This behavior is not observed for SO3LR-T-M and -L.

On OPoly26, the MAE of SO3LR-T-M lies between \num{0.65} and \num{0.88} on the fragmented polymer and lipid subsets, with the exception of the nontraditional copolymers (\num{1.02}), and between \num{1.60} and \num{1.74} on the charged and reactive subsets.

\begin{table*}[h]
\centering
\caption{\textbf{PBE0+MBD(-NL) test subsets.} Force MAE and RMSE in kcal\,mol$^{-1}$\,\AA$^{-1}$ and dipole MAE in $e$\,\AA{}. Columns 1, T-S, T-M, and T-L denote SO3LR, SO3LR-T-S, SO3LR-T-M, and SO3LR-T-L, respectively. SO3LR supports H, C, N, O, F, P, S, and Cl; structures containing other elements are excluded from its columns. SO3LR therefore covers $120{,}939$ of $173{,}961$ structures ($69.5\%$), with full coverage on seven subsets, $31\%$ on test\_qcell\_ion and $47\%$ on test\_qcml. Dashes indicate no qualifying structure (for SO3LR) or no reference label. $N$ gives the full subset size}
\label{tab:pbe0-forces}
\small
\setlength{\tabcolsep}{4pt}
\renewcommand{\arraystretch}{1.10}
\begin{tabular}{@{}lr*{12}{r}@{}}
\toprule
\multirow{3}{*}{\textbf{Dataset}} & \multirow{3}{*}{\textbf{$N$}} & \multicolumn{8}{c}{\textbf{Forces}} & \multicolumn{4}{c}{\textbf{Dipole}} \\
\cmidrule(lr){3-10} \cmidrule(lr){11-14}
 & & \multicolumn{4}{c}{\textbf{MAE}} & \multicolumn{4}{c}{\textbf{RMSE}} & \multicolumn{4}{c}{\textbf{MAE}} \\
\cmidrule(lr){3-6} \cmidrule(lr){7-10} \cmidrule(lr){11-14}
 & & \textbf{1} & \textbf{T-S} & \textbf{T-M} & \textbf{T-L} & \textbf{1} & \textbf{T-S} & \textbf{T-M} & \textbf{T-L} & \textbf{1} & \textbf{T-S} & \textbf{T-M} & \textbf{T-L} \\
\midrule
test\_des400k & 4k & 0.67 & 0.64 & 0.54 & 0.32 & 1.29 & 1.31 & 1.06 & 0.64 & 0.02 & 0.01 & 0.01 & 0.01 \\
test\_qcell\_lipids & 160 & 1.08 & 0.82 & 0.82 & 0.56 & 2.34 & 1.36 & 1.38 & 0.99 & 1.90 & 0.19 & 0.13 & 0.11 \\
test\_qcell\_sugars & 740 & 1.29 & 0.95 & 0.78 & 0.49 & 1.88 & 1.36 & 1.13 & 0.72 & 0.06 & 0.02 & 0.02 & 0.01 \\
test\_qcell\_ion & 300 & 0.81 & 1.03 & 0.94 & 0.60 & 1.22 & 1.44 & 1.38 & 1.05 & 0.03 & 0.03 & 0.02 & 0.02 \\
test\_gems\_small & 118 & 0.88 & 1.14 & 0.99 & 0.63 & 1.28 & 1.64 & 1.45 & 1.00 & -- & -- & -- & -- \\
test\_spice\_dipep & 338 & 0.83 & 1.30 & 1.26 & 0.78 & 1.20 & 1.94 & 1.91 & 1.24 & 0.04 & 0.03 & 0.03 & 0.02 \\
test\_qcell\_dna & 148 & -- & 1.30 & 1.04 & 0.68 & -- & 2.15 & 1.80 & 1.31 & -- & 0.19 & 0.16 & 0.12 \\
test\_gems\_big & 27k & 0.87 & 1.42 & 1.17 & 0.82 & 1.78 & 3.16 & 2.64 & 2.06 & -- & -- & -- & -- \\
test\_qm7x & 42k & 0.96 & 1.54 & 1.19 & 0.80 & 1.53 & 2.86 & 2.21 & 1.57 & 0.02 & 0.02 & 0.02 & 0.01 \\
test\_aqm & 597 & 0.95 & 1.58 & 1.49 & 1.36 & 1.45 & 2.61 & 2.46 & 2.28 & 0.06 & 0.07 & 0.05 & 0.04 \\
test\_qcell\_rna & 200 & 3.32 & 2.26 & 2.22 & 2.16 & 5.04 & 3.39 & 3.35 & 3.29 & 0.85 & 0.14 & 0.11 & 0.09 \\
test\_qcml & 99k & 10.23 & 2.54 & 2.06 & 1.46 & 21.43 & 5.08 & 3.98 & 2.96 & 0.26 & 0.04 & 0.03 & 0.02 \\
\bottomrule
\end{tabular}
\end{table*}

\clearpage
\begin{table*}[h]
\centering
\caption{\textbf{OMol25 test subsets.} Force MAE and RMSE in kcal\,mol$^{-1}$\,\AA$^{-1}$, sorted by the SO3LR-T-S MAE. Columns 1, T-S, T-M, and T-L denote SO3LR, SO3LR-T-S, SO3LR-T-M, and SO3LR-T-L, respectively. SO3LR supports H, C, N, O, F, P, S, and Cl; structures containing other elements are excluded from its columns. SO3LR therefore covers $1{,}429{,}970$ of $2{,}762{,}021$ structures ($51.8\%$). Eight subsets contain no structure that SO3LR supports and are dashed in its columns: noble\_gas, noble\_gas\_compounds, low\_spin\_23, ml\_mo, tm\_react, mo\_hydrides, omol\_metal\_organics, and omol\_redo\_orca6. Coverage is $5\%$ for electrolytes\_reactivity and $35\%$ for 5A\_elytes. Dashes indicate no qualifying structure. $N$ gives the full subset size}
\label{tab:omol25-forces}
\small
\setlength{\tabcolsep}{4pt}
\renewcommand{\arraystretch}{1.10}
\begin{tabular}{@{}lr*{8}{r}@{}}
\toprule
\multirow{3}{*}{\textbf{Dataset}} & \multirow{3}{*}{\textbf{$N$}} & \multicolumn{8}{c}{\textbf{Forces}} \\
\cmidrule(lr){3-10}
 & & \multicolumn{4}{c}{\textbf{MAE}} & \multicolumn{4}{c}{\textbf{RMSE}} \\
\cmidrule(lr){3-6} \cmidrule(lr){7-10}
 & & \textbf{1} & \textbf{T-S} & \textbf{T-M} & \textbf{T-L} & \textbf{1} & \textbf{T-S} & \textbf{T-M} & \textbf{T-L} \\
\midrule
noble\_gas & 154 & -- & 0.46 & 0.42 & 0.31 & -- & 0.61 & 0.58 & 0.44 \\
protein\_interface & 29k & 1.56 & 0.66 & 0.52 & 0.36 & 2.47 & 1.10 & 0.88 & 0.66 \\
protein\_core & 16k & 1.85 & 0.69 & 0.55 & 0.38 & 5.29 & 1.80 & 1.46 & 1.12 \\
dna & 22k & 2.51 & 0.90 & 0.70 & 0.47 & 4.15 & 1.79 & 1.46 & 1.12 \\
rna & 24k & 2.67 & 0.97 & 0.75 & 0.51 & 4.28 & 2.18 & 1.81 & 1.45 \\
ml\_elytes & 171k & 2.54 & 0.97 & 0.78 & 0.55 & 5.84 & 2.49 & 2.15 & 1.78 \\
ml\_protein\_interface & 110k & 1.83 & 0.97 & 0.81 & 0.61 & 3.70 & 2.46 & 2.23 & 2.06 \\
pdb\_fragments\_300K & 156k & 2.02 & 0.98 & 0.75 & 0.53 & 3.66 & 2.04 & 1.57 & 1.18 \\
pdb\_pockets\_300K & 78k & 2.02 & 0.99 & 0.78 & 0.56 & 4.30 & 2.48 & 1.99 & 1.61 \\
pdb\_pockets\_400K & 15k & 2.06 & 1.03 & 0.81 & 0.59 & 4.51 & 2.57 & 2.07 & 1.64 \\
pdb\_fragments\_400K & 155k & 2.08 & 1.06 & 0.81 & 0.57 & 3.78 & 2.13 & 1.64 & 1.25 \\
5A\_elytes & 58k & 2.46 & 1.14 & 0.94 & 0.66 & 5.88 & 2.77 & 2.43 & 1.96 \\
scaled\_separations\_exp & 34k & 3.08 & 1.19 & 1.05 & 0.85 & 7.89 & 3.58 & 3.38 & 3.08 \\
nakb & 2k & 3.71 & 1.20 & 0.95 & 0.65 & 5.64 & 2.33 & 1.93 & 1.47 \\
spice & 10k & 2.41 & 1.22 & 0.95 & 0.71 & 5.43 & 3.24 & 2.73 & 2.06 \\
droplet & 7k & 2.22 & 1.32 & 1.07 & 0.74 & 4.20 & 4.48 & 3.96 & 2.89 \\
omol\_electrolytes & 1.4M & 3.36 & 1.50 & 1.24 & 0.92 & 7.99 & 3.94 & 3.41 & 2.78 \\
omol\_solvated\_protein & 21 & 2.28 & 1.51 & 1.15 & 1.25 & 3.27 & 2.69 & 1.98 & 2.23 \\
omol\_metal\_organics & 79k & -- & 2.04 & 1.79 & 1.44 & -- & 5.18 & 4.51 & 3.89 \\
rpmd & 4k & 3.42 & 2.06 & 1.60 & 1.16 & 11.66 & 9.02 & 7.30 & 5.86 \\
omol\_redo\_orca6 & 7k & -- & 2.07 & 1.76 & 1.39 & -- & 14.37 & 4.56 & 3.82 \\
pmechdb & 38k & 3.63 & 2.36 & 1.88 & 1.43 & 7.50 & 5.71 & 4.83 & 4.02 \\
orbnet\_denali & 19k & 3.95 & 2.37 & 1.93 & 1.50 & 8.44 & 5.21 & 4.53 & 3.78 \\
electrolytes\_redox & 35k & 3.91 & 2.42 & 2.25 & 2.02 & 9.72 & 7.34 & 7.18 & 6.93 \\
geom\_orca6 & 6k & 5.32 & 2.68 & 1.97 & 1.40 & 12.98 & 5.75 & 4.23 & 2.98 \\
low\_spin\_23 & 26k & -- & 3.01 & 2.68 & 2.30 & -- & 7.58 & 6.65 & 5.92 \\
ml\_mo & 12k & -- & 3.03 & 2.62 & 2.15 & -- & 7.88 & 7.06 & 6.00 \\
tm\_react & 62k & -- & 3.08 & 2.80 & 2.40 & -- & 7.94 & 7.45 & 6.81 \\
ani1xbb & 185k & 8.83 & 4.45 & 3.78 & 3.04 & 18.58 & 10.29 & 9.33 & 8.21 \\
rmechdb & 2k & 6.42 & 4.55 & 4.04 & 3.53 & 14.39 & 11.28 & 10.63 & 9.99 \\
ani2x & 3k & 7.97 & 4.58 & 3.91 & 3.38 & 29.45 & 14.91 & 12.36 & 10.96 \\
noble\_gas\_compounds & 44 & -- & 5.10 & 11.72 & 7.59 & -- & 10.62 & 22.62 & 17.22 \\
electrolytes\_reactivity & 11k & 6.08 & 6.24 & 5.48 & 4.46 & 11.73 & 12.44 & 11.27 & 9.64 \\
electrolytes\_scaled\_sep & 13k & 14.48 & 8.76 & 7.03 & 5.71 & 34.52 & 59.21 & 17.37 & 46.07 \\
mo\_hydrides & 7k & -- & 26.36 & 5.00 & 4.34 & -- & 3166.97 & 13.64 & 13.94 \\
\bottomrule
\end{tabular}
\end{table*}
\clearpage

\begin{table*}[h]
\centering
\caption{\textbf{OPoly26 test subsets.} Force MAE and RMSE in kcal\,mol$^{-1}$\,\AA$^{-1}$, sorted by the SO3LR-T-S MAE. Columns 1, T-S, T-M, and T-L denote SO3LR, SO3LR-T-S, SO3LR-T-M, and SO3LR-T-L, respectively. SO3LR supports H, C, N, O, F, P, S, and Cl; structures containing other elements are excluded from its columns. SO3LR therefore covers $184{,}219$ of $210{,}327$ structures ($87.6\%$). The lowest coverage is $35\%$, for omers\_charged2. $N$ gives the full subset size}
\label{tab:opoly26-forces}
\small
\setlength{\tabcolsep}{4pt}
\renewcommand{\arraystretch}{1.10}
\begin{tabular}{@{}lr*{8}{r}@{}}
\toprule
\multirow{3}{*}{\textbf{Dataset}} & \multirow{3}{*}{\textbf{$N$}} & \multicolumn{8}{c}{\textbf{Forces}} \\
\cmidrule(lr){3-10}
 & & \multicolumn{4}{c}{\textbf{MAE}} & \multicolumn{4}{c}{\textbf{RMSE}} \\
\cmidrule(lr){3-6} \cmidrule(lr){7-10}
 & & \textbf{1} & \textbf{T-S} & \textbf{T-M} & \textbf{T-L} & \textbf{1} & \textbf{T-S} & \textbf{T-M} & \textbf{T-L} \\
\midrule
omers\_fragmented\_lipids & 10k & 1.99 & 0.83 & 0.67 & 0.46 & 5.50 & 2.05 & 1.87 & 1.50 \\
omers\_unfragmented & 176 & 1.64 & 0.84 & 0.65 & 0.46 & 2.69 & 1.31 & 1.02 & 0.73 \\
omers\_fragmented\_solvated & 9k & 2.30 & 0.88 & 0.68 & 0.47 & 4.74 & 1.94 & 1.54 & 1.14 \\
omers\_fragmented\_trad\_copolymers & 9k & 2.08 & 0.89 & 0.68 & 0.48 & 4.40 & 1.76 & 1.36 & 0.96 \\
omers\_fragmented\_ion\_inserted & 4k & 2.24 & 1.05 & 0.81 & 0.58 & 4.05 & 2.01 & 1.62 & 1.36 \\
omers\_fragmented\_peptoids & 32k & 1.96 & 1.10 & 0.85 & 0.61 & 3.36 & 2.16 & 1.62 & 1.23 \\
omers\_fragmented\_homopolymers & 4k & 2.46 & 1.13 & 0.84 & 0.59 & 4.83 & 2.22 & 1.59 & 1.15 \\
omers\_fragmented\_random\_copolymers & 56k & 2.68 & 1.17 & 0.88 & 0.62 & 4.99 & 2.22 & 1.70 & 1.26 \\
omers\_fragmented\_overflow & 65k & 2.76 & 1.18 & 0.88 & 0.62 & 5.18 & 2.23 & 1.68 & 1.23 \\
omers\_fragmented\_nontrad\_copolymers & 10k & 3.70 & 1.41 & 1.02 & 0.71 & 6.56 & 2.51 & 1.79 & 1.27 \\
omers\_charged2 & 150 & 3.12 & 1.89 & 1.60 & 1.35 & 6.69 & 4.19 & 3.84 & 3.63 \\
omers\_charged & 2k & 3.35 & 1.92 & 1.61 & 1.33 & 8.27 & 6.06 & 5.11 & 5.86 \\
omers\_reactivity & 11k & 2.89 & 1.98 & 1.74 & 1.49 & 6.69 & 5.86 & 5.45 & 5.11 \\
\bottomrule
\end{tabular}
\end{table*}

\vspace{-1mm}
\subsection{External benchmarks}
\vspace{-1mm}

\textbf{Interaction energy.} 
Interaction energies are computed as $E_\mathrm{int}=E_{AB}-E_{AB}(\infty)$. In the separated reference $E_{AB}(\infty)$, the monomer geometries are kept fixed and one monomer is translated by $10^{4}$~\AA, well beyond the SO3LR long-range cutoff, so that all explicit intermolecular short-range, electrostatic and dispersion contributions vanish. Both states are evaluated with the same total charge. In the separated reference, each fragment is constrained to its formal charge as an isolated monomer, which prevents net charge transfer between the distant fragments. This construction retains the global electronic-state inputs of the combined system while removing all explicit intermolecular interactions. It does not require the predicted partial charges to match those of independent monomer calculations.

Alternatively, interaction energies can be computed as $E_\mathrm{int}=E_{AB}-E_A-E_B$, where each monomer is evaluated independently in its specified charge state. For a size-consistent model, the two definitions are identical. They can differ, however, when the energy of the separated system depends on global electronic-state inputs (discussed in more detail below). We recommend the separated-reference definition for SO3LR and use it for all reported benchmarks; the differences between the two definitions are quantified in Table~\ref{tab:sapt-size-consistency}. An example script for computing interaction energies is provided in the accompanying GitHub repository.

\textbf{Size consistency.} 
We assess the size consistency of SO3LR-T-M and other MLFFs by comparing interaction energies computed from independent monomers with those computed from the separated supermolecule on SAPT10k~\cite{villot2024ab}, resolved by the formal charges of the fragments. Size consistency requires the energy of non-interacting fragments to equal the sum of their independently evaluated energies for the same fragment electronic states. For MLFFs, an atom-wise energy decomposition alone does not guarantee this property: disconnected fragments may still be coupled through global electronic-state conditioning or charge assignment. We therefore compare the interaction energies obtained from independent monomers and from the separated supermolecule, and report the asymptotic residual $\delta_\infty=E_{AB}(\infty)-E_A-E_B$ (Table~\ref{tab:sapt-size-consistency}). Under their own evaluation conventions, all models tested here are non-additive for at least some combinations of fragment charges.

The charge dependence differs substantially between models. MACE-POLAR~\cite{batatia2026mace} and AIMNet2~\cite{anstine2024aimnet2} exhibit their largest mean residuals for opposite-charge pairs, whereas UMA~\cite{wood2026family} also shows pronounced offsets for like-charge pairs. SO3LR-T-M is additive to numerical precision for neutral--neutral pairs but retains finite residuals for charged fragments. Size consistency and interaction-energy accuracy are distinct properties: the independent-monomer and separated-reference interaction energies differ by exactly the residual, $I_{\mathrm{ind}}-I_{\mathrm{sep}}=\delta(R)$, so a separation offset can either increase or cancel the error against the reference. For example, the separated definition improves SO3LR-T-M's charged--neutral MAE but worsens its opposite-charge MAE. These findings motivate further development of electronic-state conditioning and charge-assignment schemes that preserve additivity for non-interacting fragments while retaining the ability to describe charge redistribution in interacting systems. \vspace{-2mm}

\begin{table*}[h]
\centering
\caption{\textbf{Asymptotic separation residuals on SAPT10k.} Mean absolute residual $\langle|\delta_\infty|\rangle$, in kcal\,mol$^{-1}$, where $\delta_\infty=E_{AB}(\infty)-E_A-E_B$ is the difference between the energy of the separated dimer and the sum of the independently computed monomer energies.  Results are grouped by the formal charges of the two fragments, and $N$ is the number of dimers in each group}
\label{tab:sapt-size-consistency}
\small
\setlength{\tabcolsep}{4pt}
\renewcommand{\arraystretch}{1.10}
\begin{tabular}{@{}lr*{5}{r}@{}}
\toprule
\textbf{Model} & \textbf{$0/0$} & Charged/$0$ & \textbf{$+/+$} & \textbf{$+/-$} & \textbf{$-/-$} & All \\
\midrule
$N$             & $8{,}834$  & 952   & 7      & 175    & 14     & $9{,}982$ \\
\midrule
SO3LR-T-M             & $<10^{-8}$ & 1.374 & 0.710  & 1.335  & 0.153  & 0.155 \\
MACE-POLAR-1-M  & 0.011      & 1.283 & 0.003  & 5.955  & 0.001  & 0.237 \\
AIMNet2-2025    & 0.041      & 3.368 & 0.054  & 25.613 & 0.021  & 0.807 \\
UMA-S-1.2.1     & 0.024      & 1.506 & 16.785 & 36.086 & 23.495 & 0.842 \\
\bottomrule
\end{tabular}\vspace{-2mm}
\end{table*}

The residual is evaluated as a function of the monomer separation $R$, $\delta(R)=E_{AB}(R)-E_A-E_B$, where $R$ is the distance between the geometric centers of the two monomers, and $\delta_\infty$ denotes its limit at infinite separation. For SO3LR-T-M and UMA, $\delta(R)$ is constant beyond the model cutoff, and $\delta_\infty$ is taken at a separation of \qty{10000}{\angstrom}; we verified that the values at \qty{3000}{\angstrom} are identical. For MACE-POLAR and AIMNet2, which do not reach such a plateau, $\delta_\infty$ is obtained by fitting $\delta(R)=\delta_\infty+a/R+b/R^2$ to configurations in which one monomer is translated by \num{300}, \num{1000}, \num{3000} and \qty{10000}{\angstrom}. Using a linear fit or a different range of separations changes the extrapolated residuals by at most \qty{0.001}{kcal\,mol^{-1}} for MACE-POLAR and \qty{0.007}{kcal\,mol^{-1}} for AIMNet2. For AIMNet2, the isolated monomers are evaluated at their coordinates in each separated configuration, which reduces numerical artifacts caused by the translation. As described above, SO3LR-T-M constrains each fragment to its formal charge in the separated reference, whereas the other models use their standard global electronic-state inputs.

\textbf{JSCH-2005 reference energies.}
The complete 124-complex JSCH-2005 nucleobase benchmark contains 102 estimated CCSD(T)/CBS reference interaction energies and 22 MP2/CBS references~\cite{jurevcka2006benchmark}. For the former, the MP2/CBS interaction energy is supplemented by a finite-basis correction for correlation beyond MP2. The MP2-only entries are JSCH-11 through JSCH-32, all belonging to the hydrogen-bonded subset. To use a more consistent reference level in the main comparison, we retain only the 102 complexes with estimated CCSD(T)/CBS references and apply this selection identically to all models. The complete-set results are retained here for comparison with previous studies and to preserve the broader hydrogen-bonded coverage. For SO3LR-T-M, the subset selection changes the MAE/RMSE from $0.772/1.024$ to $0.778/1.045$~kcal/mol.

\begin{table}[h]
\centering
\caption{\textbf{Effect of reference-level selection on JSCH-2005.} Interaction-energy MAE and RMSE in kcal\,mol$^{-1}$ for the complete 124-complex set and its 102-complex subset with estimated CCSD(T)/CBS references~\cite{jurevcka2006benchmark}. The complete set additionally contains 22 MP2/CBS references. All models use the same selection and the separated-reference interaction-energy definition}
\setlength{\tabcolsep}{4pt}
\begin{tabular}{l cc cc}
\toprule
& \multicolumn{2}{c}{Full set} & \multicolumn{2}{c}{CCSD(T)/CBS subset}\\
\cmidrule(lr){2-3}\cmidrule(lr){4-5}
Model & MAE & RMSE & MAE & RMSE \\
\midrule
SO3LR   & 1.69 & 2.18 & 1.75 & 2.24 \\
SO3LR-T-S & 1.12 & 1.40 & 1.01 & 1.27 \\
SO3LR-T-M & 0.77 & 1.02 & 0.78 & 1.04 \\
SO3LR-T-L & 0.61 & 0.79 & 0.58 & 0.76 \\
\bottomrule
\end{tabular}
\label{tab:jsch-reference-subsets}
\end{table}

\begin{figure*}[h]
\centering
\includegraphics[width=0.99\linewidth]{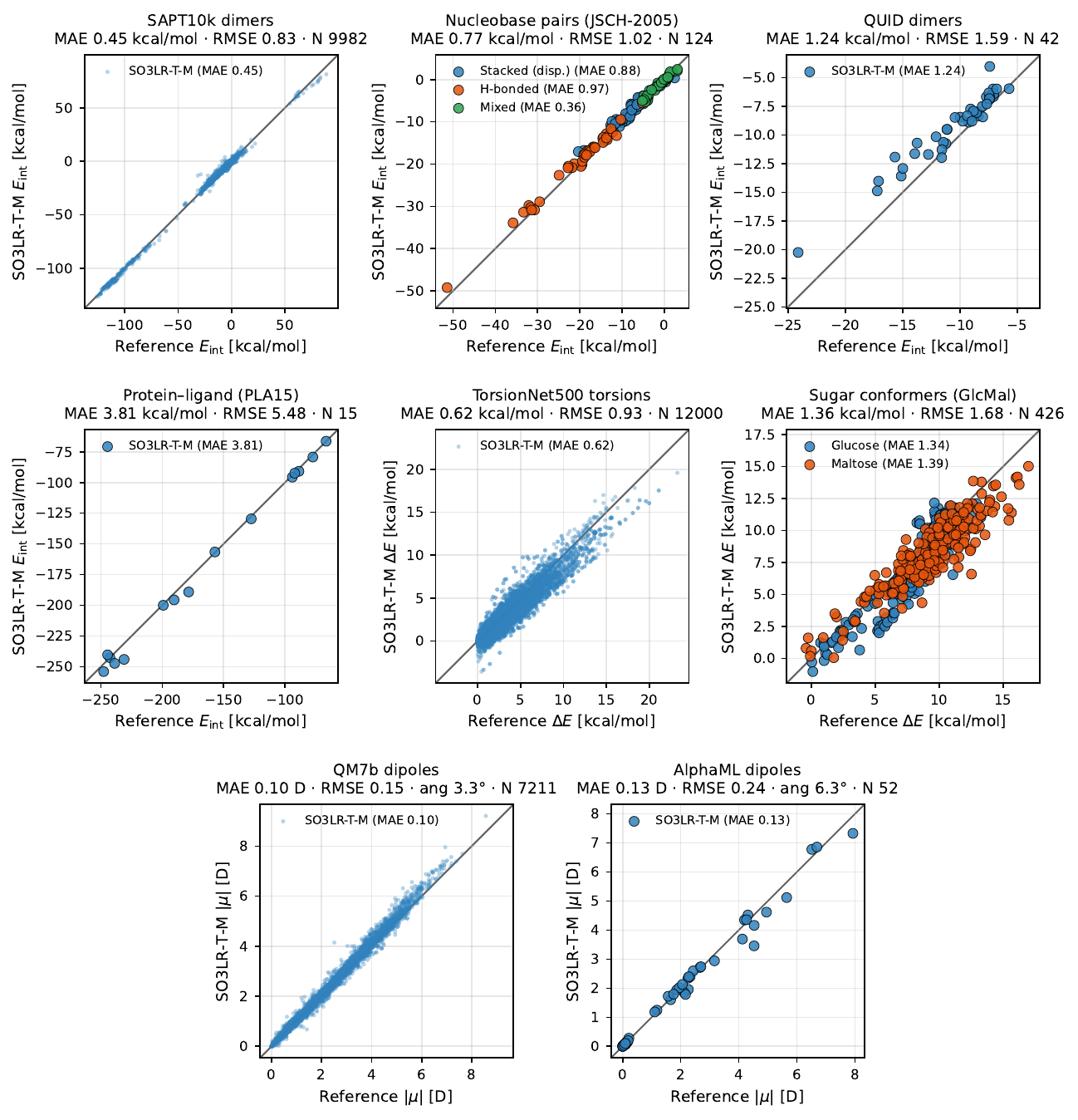}
\caption{\textbf{Interaction energies, conformational energetics, and molecular dipoles across eight benchmarks.} Parity plots of SO3LR-T-M predictions against reference values; the diagonal line marks perfect agreement. \emph{Top row}: intermolecular interaction energies (kcal/mol) for SAPT10k ($N{=}9982$)~\cite{villot2024ab}, the JSCH-2005 nucleobase pairs ($N{=}124$, colored by interaction type as H-bonded, stacked/dispersion-dominated, and mixed; Table~\ref{tab:so3lr-performance} of the main text uses the 102-complex CCSD(T)/CBS subset, Table~\ref{tab:jsch-reference-subsets})~\cite{jurevcka2006benchmark}, and the QUID dimers ($N{=}42$)~\cite{puleva2025extending}. \emph{Middle row}: the PLA15 protein--ligand active sites ($N{=}15$)~\cite{kriz2020benchmarking}, per-group relative conformer energies (kcal/mol) for TorsionNet500 torsion profiles~\cite{rai2022torsionnet, kabylda2025molecular} ($N{=}12000$, relative to each profile minimum) and the GlcMal glucose/$\alpha$-maltose conformers ($N{=}426$, relative to the designated reference conformer)~\cite{marianski2016assessing}. \emph{Bottom row}: dipole-moment magnitudes $|\mu|$ (Debye) for QM7b ($N{=}7211$) and AlphaML ($N{=}52$)~\cite{yang2019quantum}. Each panel title reports the mean absolute error (MAE) and root-mean-square error (RMSE); the dipole panels additionally give the mean angular error between predicted and reference dipole vectors}
\label{fig:tab2_overview}
\end{figure*}

\begin{table*}[h]
\caption{Interaction-energy errors on the PLA15 benchmark based on MP2-F12 and DLPNO-CCSD(T) calculations~\cite{kriz2020benchmarking}. For each complex $i$ the relative error is defined as $\delta_i = (E_i - E_i^{\rm ref})\,/\,|E_i^{\rm ref}|\times 100\,\%$, where $E_i^{\rm ref}$ is the reference interaction energy and positive $\delta_i$ indicates under-binding. Column definitions: MAE\,\% $= \tfrac{1}{N}\sum_i |\delta_i|$ (mean absolute percent error); mean\,\% $= \tfrac{1}{N}\sum_i \delta_i$; MAE $= \tfrac{1}{N}\sum_i |E_i - E_i^{\rm ref}|$ (kcal/mol); RMSE $= \sqrt{\tfrac{1}{N}\sum_i (E_i - E_i^{\rm ref})^2}$ (kcal/mol). Rows are ordered by MAE\,\%. SO3LR results are obtained in this work; other results are taken from Refs.~\citenum{kriz2020benchmarking, Pla15Repo}} 
\label{tab:pla15-extended}
\centering
\begin{tabular}{l r r r r}
\toprule
Method & MAE\,\% & mean\,\% & MAE & RMSE \\
\midrule
SO3LR-T-M      & 2.1  & -1.8 & 3.81  & 5.48  \\
SO3LR-T-S      & 2.5  & +0.8 & 3.74  & 4.50  \\
SO3LR-T-L      & 4.3  & -4.2 & 6.42  & 7.36  \\
SO3LR      & 5.8  & +4.8 & 8.75  & 9.85  \\
\midrule
g-xTB                 & 6.1           & +4.7          & 7.70           & 8.57           \\
GFN2                  & 8.2           & +8.1          & 10.62          & 13.44          \\
GFN2-xTB              & 8.2           & +8.1          & 10.65          & 13.40          \\
DFTB3-D3H5            & 8.4           & -8.4          & 12.99          & 15.20          \\
UMA-m                 & 9.6           & -9.6          & 13.66          & 15.11          \\
PM6-D3H4              & 9.7           & -9.4          & 17.01          & 21.00          \\
eSEN-s OMol25         & 10.9          & -10.9         & 14.40          & 15.64          \\
DFTB3-D3H4            & 11.1          & -11.1         & 18.25          & 21.40          \\
UMA-s                 & 12.7          & -12.7         & 16.27          & 18.58          \\
GFNFF                 & 21.7          & +6.3          & 41.16          & 68.05          \\
AIMNet2 (DSF)         & 22.1          & +13.3         & 39.69          & 52.11          \\
PM7                   & 24.1          & -24.1         & 35.35          & 38.80          \\
Egret-1               & 24.3          & +24.2         & 48.71          & 63.69          \\
AIMNet2               & 27.4          & +27.4         & 35.84          & 37.90          \\
PM6                   & 31.5          & +31.5         & 41.19          & 42.10          \\
ANI-2x                & 38.8          & +37.2         & 73.03          & 88.12          \\
Orb-v3                & 46.6          & +46.6         & 81.63          & 94.63          \\
AM1                   & 55.9          & +55.9         & 77.79          & 79.10          \\
MACE-MP-0b2-L         & 67.3          & +67.3         & 112.19         & 122.57         \\
\bottomrule
\end{tabular}
\end{table*}

\begin{table*}[h]
\caption{Relative conformer-energy accuracy on the GlcMal glucose / $\alpha$-maltose sets: MAE (kcal/mol) and Spearman rank correlation $\rho$ against the CCSD(T)/CBS (glucose) and DLPNO-CCSD(T)/CBS ($\alpha$-maltose) references. All methods are evaluated on the same conformers and relative to the same anchor conformer; SO3LR values are from this work, and all others are computed from the energies tabulated in
Ref.~\citenum{marianski2016assessing}}
\centering
\setlength{\tabcolsep}{4pt}
\begin{tabular}{l cc cc cc}
\toprule
& \multicolumn{2}{c}{Glucose} & \multicolumn{2}{c}{$\alpha$-Maltose}
& \multicolumn{2}{c}{Combined}\\
\cmidrule(lr){2-3}\cmidrule(lr){4-5}\cmidrule(lr){6-7}
Method & MAE & $\rho$ & MAE & $\rho$ & MAE & $\rho$ \\
\midrule
GLYCAM06       & 2.69 & 0.76 & 3.17 & 0.67 & 2.96 & 0.74 \\
CHARMM36       & 3.33 & 0.80 & 3.38 & 0.78 & 3.34 & 0.77 \\
PM7            & 4.24 & 0.49 & 5.60 & 0.32 & 4.93 & 0.38 \\
DFTB+D3        & 5.00 & 0.42 & 6.10 & 0.60 & 5.55 & 0.54 \\
\midrule
SO3LR        & 2.70 & 0.50 & 1.89 & 0.69 & 2.28 & 0.63 \\
SO3LR-T-S        & 1.39 & 0.89 & 1.35 & 0.80 & 1.37 & 0.84 \\
SO3LR-T-M        & 1.34 & 0.91 & 1.39 & 0.88 & 1.36 & 0.89 \\
SO3LR-T-L        & 1.02 & 0.93 & 0.85 & 0.93 & 0.93 & 0.92 \\
\midrule
B3LYP+D3       & 1.24 & 0.75 & 0.60 & 0.97 & 0.90 & 0.88 \\
PBE0+MBD       & 0.72 & 0.96 & 0.77 & 0.98 & 0.74 & 0.95 \\
M06-2X         & 0.48 & 0.97 & 0.69 & 0.98 & 0.59 & 0.96 \\
SCAN           & 0.41 & 0.98 & 0.50 & 0.98 & 0.46 & 0.98 \\
MP2/CBS        & 0.31 & 1.00 & 0.35 & 0.99 & 0.33 & 0.99 \\
\bottomrule
\end{tabular}
\label{tab:marianski-methods}
\end{table*}

\clearpage 
\subsection{Metal--organic frameworks}\label{sec:mof}
\vspace{-2mm}

Metal--organic frameworks (MOFs) are periodic, metal-containing solids and therefore test SO3LR-T on a class of systems that differs from its predominantly molecular training data. We evaluate the models in two ways. First, we assess the accuracy of forces and pressures on snapshots of two frameworks. Second, we evaluate relaxed structures and bulk moduli on the MOFSimBench set.

\begin{figure*}[h]
\centering
\includegraphics[width=0.99\linewidth]{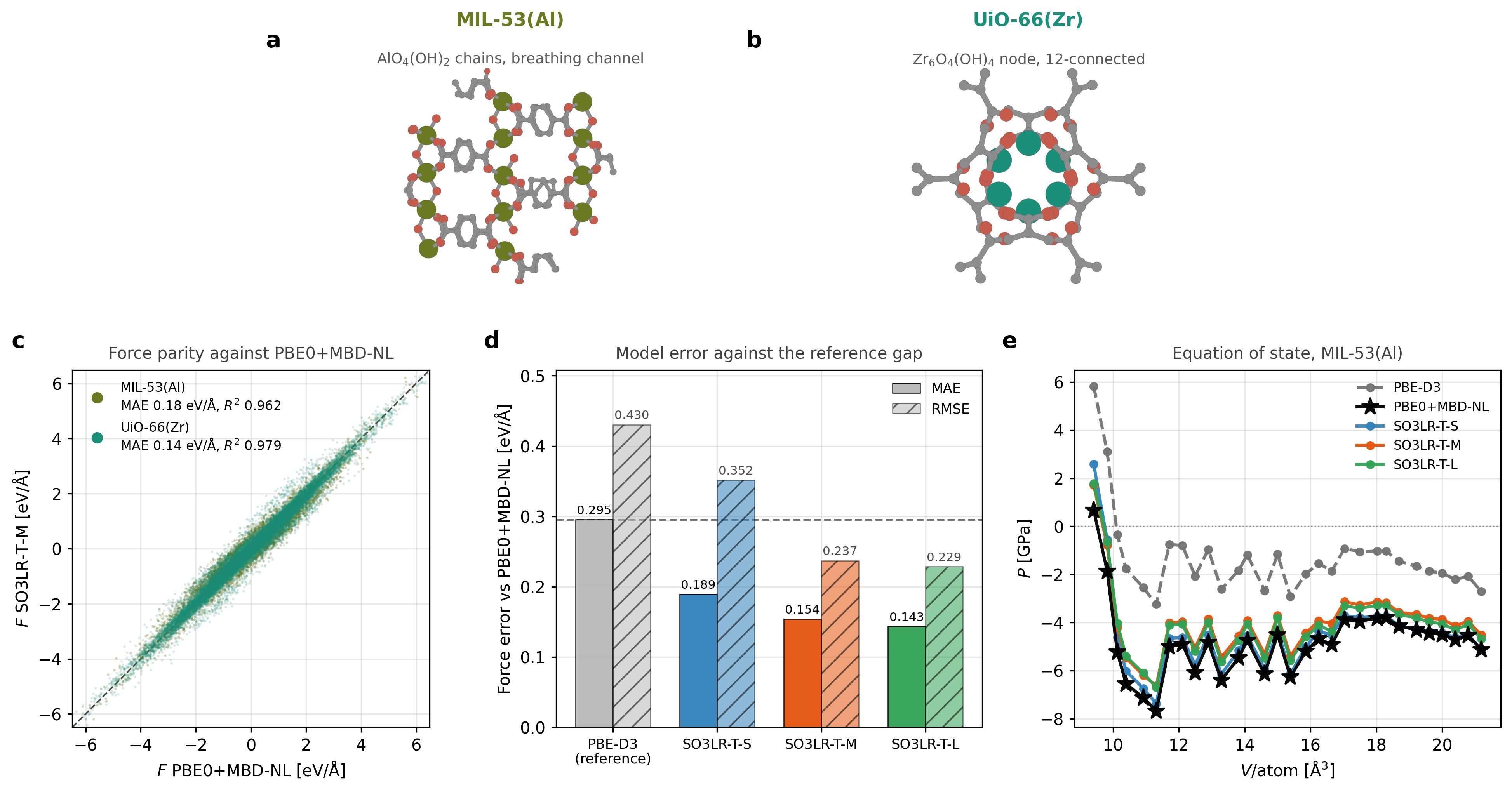}
\caption{\textbf{SO3LR-T on metal--organic frameworks.} \textbf{(a,b)}~MIL-53(Al) and UiO-66(Zr); Al olive, Zr teal-green, O red, C grey, H omitted. \textbf{(c)}~Force parity of SO3LR-T-M against PBE0+MBD-NL on $50$ snapshots. \textbf{(d)}~Force MAE and RMSE against PBE0+MBD-NL for the PBE-D3 reference and the three SO3LR-T models; the dashed line marks the PBE-D3 MAE. \textbf{(e)}~Hydrostatic pressure against volume for MIL-53(Al), with all levels evaluated on the same $30$ MD snapshots}\vspace{-2mm}
\label{fig:mof}
\end{figure*}

The two frameworks, UiO-66(Zr) and MIL-53(Al), are taken from a published CP2K PBE-D3 reference set~\cite{vandenhaute2023mof}. UiO-66(Zr) is built from \ce{Zr6O4(OH)4} clusters, and MIL-53(Al) is a flexible (breathing) framework composed of \ce{AlO4(OH)2} chains (Fig.~\ref{fig:mof}a,b). All calculations use particle--mesh Ewald electrostatics and double (float64) precision. Snapshots from the reference set are replicated until every cell height exceeds twice the \qty{12}{\angstrom} long-range cutoff plus \qty{3}{\angstrom}, and are evaluated under the minimum-image convention. MOFSimBench frameworks are relaxed as unit cells, following the published protocol, using a neighbor list that includes all periodic images within the cutoff.

\textbf{Reference level.} Since SO3LR-T is trained on PBE0+MBD-NL, we recomputed \num{50} snapshots at this level with FHI-aims~\cite{blum2009aims, hermann2020mbdnl}. The snapshots were selected to span the available range of cell volumes and comprise \num{30} MIL-53(Al) structures, covering both phases, and \num{20} UiO-66(Zr) structures. Relative to this reference, the published PBE-D3 forces have an MAE of \qty{0.295}{eV/\angstrom}, whereas SO3LR-T-S, -M and -L achieve \num{0.189}, \num{0.154} and \qty{0.143}{eV/\angstrom}, respectively (Fig.~\ref{fig:mof}c,d). PBE-D3 overestimates the pressure by \qty{3.2}{GPa} on average, compared with \qtyrange{0.6}{1.0}{GPa} for the models (Fig.~\ref{fig:mof}e). The PBE-D3 offset depends on volume and decreases from \num{5.2} to \qty{2.4}{GPa} across the breathing range of MIL-53(Al). Control calculations on two MIL-53(Al) snapshots show that most of this difference arises from exact exchange. Changing only the exchange treatment, with the code, basis set and dispersion correction kept fixed, accounts for \qty{0.226}{eV/\angstrom} of the \qty{0.285}{eV/\angstrom} difference in forces and for \qty{2.68}{GPa} of the \qty{2.99}{GPa} difference in pressure. The basis set is converged, since increasing the basis settings from \texttt{intermediate} to \texttt{tight} changes the forces by only \qty{0.021}{eV/\angstrom} and the pressure by only \qty{0.005}{GPa}.

\textbf{Error by site.} Measured relative to the local root-mean-square reference force, the PBE-D3 error concentrates on the carboxylate oxygens (${\sim}50\%$, against ${<}10\%$ at the metals). The model error concentrates on the Zr node instead, falling with model size from $87\%$ ($R^2 = 0.63$) for SO3LR-T-S to $34\%$ and $26\%$ for -M and -L; the Al node and the linker stay within ${\sim}20\%$ and $11\%$ at every size.

\textbf{MOFSimBench.} Following the published protocols~\cite{krass2025mofsimbench} on $100$ nanoporous frameworks with PBE-D3(BJ) reference geometries, SO3LR-T-S, -M, and -L give median $|\Delta V|$ of $2.78$, $2.27$, and $2.63\%$ and median RMSD of $0.299$, $0.257$, and $0.227$~\AA{} (range of the $22$ published models: $1.06$--$26.2\%$ and $0.204$--$1.49$~\AA); $97$--$98$ frameworks reach $f_\mathrm{max} \leq 10^{-3}$~eV/\AA{}. The volume error is a systematic contraction, with a median signed $\Delta V$ of $-1.5$ to $-1.9\%$, and the frameworks stay intact: the median fraction of atoms that change coordination is zero. Bulk moduli give median errors of $2.74$, $1.83$, and $1.89$~GPa ($r = 0.78$, $0.67$, $0.86$; published range $1.23$--$10.0$~GPa), with no systematic softening. We note that references are computed at PBE-D3(BJ), a lower level of theory than the model's training data. For example, at fixed geometry, PBE-D3 gives higher pressures than PBE0+MBD-NL (Fig.~\ref{fig:mof}e), which favors larger cells, so part of the systematic contraction may reflect the reference level.

\section{Reactive chemistry}\label{sec:si-reactions}

The following section presents the results of the reactivity studies described in the main manuscript. The reactivity results in Fig.~\ref{fig:reactivity} cover 17 transition states (TSs): nine in the proline aldol enamine cycle, four aldol diastereomers ($6a$--$6d$), three in the benzoin condensation, and one in the cobalamin methyl transfer. All results are reported for SO3LR-T-L, with SO3LR-T-M discussed in Sec.~\ref{sec:reactions-medium}. Geometries, energies, Hessian summaries, analysis scripts, and renderings of all transition states are provided in the Electronic Supplementary Material (ESM).

\subsection{Settings, transition-state search and validation}\label{sec:si-react-search}

\noindent
All calculations use total charge $0$, a closed-shell singlet, and float64 precision. Energies are reported as gas-phase electronic energies without zero-point, thermal, or implicit-solvation corrections. The activation and reaction energies are defined as $\Delta E^\ddagger = E(\mathrm{TS})-E(\mathrm{reactant})$ and $\Delta E_\mathrm{rxn} = E(\mathrm{product})-E(\mathrm{reactant})$, respectively. All searches used the automated \textsc{mars-rover} workflow~\cite{gallegos2026marsrover}, with SO3LR-T-L as the sole source of energies and forces.
Further details regarding the system-specific TS searches discussed in the main manuscript are provided below:

\begin{itemize}

\item For the aldol systems, the search was blind. Reactant and product minima were optimized using L-BFGS until the maximum force fell below $0.02$~eV/\AA. A climbing-image nudged elastic band (CI-NEB)~\cite{henkelman2000cineb} with 12 images, densified to ${\sim}60$ where needed, was converged using alternating L-BFGS and FIRE optimization, with the force threshold tightened from $0.05$ to $0.02$~eV/\AA. The highest-energy image was then refined with Sella~\cite{hermes2022sella} until the maximum force fell below $0.01$~eV/\AA.

\item For benzoin, which involves a flexible catalyst, and the 207-atom solvated cobalamin cluster, Sella was initialized from published TS geometries~\cite{huang2022benzoin,jensen2003meth}. 

\end{itemize}

\noindent
As for the validation, each valid saddle has a single imaginary frequency above $50$~cm$^{-1}$, obtained from a numerical Hessian with a $0.01$~\AA{} displacement, and was validated by an intrinsic reaction coordinate (IRC) calculation (arc-length step $0.05$~\AA, force tolerance $0.1$~eV/\AA, maximum 100 steps). A TS is considered validated if each IRC terminus lies within $0.5$~\AA{} of the corresponding reactant or product minimum, based on Kabsch-aligned heavy-atom RMSD.

\subsection{Reference data}\label{sec:si-react-refs}

For comparison, reference data were used to assess the validity and quality of the reactive chemistry results. These data were obtained either from in-house ab initio calculations or from the literature, including both computational and experimental studies whenever available:

\begin{itemize}
\item For the aldol systems, the reference method is PBE0-D3(BJ)~\cite{adamo1999pbe0, grimme2010d3, grimme2011d3bj}, implemented in ORCA 6.0.0~\cite{neese2022orca} (tight SCF, DEFGRID3). The stationary points, referred to below as DFT-optimized, were optimized with def2-SVP~\cite{weigend2005def2}, using an exact Hessian for the transition states, and verified by analytical frequency calculations. All energies are def2-TZVP single points evaluated on either the DFT-optimized or SO3LR-T-L geometries. The diastereomers are also compared with the relative TS energies at the B3LYP/6-31G(d) level reported by Bahmanyar \emph{et al.}~\cite{houk2003transition}.

\item For benzoin, we compare our electronic energies with the free-energy barriers reported by Huang \emph{et al}.~\cite{huang2022benzoin}. Because their values are free energies and ours are electronic energies, and because the rate-determining proton migration is relayed by a second aldehyde molecule, the agreement should be taken as indicative rather than as a matched-level benchmark.

\item For cobalamin, we use the truncated-cluster barrier band of Kozlowski \emph{et al.}~\cite{kumar2012jbic} and the forming \mbox{S$\cdots$C} distance of the Jensen--Ryde TS~\cite{jensen2003meth}. Since the Jensen--Ryde value comes from the dispersion-free B3LYP/6-31G(d) level, we re-optimized their 63-atom model at PBE0-D3(BJ)/def2-TZVP, in the gas phase and in implicit water (CPCM~\cite{barone1998cpcm}; Sec.~\ref{sec:si-react-jr}).

\end{itemize}

  In addition, given the potentially substantial influence of dispersion interactions on reaction barriers and selectivities, we evaluated many-body dispersion corrections for all in-house gas-phase reference structures. PBE0+MBD and PBE0+MBD-NL single-point calculations were performed with FHI-aims~\cite{blum2009aims,ambrosetti2014mbd, hermann2020mbdnl} (tight \texttt{defaults\_2020} basis, scalar-relativistic ZORA) on the DFT-optimized aldol and cobalamin structures. Additional PBE0+MBD-NL single points were computed on the SO3LR-T-L stationary points (reactants and transition states) of the aldol cycle. These two sets of calculations serve distinct purposes. The single points on the DFT-optimized structures assess the sensitivity of the reference data to the choice of dispersion correction, whereas those on the SO3LR-T-L geometries provide a consistent basis for evaluating the SO3LR-T-L model.

  \bigbreak
  \noindent
  \textit{Sensitivity of the reference data to the dispersion correction.} The reference transition states were optimized with PBE0-D3(BJ), which employs a pairwise dispersion correction, whereas the SO3LR-T production head was trained on MBD(-NL)-corrected data. To assess whether this difference affects the reference barriers, we evaluated MBD and MBD-NL corrections at the DFT-optimized geometries. Replacing D3(BJ) with MBD or MBD-NL shifts the 13 aldol barriers by an average of $0.4$ (MBD) and $0.6$~kcal/mol (MBD-NL), with a maximum shift of $1.3$~kcal/mol (Tables~\ref{tab:si_react_cycle} and~\ref{tab:si_aldol_val1}). These shifts are nearly uniform across the four diastereomers, leaving their relative energies largely unchanged (differences below $0.3$~kcal/mol). For the gas-phase 63-atom cobalamin model, the barrier increases by $1.9$~kcal/mol with MBD and by $1.7$~kcal/mol with MBD-NL (Table~\ref{tab:si_react_cobalamin_recalc}). The two many-body dispersion schemes differ from each other by no more than $0.6$~kcal/mol in any case. Although the absolute barriers show some sensitivity to the dispersion treatment, the conclusions drawn from the original D3(BJ)-corrected reference data remain unchanged.

  \bigbreak
  \noindent
  \textit{Evaluation of the SO3LR-T-L model.} A separate question is how accurately the model predicts energies at a given geometry. Comparing barriers obtained at the respective stationary-point geometries of each method conflates errors in the predicted energies with errors in the optimized geometries (Sec.~\ref{sec:si-react-aldol}). To isolate the energy error, we evaluated the SO3LR-T-L geometries using single-point calculations. %

\subsection{Proline aldol enamine cycle}\label{sec:si-react-aldol}

\begin{figure*}[h!]
\centering
\includegraphics[width=0.6\linewidth]{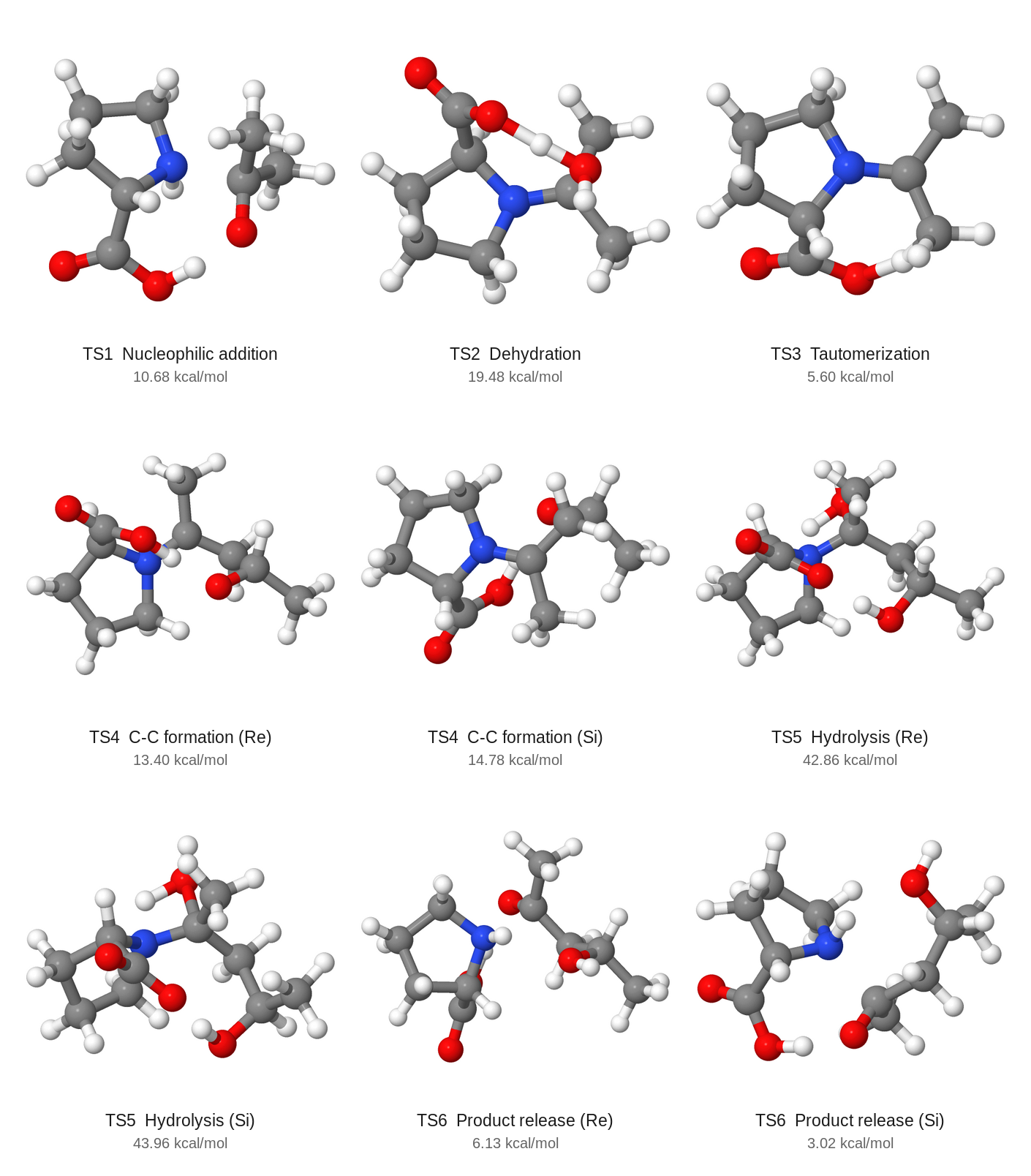}
\caption{\textbf{Proline aldol enamine cycle: SO3LR-T-L transition states.} The nine saddles (Table~\ref{tab:si_react_cycle}), each labeled with its forward barrier and shown in its principal-axis orientation}
\label{fig:si_react_cycle}
\end{figure*}

We now proceed to the proline aldol enamine cycle, whose nine saddles span the mechanism from carbinolamine formation to hydrolytic release. The saddles were located blindly (Sec.~\ref{sec:si-react-search}) and are compared with the PBE0 reference in geometry, barrier height and, for the C--C forming step, stereoselectivity. SO3LR-T-L locates all nine (Table~\ref{tab:si_react_cycle}, Fig.~\ref{fig:si_react_cycle}).

\begin{table*}[h!]
\caption{\textbf{Proline aldol enamine cycle on SO3LR-T-L.} Forward barriers $\Delta E^\ddagger$ against PBE0 at two sets of geometries: single points on the SO3LR-T-L structures (PBE0-D3(BJ), ORCA; PBE0+MBD-NL, FHI-aims), and the PBE0-D3(BJ)-optimized stationary points with PBE0+MBD and PBE0+MBD-NL single points. $\Delta E_\mathrm{rxn}$ and $\nu^\ddagger$ are SO3LR-T-L values; RMSD is the heavy-atom deviation of each SO3LR-T-L saddle from the PBE0-D3(BJ) saddle. Energies in kcal/mol, $\nu^\ddagger$ in cm$^{-1}$, RMSD in \AA} \label{tab:si_react_cycle} \label{tab:si_react_mbd}
\centering
\begin{tabular}{l r r r @{\hspace{12pt}} r r r @{\hspace{12pt}} r r r}
\toprule & \multicolumn{3}{c}{SO3LR-T-L geometry} & \multicolumn{3}{c}{PBE0-D3(BJ)-optimized geometry} & & & \\
\cmidrule(lr){2-4}\cmidrule(lr){5-7}
Step & SO3LR-T-L & \makecell[c]{PBE0-\\D3(BJ)} & \makecell[c]{PBE0+\\MBD-NL} & \makecell[c]{PBE0-\\D3(BJ)} &
\makecell[c]{PBE0+\\MBD} & \makecell[c]{PBE0+\\MBD-NL} & $\Delta E_\mathrm{rxn}$ & $\nu^\ddagger$ & RMSD \\
\midrule
TS1 Nucleophilic addition & 10.68 & 7.53  & 6.73  & 7.93  & 7.40  & 7.17  & $-5.27$  & $-162$  & 0.07 \\
TS2 Dehydration           & 19.48 & 18.33 & 18.49 & 17.99 & 18.32 & 18.26 & $+8.60$  & $-471$  & 0.21 \\
TS3 Tautomerization       & 5.60  & 3.24  & 3.38  & 7.19  & 7.11  & 7.07  & $-12.02$ & $-1331$ & 0.09 \\
TS4 C--C formation ($Re$) & 13.40 & 7.02  & 6.28  & 6.13  & 5.99  & 5.64  & $-8.40$  & $-737$  & 0.08 \\
TS4 C--C formation ($Si$) & 14.78 & 7.87  & 7.15  & 7.07  & 6.89  & 6.46  & $-6.91$  & $-718$  & 0.08 \\
TS5 Hydrolysis ($Re$)     & 42.86 & 42.74 & 43.24 & 51.27 & 51.75 & 51.56 & $+12.00$ & $-1329$ & 0.17 \\
TS5 Hydrolysis ($Si$)     & 43.96 & 44.40 & 44.82 & 44.79 & 45.35 & 45.09 & $+13.84$ & $-1335$ & 0.18 \\
TS6 Product release ($Re$)& 6.13  & 4.41  & 4.98  & 3.62  & 3.96  & 4.11  & $-6.64$  & $-163$  & 0.13 \\
TS6 Product release ($Si$)& 3.02  & 0.75  & 1.19  & 2.11  & 2.29  & 2.35  & $-7.42$  & $-277$  & 0.05 \\
\bottomrule
\end{tabular}
\end{table*}

  \bigbreak
  \noindent
  \textit{Transition-state geometries.} As for the quality of the TS geometries, SO3LR-T-L saddles deviate from the DFT-optimized saddles by $0.12$~\AA{} on average (Kabsch-aligned heavy-atom RMSD) and by at most $0.21$~\AA{} (TS2).

  \bigbreak
  \noindent
  \textit{Energy error at fixed geometry.} A barrier predicted by the model carries two kinds of error, one from its energies and one from its geometries. To look at the energies alone, we recomputed the PBE0 barriers on the SO3LR-T-L structures themselves (Sec.~\ref{sec:si-react-refs}), so that any remaining difference comes from the energy model. Over the nine barriers, the MAE is $2.72$~kcal/mol against PBE0-D3(BJ) and $2.90$~kcal/mol against PBE0+MBD-NL. In both cases the largest deviation is at TS4 \emph{Si} ($6.91$ and $7.63$~kcal/mol).

  \bigbreak
  \noindent
  \textit{Barriers from independently optimized saddles.} In practice, it becomes useful to evaluate the performance of the MLFF by comparing the barrier it predicts with the DFT analog, each computed at its own optimized saddle, so that geometry and energy errors add up. In this comparison the MAE against PBE0+MBD-NL rises to $3.87$~kcal/mol. The largest deviation, $8.70$~kcal/mol at TS5 \emph{Re}, is a geometry effect, since at the common SO3LR-T-L geometry the two barriers differ by only $0.38$~kcal/mol.

  \bigbreak
  \noindent
  \textit{Stereoselectivity of the C--C forming step.} The stereochemical outcome of the reaction is determined at TS4, where the \emph{Re} and \emph{Si} pathways compete. SO3LR-T-L favors the \emph{Re} pathway by $\Delta\Delta E^\ddagger = 1.38$~kcal/mol, corresponding to an estimated enantiomeric excess (ee) of $82\%$ at $298$~K. At the DFT-optimized saddle geometries, PBE0+MBD-NL predicts a smaller gap of $0.82$~kcal/mol, corresponding to an ee of $60\%$. Both predictions are reasonably close to the measured ee of $76\%$ for the analogous reaction in DMSO~\cite{list2000proline,allemann2010stereoselectivities}.

\subsection{Aldol diastereomers}\label{sec:si-react-diast}

The ability to distinguish between closely related transition states is particularly important for predicting stereoselectivity. We therefore turn now to the four diastereomeric C--C saddles on the larger substrate of Bahmanyar \emph{et al.}~\cite{houk2003transition}, which test whether SO3LR-T-L resolves small energy differences between related TSs (Table~\ref{tab:si_aldol_val1}; structures in the ESM).

\begin{table*}[h!]
\caption{\textbf{Proline aldol diastereomers on SO3LR-T-L.} Forward barriers (upper block) and TS energies relative to $6a$ (lower block). PBE0 columns as in Table~\ref{tab:si_react_cycle}; Bahmanyar: B3LYP/6-31G(d) relative energies~\cite{houk2003transition}. Relative energies are taken between the TSs themselves and differ from differences of the barriers, since each channel starts from its own reactant complex. $\nu^\ddagger$ and the heavy-atom RMSD to the Bahmanyar saddle are SO3LR-T-L values. Energies in kcal/mol, $\nu^\ddagger$ in cm$^{-1}$, RMSD in \AA} \label{tab:si_aldol_val1} \centering
\begin{tabular}{l r r @{\hspace{12pt}} r r r @{\hspace{12pt}} c r r}
\toprule
& \multicolumn{2}{c}{\makecell[c]{SO3LR-T-L\\geometry}} & \multicolumn{3}{c}{\makecell[c]{PBE0-D3(BJ)-optimized\\geometry}} & & & \\
\cmidrule(lr){2-3}\cmidrule(lr){4-6}
Diastereomer & SO3LR-T-L & \makecell[c]{PBE0-\\D3(BJ)} & \makecell[c]{PBE0-\\D3(BJ)} & \makecell[c]{PBE0+\\MBD} & \makecell[c]{PBE0+\\MBD-NL} & Bahmanyar & $\nu^\ddagger$ & RMSD \\
\midrule
\multicolumn{9}{l}{\emph{Forward barrier $\Delta E^\ddagger$}} \\
6a & 16.98 & 8.38  & 7.44  & 6.70  & 6.31  & --   & $-741$ & 0.10 \\
6b & 21.59 & 14.13 & 12.91 & 12.10 & 11.60 & --   & $-799$ & 0.05 \\
6c & 23.30 & 16.56 & 14.98 & 14.31 & 13.83 & --   & $-721$ & 0.08 \\
6d & 19.40 & 14.27 & 12.02 & 11.56 & 11.07 & --   & $-690$ & 0.14 \\
\midrule
\multicolumn{9}{l}{\emph{Transition-state energy relative to $6a$}} \\
6a & 0.00 & 0.00 & 0.00 & 0.00 & 0.00 & 0.00 & & \\
6b & 5.10 & 6.76 & 6.76 & 6.62 & 6.56 & 6.70 & & \\
6c & 6.20 & 8.49 & 8.36 & 8.46 & 8.41 & 7.50 & & \\
6d & 2.05 & 5.75 & 5.24 & 5.50 & 5.44 & 4.43 & & \\
\bottomrule
\end{tabular}
\end{table*}

\bigbreak
\noindent
\textit{Forward barriers of the four channels.} The four transition states correspond to the same C--C forming step as TS4 in the reaction cycle, but involve a larger substrate. SO3LR-T-L overestimates all four forward barriers by $5.1$--$8.6$~kcal/mol relative to PBE0-D3(BJ) at fixed geometry, giving a mean absolute error (MAE) of $6.98$~kcal/mol. When each method is evaluated at its own optimized saddle, the MAE increases to $8.48$~kcal/mol relative to PBE0-D3(BJ), and to $9.61$~kcal/mol relative to PBE0+MBD-NL. Notably, the overestimation is comparable to that observed for the two TS4 pathways, \emph{Re} and \emph{Si}, for which the errors at fixed geometry are $6.38$ and $6.91$~kcal/mol, respectively. This suggests that the systematic error is associated primarily with the C--C forming step itself, rather than with the increased substrate size. Furthermore, because the overestimation is similar across all four channels, much of the error cancels when comparing their relative energies.

\bigbreak
\noindent
\textit{Transition-state energies relative to $6a$.} Because the four transition states have the same composition, their relative energies can be compared directly, independently of the energies of their respective reactant complexes. These energy differences are particularly relevant for predicting the stereochemical outcome of the reaction. Taking $6a$ as the reference ($\Delta E = 0$), the MAE of SO3LR-T-L for the energies of $6b$--$6d$ is $1.76$~kcal/mol relative to Bahmanyar, $2.55$~kcal/mol relative to PBE0-D3(BJ) evaluated at the SO3LR-T-L geometries, and $2.34$~kcal/mol relative to PBE0-D3(BJ) at the DFT-optimized geometries. Importantly, SO3LR-T-L reproduces the ordering $6a < 6d < 6b < 6c$ for all three references. The same ordering is also obtained when comparing forward barriers, with one exception: PBE0-D3(BJ) at the SO3LR-T-L geometries places $6b$ marginally ($0.14$~kcal/mol) below $6d$. Thus, despite the systematic overestimation of the absolute forward barriers, SO3LR-T-L correctly captures the relative stability of the four diastereomeric transition states, although it underestimates their energy differences with respect to $6a$, particularly for $6d$.

\subsection{Benzoin condensation}\label{sec:si-react-benzoin}

 We now turn to the SIPr-catalyzed benzoin condensation, for which three transition states were refined starting from the published geometries of Huang \emph{et al.}~\cite{huang2022benzoin} (seeded refinement, Sec.~\ref{sec:si-react-search}; Table~\ref{tab:si_react_benzoin}; structures in the ESM). These correspond to carbene addition (TS1), the rate-determining proton migration relayed by a second aldehyde (TS3), and the direct proton shift (TS3$'$). All three refined structures remain within $0.34$~\AA{} heavy-atom RMSD of their respective starting geometries. For the two transition states with published barriers, the MAE relative to Huang is $0.56$~kcal/mol, and the large energy gap between the relayed migration and the direct shift is reproduced. Since Huang's barriers are free energies whereas ours are electronic energies, this agreement is only indicative (Sec.~\ref{sec:si-react-refs}). Nevertheless, SO3LR-T-L retains all three published transition states and reproduces the energetic separation between the two proton-transfer pathways.

\begin{table*}[h!]
\caption{\textbf{Benzoin condensation on SO3LR-T-L.} Forward barriers against the Huang free-energy reference~\cite{huang2022benzoin}, imaginary frequency, heavy-atom RMSD of each refined saddle to the reference saddle it was seeded from, and key forming-bond length $d$. Energies in kcal/mol, $\nu^\ddagger$ in cm$^{-1}$, RMSD and $d$ in \AA}
\label{tab:si_react_benzoin}
\centering
\begin{tabular}{l r r r r r r}
\toprule
Transition state & $\Delta E^\ddagger$ (SO3LR-T-L) & $\Delta E^\ddagger$ (Huang) & $\nu^\ddagger$ & RMSD & $d$ (SO3LR-T-L) & $d$ (ref.) \\
\midrule
TS1 carbene addition   & 11.56 & --    & $-539$  & 0.23 & 1.93 & 2.16 \\
TS3 proton migration (RDS) & 5.93  & 6.28  & $-680$  & 0.14 & 1.33 & 1.32 \\
TS3$'$ direct shift    & 41.30 & 40.53 & $-1939$ & 0.34 & 1.33 & 1.32 \\
\bottomrule
\end{tabular}
\end{table*}

\subsection{Cobalamin methyl transfer}\label{sec:si-react-cobalamin}

\begin{figure}[h!]
\centering
\includegraphics[width=0.32\linewidth]{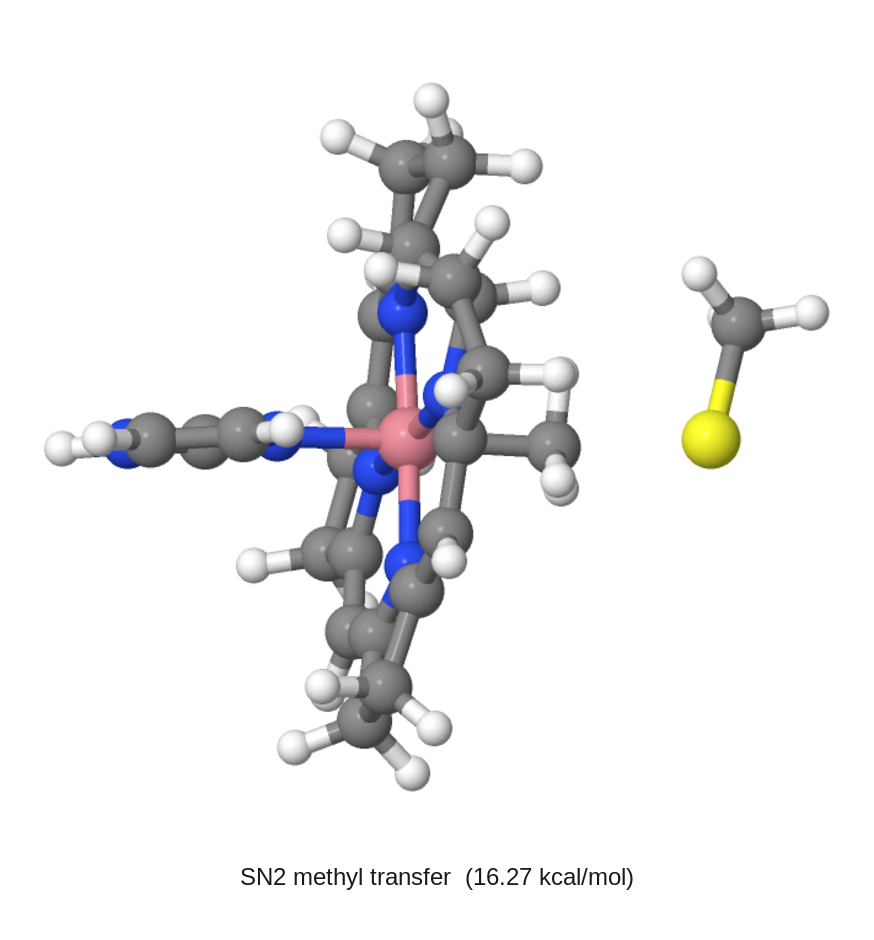}
\caption{\textbf{Cobalamin \mbox{S$_\mathrm{N}$2} methyl transfer: SO3LR-T-L transition state.} The 63-atom solute of the 207-atom cluster (Table~\ref{tab:si_react_cobalamin_recalc}), waters omitted. Co pink, N blue, S yellow, C grey, H white}
\label{fig:si_react_cobalamin}
\end{figure}

  The final system considered is the cobalamin methyl transfer, modeled using a 207-atom base-off/His-on cluster containing the corrin, an axial imidazole representing the His ligand, the transferring methyl group, CH$_3$S$^-$, and 48 water molecules (Fig.~\ref{fig:si_react_cobalamin}). As in the benzoin condensation, the search was initiated from a published transition-state geometry (Sec.~\ref{sec:si-react-search}). SO3LR-T-L refines this structure into a first-order saddle, with a single imaginary frequency of $-498$~cm$^{-1}$. The resulting transition state describes backside methyl transfer along the \mbox{Co$\cdots$C$\cdots$S} axis, with a forward barrier of $\Delta E^\ddagger = 16.27$~kcal/mol and a forming \mbox{S$\cdots$C} distance of $2.35$~\AA{} (Table~\ref{tab:si_react_cobalamin_recalc}). Previous studies reported lower barriers, ranging from $4.9$ to $10.5$~kcal/mol for truncated clusters~\cite{kumar2012jbic}, as well as a longer \mbox{S$\cdots$C} distance of $2.88$~\AA{} for the Jensen--Ryde transition state~\cite{jensen2003meth}. We revisit the Jensen--Ryde model in Sec.~\ref{sec:si-react-jr} and show that its low barrier and long distance largely arise from the level of theory and model environment used in the 2003 study.

\bigbreak  
\noindent
  \textit{Role of the axial base.} In the enzyme, the His ligand coordinated trans to the transferring methyl group is believed to assist the methyl transfer. The cluster model allows this effect to be tested directly by removing the axial imidazole. In its absence (198 atoms), SO3LR-T-L still locates an \mbox{S$_\mathrm{N}$2} saddle with a single imaginary frequency of $-506$~cm$^{-1}$, but the barrier increases to $22.51$~kcal/mol, corresponding to $\Delta\Delta E^\ddagger = +6.24$~kcal/mol with respect to the full cluster. The lowering of the barrier by the axial base is therefore consistent with the trans-base assistance ascribed to the enzyme.

\begin{table*}[h!]
\caption{\textbf{Cobalamin \mbox{S$_\mathrm{N}$2} methyl transfer: SO3LR-T-L and references.} Forming \mbox{S$\cdots$C} distance and electronic forward barrier $\Delta E^\ddagger$ (kcal/mol). The original 63-atom Jensen--Ryde model (corrin, axial imidazole, methyl, CH$_3$S$^-$; charge $0$, singlet) is recomputed at the present level; MBD and MBD-NL values are single points at the gas-phase PBE0-D3(BJ) geometry. SO3LR-T-L saddles are seeded from the reference saddle. The bare 63-atom references and the solvated SO3LR-T-L clusters do not give a like-for-like barrier comparison (Sec.~\ref{sec:si-react-jr})}
\label{tab:si_react_cobalamin_recalc}
\centering
\begin{tabular}{l l c c}
\toprule
Source & Level of theory / environment & \mbox{S$\cdots$C} (\AA) & $\Delta E^\ddagger$ \\
\midrule
Kozlowski \emph{et al.}~\cite{kumar2012jbic} & truncated clusters & -- & 4.9 to 10.5 \\
Jensen--Ryde 2003~\cite{jensen2003meth} & B3LYP/6-31G(d), gas, 63 atoms & 2.880 & 4.9 \\
This work & PBE0-D3(BJ)/def2-TZVP, gas, 63 atoms & 2.682 & 16.30 \\
This work & PBE0+MBD/tight (FHI-aims)~\cite{ambrosetti2014mbd, blum2009aims}, gas, 63 atoms & -- & 18.18 \\
This work & PBE0+MBD-NL/tight (FHI-aims)~\cite{hermann2020mbdnl}, gas, 63 atoms & -- & 18.03 \\
This work & PBE0-D3(BJ)/def2-TZVP, CPCM(water)~\cite{barone1998cpcm}, 63 atoms & 2.399 & 24.61\textsuperscript{a} \\
\midrule
SO3LR-T-L & 207 atoms, 48 explicit waters & 2.350 & 16.27 \\
SO3LR-T-L & 198 atoms, without the axial imidazole & -- & 22.51 \\
\bottomrule
\end{tabular}

\smallskip
\begin{minipage}{0.9\linewidth}
\footnotesize \textsuperscript{a} The CPCM barrier runs from a base-on reactant complex (Co--N$_\mathrm{ax}$ $2.11$~\AA) to a base-off saddle (Co--N$_\mathrm{ax}$ $3.48$~\AA), so it mixes base states and is inflated.
\end{minipage}
\end{table*}

\renewcommand{\thesubsubsection}{\thesubsection.\arabic{subsubsection}}
\subsubsection{Re-examination of the Jensen--Ryde reference}\label{sec:si-react-jr}

  The only structural benchmark available for this system is the 63-atom model of Jensen and Ryde, which was optimized at the dispersion-free B3LYP/6-31G(d) level~\cite{jensen2003meth}. To separate the effects of the level of theory from those of the environment, we re-optimized this model at PBE0-D3(BJ)/def2-TZVP (ORCA~6.0.0), both in the gas phase and in implicit water (CPCM~\cite{barone1998cpcm}), starting the reactant, transition-state and product searches from several initial geometries (Table~\ref{tab:si_react_cobalamin_recalc}). In water, two unrelated starting guesses converged to saddles whose \mbox{S$\cdots$C} distances differ by only $0.003$~\AA. In both environments, the resulting saddle has a single imaginary mode, predominantly along the Co$\cdots$S coordinate.

  \bigbreak
  \noindent
  \textit{Forming \mbox{S$\cdots$C} distance.} The re-optimized references make it possible to decompose the $0.53$~\AA{} difference between the forming-bond distance of the 2003 transition state ($2.88$~\AA) and that of SO3LR-T-L ($2.35$~\AA). Re-optimizing the reference at PBE0-D3(BJ)/def2-TZVP shortens the \mbox{S$\cdots$C} distance by about $0.20$~\AA, and including implicit solvation shortens it by a further $0.28$~\AA. Most of the discrepancy therefore originates from the reference itself, and SO3LR-T-L lies within $0.05$~\AA{} of the
  implicit-water value, as shown in Fig.~\ref{fig:reactivity}B.

  \bigbreak
  \noindent
  \textit{Barrier of the 63-atom model.} The same re-optimization reveals how much of the low barrier reported in 2003 is attributable to the level of theory. For the same 63-atom model, the barrier increases from $4.9$~kcal/mol at the original level to $16.30$~kcal/mol at PBE0-D3(BJ)/def2-TZVP, and to about $18$~kcal/mol when many-body dispersion is included (Sec.~\ref{sec:si-react-refs}). The 2003 geometry is more reliable than its energy: the Co--C and Co--N$_\mathrm{ax}$ distances change from $2.080$ and $2.700$~\AA{} to $2.069$ and $2.589$~\AA{} upon re-optimization. Because the reference is a bare 63-atom model whereas SO3LR-T-L describes a 207-atom solvated cluster, the two barrier heights cannot be compared directly. The meaningful comparison is instead the sensitivity of the reference barrier to the level of theory, which shows that the 2003 value is largely an artifact of the level at which it was computed.

\clearpage
\subsection{Comparison with SO3LR-T-M}\label{sec:reactions-medium}

  To test whether the largest model is required for reactive chemistry, all 17 transition states were re-evaluated with SO3LR-T-M in three ways (Table~\ref{tab:si_react_M_barriers}): single-point energies on the SO3LR-T-L geometries, Sella refinements seeded from the SO3LR-T-L saddles, and an independent de-novo search (CI-NEB for the aldol systems, reference-seeded refinement for benzoin and cobalamin). Every system retains a first-order saddle in at least one of the three evaluations. The missing entries in Table~\ref{tab:si_react_M_barriers} correspond to the de-novo benzoin searches, in which the catalyst relaxes to a minimum or the refinement does not converge, and to the seeded cobalamin refinement.

\bigbreak
\noindent
  SO3LR-T-M locates the same saddles as SO3LR-T-L, with the aldol saddles within $0.12$ to $0.14$~\AA{} (mean heavy-atom RMSD) of the reference geometries, but it predicts systematically higher aldol-cycle barriers, by roughly $3$~kcal/mol on average. As a result, the MAE against PBE0-D3(BJ) at the SO3LR-T-L geometries increases from $2.72$ to $5.5$--$5.8$~kcal/mol, and the estimated ee at TS4 falls from $82\%$ to $50$--$56\%$ (Table~\ref{tab:si_react_M_summary}). The medium model therefore reproduces the transition-state geometries but not the barrier heights, and the larger model is needed for quantitatively accurate barriers.

\begin{table*}[h!]
\caption{\textbf{Forward barriers of SO3LR-T-L and SO3LR-T-M} (kcal/mol) for the 17 TSs. M s.p.: SO3LR-T-M single point on the SO3LR-T-L geometry; M seeded: Sella refinement from the SO3LR-T-L saddle; M de-novo: independent search. All seeded and de-novo values are characterized first-order saddles (single imaginary mode). ``min'': the search relaxed to a minimum; ``--'': no converged saddle}
\label{tab:si_react_M_barriers}
\centering
\small
\begin{tabular}{l l r r r r r}
\toprule
System & Transition state & atoms & SO3LR-T-L & M s.p. & M seeded & M de-novo \\
\midrule
Aldol cycle  & TS1 Nucleophilic addition & 27 & 10.68 & 11.72 & 12.41 & 12.41 \\
             & TS2 Dehydration           & 27 & 19.48 & 26.04 & 25.80 & 25.79 \\
             & TS3 Tautomerization       & 24 & 5.60  & 8.84  & 9.74  & 9.74  \\
             & TS4 C--C formation ($Re$) & 31 & 13.40 & 15.65 & 15.56 & 15.58 \\
             & TS4 C--C formation ($Si$) & 31 & 14.78 & 16.30 & 16.31 & 16.33 \\
             & TS5 Hydrolysis ($Re$)     & 34 & 42.86 & 44.92 & 45.37 & 45.35 \\
             & TS5 Hydrolysis ($Si$)     & 34 & 43.96 & 46.47 & 46.57 & 46.57 \\
             & TS6 Product release ($Re$)& 34 & 6.13  & 9.78  & 9.51  & 9.47  \\
             & TS6 Product release ($Si$)& 34 & 3.02  & 6.33  & 7.48  & 7.09  \\
\midrule
Diastereomers  & 6a & 44 & 16.98 & 16.31 & 16.20 & 16.44 \\
               & 6b & 44 & 21.59 & 22.71 & 21.23 & 21.28 \\
               & 6c & 44 & 23.30 & 22.33 & 22.28 & 22.30 \\
               & 6d & 44 & 19.40 & 20.58 & 20.02 & 20.00 \\
\midrule
Benzoin       & TS1 carbene addition        & 81  & 11.56 & 5.32  & 2.82 & min \\
              & TS3 proton migration (RDS)  & 95  & 5.93  & 7.70  & 9.13 & --  \\
              & TS3$'$ direct shift         & 81  & 41.30 & 43.72 & 41.58 & 42.70 \\
\midrule
Cobalamin     & S$_\mathrm{N}$2 methyl transfer & 207 & 16.27 & 19.44 & -- & 17.14 \\
\bottomrule
\end{tabular}
\end{table*}

\begin{table*}[h!]
\caption{\textbf{Reactivity accuracy of SO3LR-T-L and SO3LR-T-M.} MAE of the nine aldol-cycle barriers against PBE0-D3(BJ) at the SO3LR-T-L geometries (kcal/mol), mean heavy-atom RMSD of the aldol saddles to the reference (\AA; the single points reuse the SO3LR-T-L geometry), and estimated ee of the TS4 step (\%)}
\label{tab:si_react_M_summary}
\centering
\begin{tabular}{l r r r}
\toprule
Model / evaluation & Aldol-cycle MAE vs PBE0-D3(BJ) & Mean TS RMSD & Estimated ee \\
\midrule
SO3LR-T-L                    & 2.72 & 0.12 & 82 \\
SO3LR-T-M, single point      & 5.53 & 0.12 & 50 \\
SO3LR-T-M, seeded refine     & 5.83 & 0.12 & 56 \\
SO3LR-T-M, de-novo search    & 5.78 & 0.14 & 56 \\
\bottomrule
\end{tabular}
\end{table*}

\clearpage

\section{Liquid simulations}\label{sec:liquids}

\textbf{Software and hardware.} All (bio)molecular dynamics simulations (Sec.~\ref{sec:liquids}-\ref{sec:md-biomol}) in this work were run with the \texttt{so3lr} package (\url{https://github.com/general-molecular-simulations/so3lr}), which integrates the equations of motion with JAX-MD~\cite{schoenholz2020jax}, on single NVIDIA A100 40~GB GPUs. The timing benchmarks of Sec.~\ref{sec:si-perf} are the exception and additionally employed AMD MI355X GPUs. %

\subsection{Neat liquids}\label{sec:si-neat}

\textbf{Initial structures.} Initial configurations for each of the neat molecular liquids were generated with Packmol 21.2.1~\cite{Martinez2009}, placing 254--1024 molecules (\numrange{1995}{5080} atoms) in a cubic box of 29--39~\AA{} sized from the experimental density. Intramolecular and intermolecular force-field parameters were taken from the OPLS-AA force field with 1.14*CM1A-LBCC partial atomic charges~\cite{jorgensen1996development, Dodda2017}. Initial equilibration was performed with GROMACS 2024.4~\cite{Abraham2015}. First, the energy was minimized until the maximum force fell below 1000~kJ~mol$^{-1}$~nm$^{-1}$ (up to \num{50000} steps). Second, a \qty{10}{\pico\second} NVT run at \qty{300}{\kelvin} was performed with the V-rescale thermostat and a time constant of \qty{100}{\femto\second}. Third, a \qty{500}{\pico\second} NPT run was performed using the same thermostat and the stochastic C-rescale barostat at \qty{1}{\bar} with a time constant of \qty{2}{\pico\second}. Four uncorrelated snapshots were extracted from the NPT trajectory at intervals of \qty{100}{\pico\second} and served as initial structures for the subsequent SO3LR simulations.

The above procedure was adopted for all organic liquids except CCl$_4$, for which no suitable force-field parameters were found. Here initial structures were generated with a custom packing procedure on a cubic grid. Each CCl$_4$ molecule was assigned a uniformly random orientation. For each molecule, up to 80 candidate placements were generated, and the orientation maximizing the minimum-image intermolecular distance to all previously placed atoms was retained, subject to a minimum-distance threshold of 1.9~\AA{}.

\textbf{SO3LR simulations.}  All SO3LR-T-M liquid simulations were run in double precision, with a 0.5~fs time step, a 12~\AA{} real-space long-range cutoff, and particle-mesh Ewald summation for both electrostatics and dispersion (smearing 2.0~\AA{}, grid spacing 1.55~\AA{}). Each of the four replicas was first equilibrated for 200~ps in the NPT ensemble at 300~K and 1.01325~bar with a Nos\'e--Hoover chain thermostat and barostat (time constants 50~fs and 500~fs). Production runs followed with time constants of 2~ps and 4~ps. Production lasted 1~ns per replica for water, acetonitrile, acetone, dichloromethane, and THF, and 2~ns for methanol, ethanol, formamide, DMSO, and hexane. For the more slowly converging liquids it was extended to 3.0--3.6~ns (DMF), 3.2--3.9~ns (benzene), 4.2--5.7~ns (ethylene glycol), 5.6--7.5~ns (acetic acid), and 9.5--11.2~ns (CCl$_4$). N-methylacetamide, which melts at about 301~K, was simulated at 320~K to match its experimental references. %

\textbf{Self-diffusion.}  For each replica, the MSD was computed from the NPT production trajectory and $D=\text{slope}/6$ obtained from a linear fit (ten log-spaced lag times, unweighted) to its diffusive regime. The fit window ends at $30\%$ of the trajectory length, and its start is detected per replica as the first lag time (after the minimum of the local log--log slope, and not before 1~ps) at which the local slope stays within $1\pm0.07$ for the following $0.3$ decades. The reported $D$ is the mean over the four replicas. The uncertainty is the larger of two estimates: the standard error of the four per-replica $D$ values (after finite-size correction), and the block-bootstrap error. For the latter, each replica's trajectory is split into five contiguous blocks, each block's MSD is fitted from the start time found for the full trajectory to the end of the block, the five block values are resampled with replacement $500$ times, and the standard deviation of the resampled means is averaged over replicas. This error is that of the block mean, whereas the point estimate uses the full-trajectory window, so the two are not strictly the same quantity.

\textbf{Finite-size correction.} Additionally, to mitigate the effects of hydrodynamic self-interaction of diffusing particles with their own periodically replicated images in finite MD boxes, the following correction by \citet{yeh2004system} is considered in post processing:
\begin{align}
D_{\text{corr}} = D_{\text{PBC}} + \frac{\xi\,k_B T}{6\pi\eta L},
\end{align}
where $D_{\text{PBC}}$ is the diffusion coefficient obtained from the mean-squared displacement in the simulated (finite) box, $\xi = 2.837297$ is the dimensionless self-term for a cubic lattice under periodic boundary conditions, $k_B T$ is the thermal energy, $\eta$ is the shear viscosity of the solvent, and $L$ is the cubic simulation box length. We take $L$ as the cube root of the mean simulation-cell volume over the equilibrated production trajectories, and use experimental shear viscosities at 298 K as $\eta$~\cite{Lide2003} (N-methylacetamide: 320 K~\cite{Williams1957}, with $T = 320$\,K). The same correction was applied to the reference OPLS4 diffusion coefficients using their reported NPT box dimensions; GAFF values lack corresponding box-size data and are reported uncorrected.

\textbf{Error metrics.} Simulated properties $X$ (densities or diffusion coefficients) are compared with experiment through the mean relative error over the $N$ solvents a method reports,
\begin{equation}
  \mathrm{MRE} = \exp\!\Big(\frac{1}{N}\sum_{i=1}^{N}\Big|\ln\frac{X_i}{X_i^\mathrm{exp}}\Big|\Big) - 1 .
  \label{eq:si-mre}
\end{equation}
Being based on log ratios, the MRE weighs over- and underestimation equally: a value twice and a value half the experimental one give the same error. An MRE of $x\,\%$ corresponds to a typical deviation from experiment by a factor of $1 + x/100$.

\begin{figure*}
\centering
\includegraphics[width=0.99\linewidth]{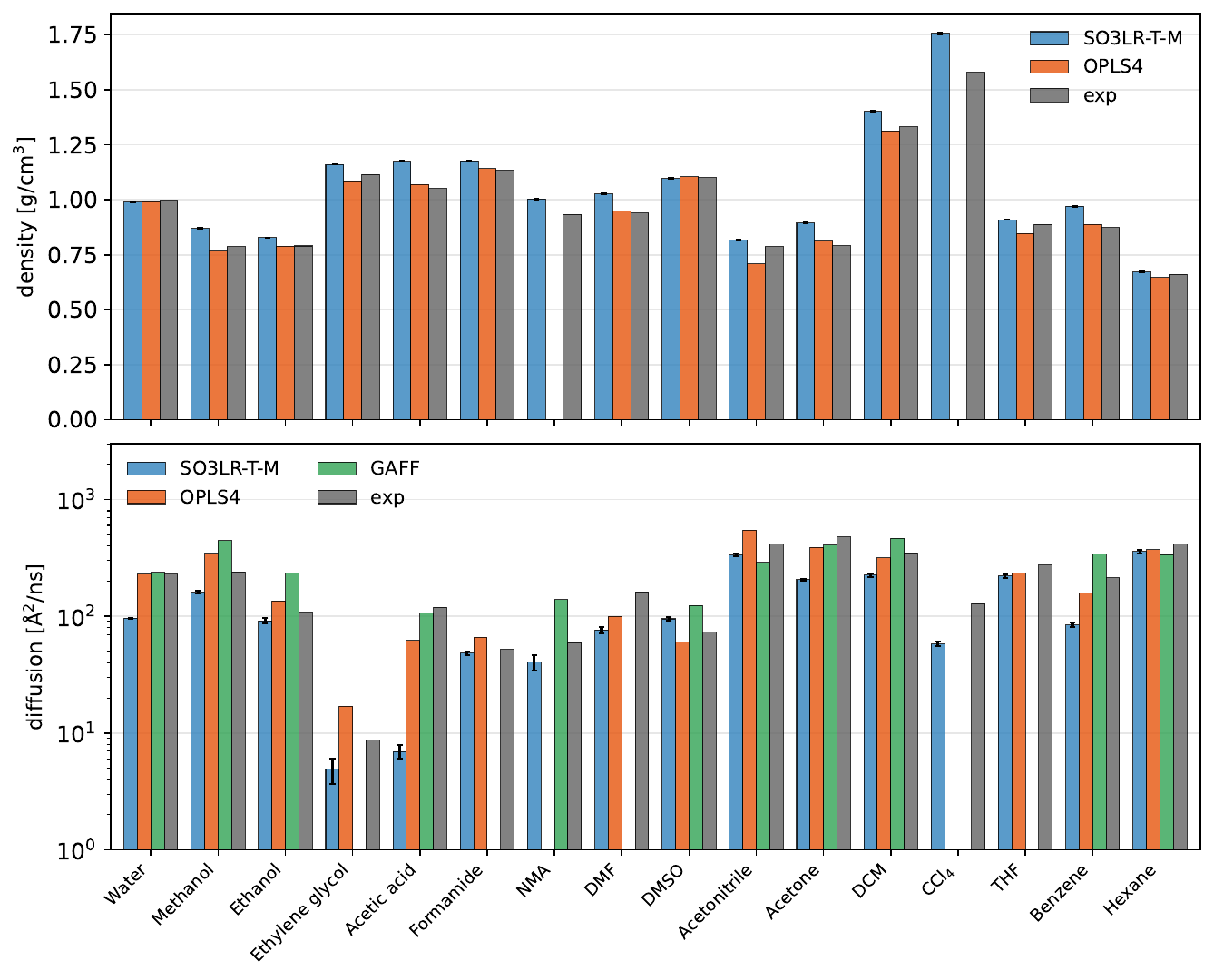}
\caption{\textbf{Bulk liquid densities and self-diffusion coefficients of 16 solvents from SO3LR-T-M, compared with classical force fields and experiment.} \emph{Top}: equilibrium mass densities (g/cm$^3$). \emph{Bottom}: self-diffusion coefficients without the Yeh--Hummer correction ($\mathrm{\AA^2/ns}$, logarithmic axis). Bars are colored by method: SO3LR-T-M (this work), OPLS4, GAFF, and experiment; GAFF is shown only for diffusion and OPLS4 only where parameters are available, so missing bars indicate values not reported for that method. Error bars on the SO3LR-T-M values show the standard error over replicas (top) and the uncertainty of $D$ defined in Sec.~\ref{sec:si-neat} (bottom). Solvents are ordered from polar protic through polar aprotic to nonpolar. Experimental values and the corresponding errors are listed in Tables~\ref{tab:bulk_density} and~\ref{tab:bulk_diffusion}}
\label{fig:bulk_bar}
\end{figure*}

\begin{figure*}
\centering
\includegraphics[width=0.99\linewidth]{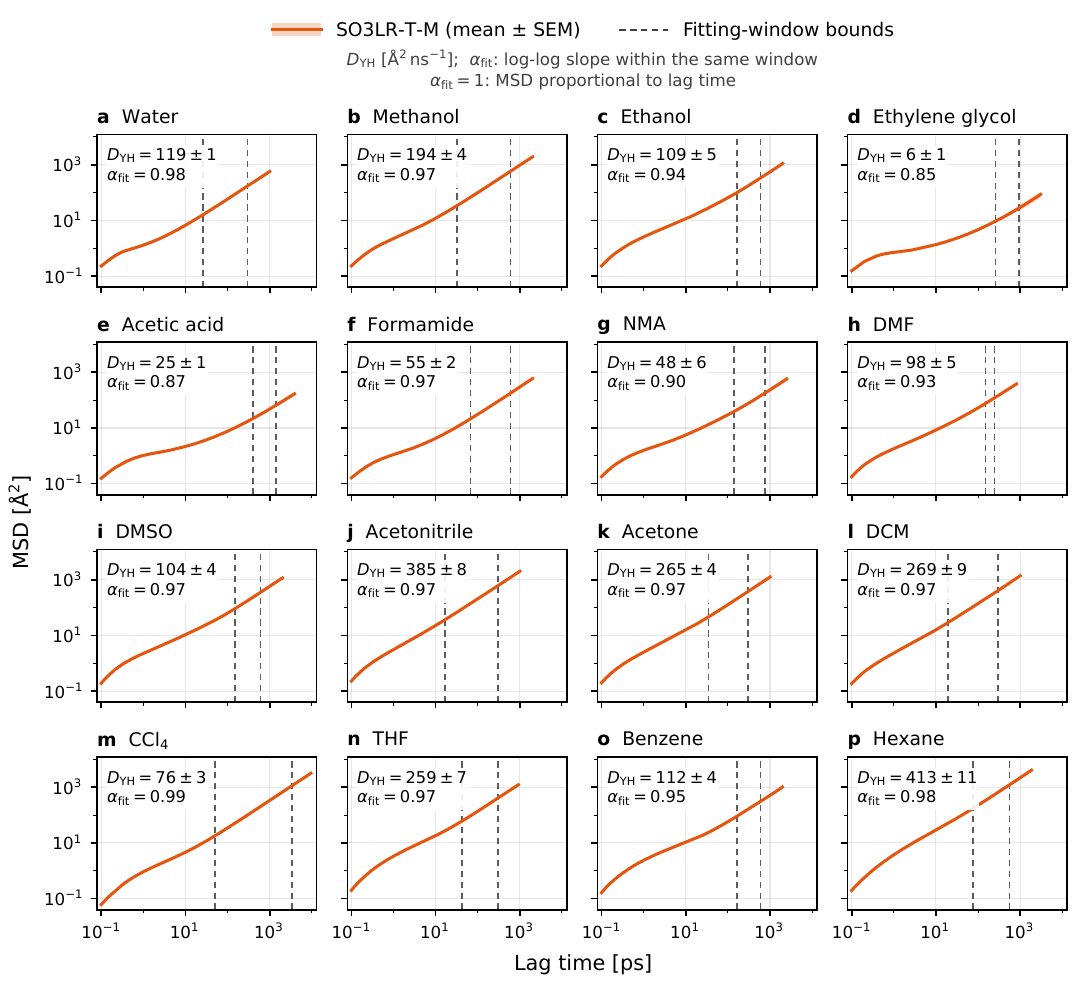}
\caption{\textbf{Mean-squared displacements of 16 molecular liquids simulated with SO3LR-T-M.} Solid orange curves show the mean MSDs from four independent trajectories; shaded bands indicate $\pm 1$ standard error of the mean. All panels use the same logarithmic axes, with each curve retaining its solvent-specific lag-time sampling and extent. Dashed vertical lines delimit the fitting intervals used to estimate the finite-box self-diffusion coefficient, $D_{\mathrm{PBC}}$, from $\mathrm{MSD}(t)=b+6D_{\mathrm{PBC}}t$. Panel annotations report the corresponding Yeh--Hummer-corrected diffusion coefficient, $D_{\mathrm{YH}}$, in \AA$^2$\,ns$^{-1}$; the MSD curves themselves are uncorrected. The uncertainties on $D_{\mathrm{YH}}$ are the larger of the seed-to-seed standard error and the block-bootstrap error (Sec.~\ref{sec:si-neat}); the fit interval is detected per replica, so the dashed lines mark that of one representative replica. The exponent $\alpha_{\mathrm{fit}}$, shown below each diffusion coefficient, is the slope of an unweighted linear regression of $\log\mathrm{MSD}$ against $\log t$ using the mean-MSD values within the same fitting interval. $\alpha_{\mathrm{fit}}\approx 1$ is consistent with a diffusive regime, in which the MSD grows linearly with lag time. NMA, N-methylacetamide; DMF, $N,N$-dimethylformamide; DMSO, dimethyl sulfoxide; DCM, dichloromethane; THF, tetrahydrofuran} 
\label{fig:liquid_msds}
\end{figure*}

\begin{table*}[htbp]
\setlength{\tabcolsep}{6pt}
\centering
\caption{\textbf{Bulk mass densities of 16 solvents.} Densities in g\,cm$^{-3}$. SO3LR-T-M was simulated at 300\,K and 1.01325\,bar, experimental values refer to $\approx$298\,K, except N-methylacetamide (simulation and reference at 320\,K). SO3LR-T-M values are averages over the last 1000 frames of the NPT production trajectories, taken per replica, and the $\pm$ is the standard error over the four replica means; OPLS4 values are reference simulation data. MAE $=\tfrac{1}{N}\sum_i|\rho_i-\rho_i^{\rm exp}|$ and MRE (Eq.~\ref{eq:si-mre}) are taken over the solvents each method reports (SO3LR-T-M $N{=}16$, OPLS4 $N{=}14$). Solvents are ordered from polar protic through polar aprotic to nonpolar; dashes mark solvents without published parameters}
\label{tab:bulk_density}
\begin{tabular}{l r@{\,}l r r}
\toprule
 & \multicolumn{4}{c}{Density (g\,cm$^{-3}$)} \\
\cmidrule(lr){2-5}
\scriptsize Solvent & \multicolumn{2}{c}{\scriptsize Exp.} & \scriptsize SO3LR-T-M & \scriptsize OPLS4 \\
\midrule
Water & 0.997 & \cite{Lemmon2023} & 0.990 $\pm$ 0.002 & 0.991 \\
Methanol & 0.786 & \cite{Lemmon2023} & 0.871 $\pm$ 0.002 & 0.766 \\
Ethanol & 0.790 & \cite{Hurle1985} & 0.829 $\pm$ 0.000 & 0.789 \\
Ethylene glycol & 1.115 & \cite{CAMEO2026} & 1.161 $\pm$ 0.001 & 1.083 \\
Acetic acid & 1.051 & \cite{CAMEO2026} & 1.175 $\pm$ 0.002 & 1.070 \\
Formamide & 1.133 & \cite{CAMEO2026} & 1.176 $\pm$ 0.001 & 1.142 \\
NMA & 0.933 & \cite{Williams1957} & 1.003 $\pm$ 0.001 & -- \\
DMF & 0.940 & \cite{Easteal1985} & 1.027 $\pm$ 0.001 & 0.951 \\
DMSO & 1.100 & \cite{CAMEO2026} & 1.097 $\pm$ 0.002 & 1.105 \\
Acetonitrile & 0.787 & \cite{CAMEO2026} & 0.817 $\pm$ 0.003 & 0.709 \\
Acetone & 0.791 & \cite{CAMEO2026} & 0.894 $\pm$ 0.002 & 0.811 \\
DCM & 1.332 & \cite{CAMEO2026} & 1.403 $\pm$ 0.003 & 1.314 \\
CCl$_4$ & 1.580 & \cite{McCool1972} & 1.756 $\pm$ 0.003 & -- \\
THF & 0.888 & \cite{CAMEO2026} & 0.909 $\pm$ 0.001 & 0.846 \\
Benzene & 0.874 & \cite{Lemmon2023} & 0.971 $\pm$ 0.002 & 0.889 \\
Hexane & 0.660 & \cite{Harris1982} & 0.672 $\pm$ 0.001 & 0.647 \\
\midrule
MAE & \multicolumn{2}{c}{} & 0.063 & 0.021 \\
MRE\,\% & \multicolumn{2}{c}{} & 6.3 & 2.4 \\
\bottomrule
\end{tabular}
\end{table*}

\begin{table*}[htbp]
\setlength{\tabcolsep}{4pt}
\centering
\caption{\textbf{Self-diffusion coefficients of 16 solvents.} SO3LR-T-M was simulated at 300\,K, experimental $D$ and $\eta$ refer to $\approx$298\,K, except N-methylacetamide (simulation and references at 320\,K~\cite{Williams1957}). Diffusion coefficients in \AA$^2$\,ns$^{-1}$, experimental shear viscosities $\eta$ in mPa\,s~\cite{Lide2003}. Columns marked $+$YH carry the Yeh--Hummer finite-size correction evaluated with the experimental $\eta$ and the mean simulation-cell length; GAFF lacks the box-size data the correction needs and is reported uncorrected, so the like-for-like comparison with GAFF is the raw column. SO3LR-T-M values are means over four independent replicas, and the $\pm$ is the larger of the seed-to-seed standard error and the block-bootstrap error (Sec.~\ref{sec:si-neat}); it belongs to the raw value, and the YH term carries no separate uncertainty. MAE $=\tfrac{1}{N}\sum_i|D_i-D_i^{\rm exp}|$ and MRE (Eq.~\ref{eq:si-mre}) are taken over the solvents each method reports ($N$ in the last row). Solvents are ordered from polar protic through polar aprotic to nonpolar; dashes mark solvents without published parameters}
\label{tab:bulk_diffusion}
\begin{tabular}{l r@{\,}l c cc cc c}
\toprule
 & \multicolumn{2}{c}{} & & \multicolumn{2}{c}{SO3LR-T-M} & \multicolumn{2}{c}{OPLS4} & \\
\cmidrule(lr){5-6} \cmidrule(lr){7-8}
\scriptsize Solvent & \multicolumn{2}{c}{\scriptsize Exp.} & \scriptsize $\eta$ & \scriptsize raw & \scriptsize $+$YH & \scriptsize raw & \scriptsize $+$YH & \scriptsize GAFF \\
\midrule
Water & 230 & \cite{Das1987} & 0.89 & 96 $\pm$ 1 & 119 & 229 & 241 & 239 \\
Methanol & 240 & \cite{Derlacki1985} & 0.544 & 161 $\pm$ 4 & 194 & 346 & 366 & 446 \\
Ethanol & 108 & \cite{Hurle1985} & 1.09 & 91 $\pm$ 5 & 109 & 135 & 145 & 236 \\
Ethylene glycol & 8.8 & \cite{spees2012use} & 16.2 & 5 $\pm$ 1 & 6 & 17 & 18 & -- \\
Acetic acid & 120 & \cite{Kida1977} & 1.06 & 7 $\pm$ 1 & 25 & 62 & 72 & 106 \\
Formamide & 52 & \cite{Das1987} & 3.34 & 48 $\pm$ 2 & 55 & 66 & 69 & -- \\
NMA & 59 & \cite{Williams1957} & 2.41 & 41 $\pm$ 6 & 48 & -- & -- & 139 \\
DMF & 162 & \cite{Das1987} & 0.794 & 77 $\pm$ 5 & 98 & 100 & 114 & -- \\
DMSO & 73 & \cite{Holz2000} & 1.99 & 95 $\pm$ 4 & 104 & 60 & 66 & 124 \\
Acetonitrile & 417 & \cite{Holz1996} & 0.369 & 335 $\pm$ 8 & 385 & 549 & 578 & 291 \\
Acetone & 477 & \cite{McCall1959} & 0.306 & 207 $\pm$ 4 & 265 & 389 & 425 & 407 \\
DCM & 350 & \cite{Prielmeier1986} & 0.413 & 225 $\pm$ 9 & 269 & 316 & 342 & 465 \\
CCl$_4$ & 129 & \cite{McCool1972} & 0.908 & 58 $\pm$ 3 & 76 & -- & -- & -- \\
THF & 275 & \cite{Baba2022} & 0.456 & 221 $\pm$ 7 & 259 & 234 & 259 & -- \\
Benzene & 214 & \cite{McCall1959} & 0.604 & 84 $\pm$ 4 & 112 & 158 & 177 & 342 \\
Hexane & 420 & \cite{Harris1982} & 0.3 & 358 $\pm$ 11 & 413 & 371 & 405 & 337 \\
\midrule
MAE & \multicolumn{2}{c}{} & & 79 & 54 & 49 & 42 & 92 \\
MRE\,\% & \multicolumn{2}{c}{} & & 86.1 & 46.6 & 33.0 & 28.0 & 49.7 \\
$N$ & \multicolumn{2}{c}{} & & 16 & 16 & 14 & 14 & 11 \\
\bottomrule
\end{tabular}
\end{table*}

\clearpage

\subsection{Water}\label{sec:si-water}

Water structure and its temperature dependence were computed from separate NPT simulations of 900 \ce{H2O} molecules in a cubic box of initial side 30~\AA{}, one 2~ns trajectory per state point. The target pressure was 1~bar at 300 and 330~K and 25~MPa at 400--673~K. These runs used mixed precision, a 0.5~fs time step, a 12~\AA{} long-range cutoff, PME for electrostatics and dispersion, and Nos\'e--Hoover chain thermostat and barostat time constants of 50~fs and 500~fs. Radial distribution functions are averaged over the final 75\% of each trajectory.

\begin{figure*}[h]
\centering
\includegraphics[width=1.0\linewidth]{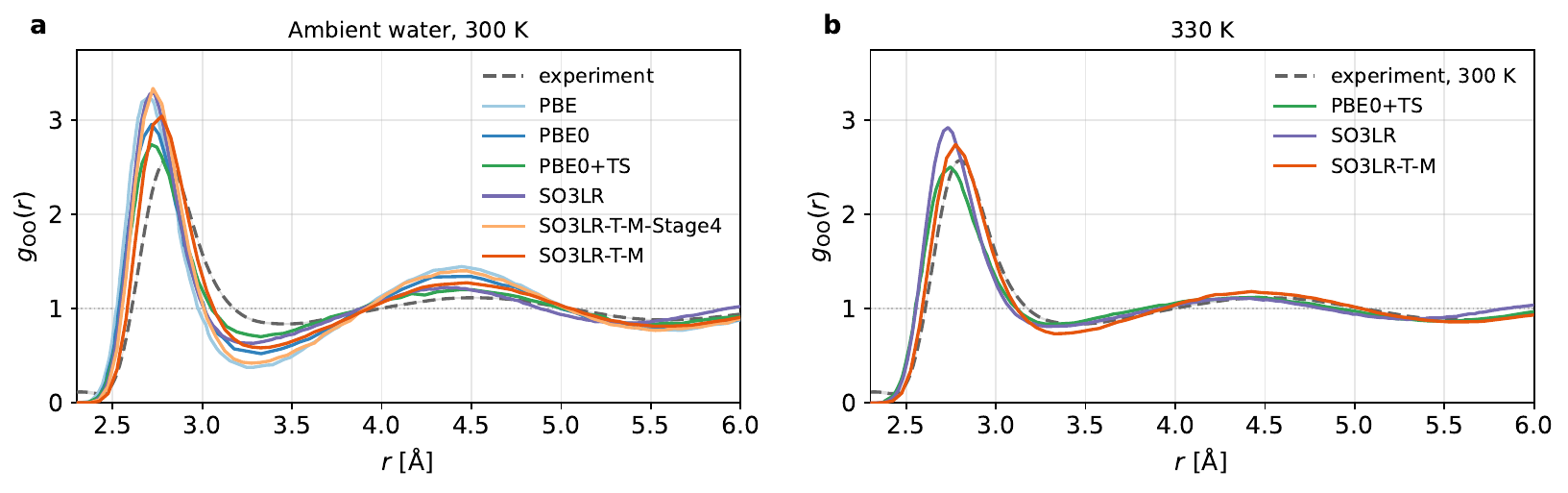}
\caption{\textbf{Oxygen--oxygen radial distribution function of liquid water: SO3LR against experiment and against its own reference functionals.} \textbf{(a)}~Ambient water at \SI{300}{\kelvin} and \SI{1}{\bar}. \textbf{(b)}~The same comparison at \SI{330}{\kelvin}. The dashed grey curve is the experimental $g_\mathrm{OO}(r)$, measured at \SI{295.15}{\kelvin}~\cite{skinner2013benchmark}; the shaded band shows the reported pointwise uncertainty. The same ambient curve is repeated in~(b) as a fixed visual reference, not as a \SI{330}{\kelvin} measurement. Solid curves are, in order of increasing first-peak height at \SI{300}{\kelvin}, PBE0+TS ($2.74$), PBE0 ($2.96$), SO3LR-T-M ($3.05$), PBE ($3.24$), SO3LR ($3.29$) and an earlier SO3LR-T-M checkpoint ($3.34$), against an experimental value of $2.58$ at \SI{2.80}{\angstrom}. The DFT curves are digitized from Ref.~\citenum{distasio2014individual}. SO3LR curves are from NPT molecular dynamics of \num{900} \ce{H2O} in a \SI{30}{\angstrom} cubic box with particle--mesh Ewald electrostatics, a \SI{0.5}{\femto\second} timestep and a \SI{12}{\angstrom} long-range cutoff, averaged over the final \SI{75}{\percent} of a \SI{2}{\nano\second} trajectory. At \SI{330}{\kelvin} the same offset persists, $2.74$ against $2.51$ for PBE0+TS} \label{fig:water_rdf_levels}
\end{figure*}

\begin{figure*}[h]
\centering
\includegraphics[width=1.0\linewidth]{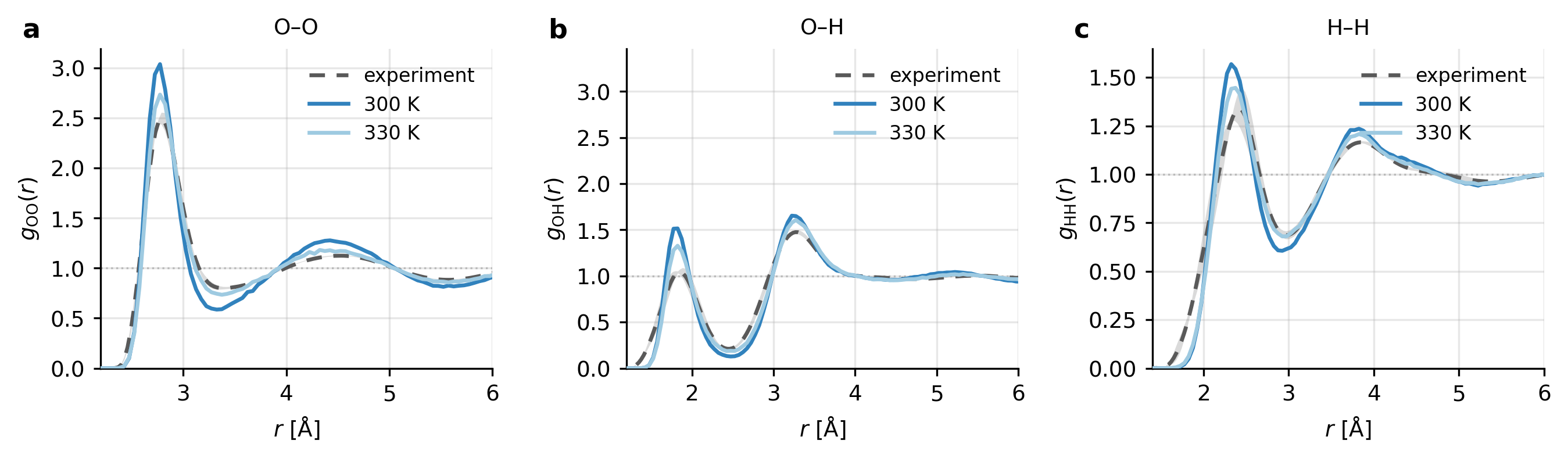}
\caption{\textbf{Water radial distribution functions at $300$~K and $330$~K.} \textbf{(a)}~$g_\text{OO}$, \textbf{(b)}~$g_\text{OH}$ and \textbf{(c)}~$g_\text{HH}$ from the NPT trajectories of Fig.~\ref{fig:water_high_tp}. Dashed curves show the scattering-derived intermolecular RDFs at \SI{298}{\kelvin} tabulated in Ref.~\citenum{soper2013radial}. Shaded bands show the reported root-mean-square deviations of the individual RDF determinations from their mean. The O--H and H--H functions are intermolecular: the two O--H and one H--H intramolecular pairs per molecule are subtracted exactly rather than removed by a cutoff in $r$. Hydrogen-bond geometry is reproduced closely. The $g_\text{OH}$ first peak sits at $1.83$~\AA{} against an experimental $1.86$~\AA{}, and the $g_\text{HH}$ first peak at $2.33$~\AA{} against $2.43$~\AA{}. Both peaks weaken with temperature as expected. The $g_\text{OO}$ first peak is over-structured, $3.04$ against $2.50$ at $300$~K, while its position is essentially exact at $2.78$~\AA{}}
\label{fig:water_partials}
\end{figure*}

\begin{figure*}[h]
\centering
\includegraphics[width=1.0\linewidth]{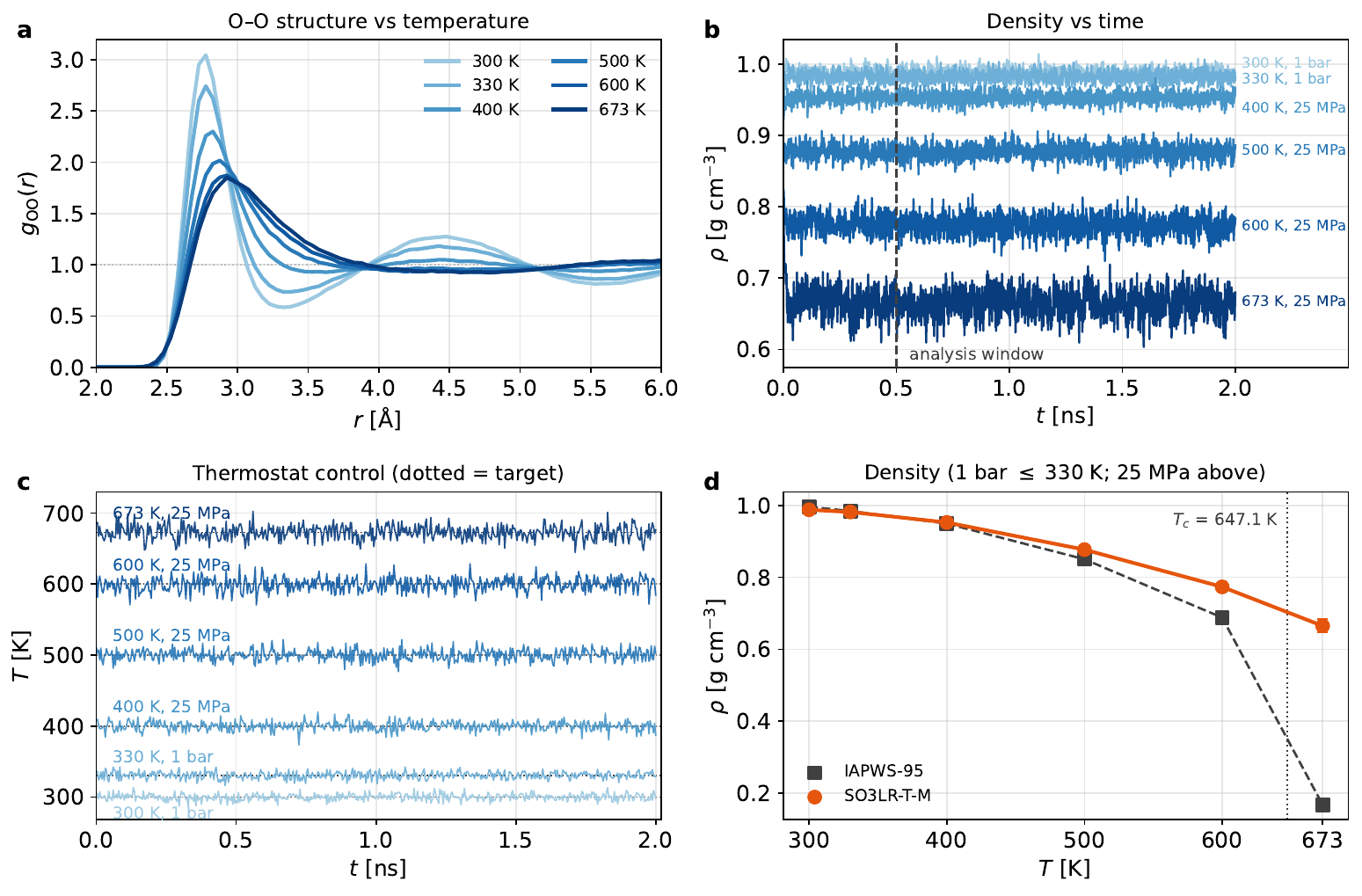}
\caption{\textbf{Water structure and density across six temperature--pressure state points with SO3LR-T-M.} Each NPT simulation contains $900$ \ce{H2O} molecules, starts from a cubic box of side $30$~\AA{}, and runs for $2$~ns with a $0.5$~fs timestep and particle--mesh Ewald electrostatics and dispersion. The target pressure is $1$~bar at $300$ and $330$~K and $25$~MPa at $400$, $500$, $600$ and $673$~K. \textbf{(a)}~Oxygen--oxygen radial distribution functions from the final $75\%$ of the saved trajectories. The first peak decreases from $3.05$ to $1.85$ and shifts from $2.78$ to $2.93$~\AA{} across the state points. \textbf{(b)}~Density time series calculated from the instantaneous cell volumes; the dashed line marks the start of the $0.5$--$2$~ns density-analysis window. \textbf{(c)}~Temperature time series, with dotted lines indicating the targets. \textbf{(d)}~Mean densities over $0.5$--$2$~ns compared with IAPWS-95 values evaluated at each target temperature and pressure~\cite{wagner2002iapws}. Error bars show the standard deviation of the instantaneous densities, not the uncertainty of the mean. Lines connecting the state points are guides to the eye; the experimental critical temperature, $T_c = 647.096$~K, is marked for context. Density deviations are below $1\%$ in magnitude at $300$, $330$ and $400$~K, and are $+3.1\%$ at $500$~K and $+12.4\%$ at $600$~K. At $673$~K and $25$~MPa, the simulated density is $0.665$~g/cm$^3$, compared with $0.167$~g/cm$^3$ from IAPWS-95, approximately a fourfold overestimate. The time series document the sampled behavior but do not by themselves establish equilibrium or exclude pressure-control errors}
\label{fig:water_high_tp}
\end{figure*}

\clearpage 

\subsection{Aqueous ions}\label{sec:si-ions} 

Each salt solution contains one cation and one or two \ce{Cl-} anions in 897--898 water molecules (2694--2696 atoms, $\approx$0.06~M), with the ions placed by farthest-point sampling so that they do not start in contact. Each system was run for 2~ns in the NPT ensemble at 300~K and 1.01325~bar under the protocol used for water. Peak positions and coordination numbers are taken from the final 75\% of each trajectory (1.5~ns). Coordination numbers are per-frame counts of water oxygens within a cutoff placed at the end of the first shell of each ion--O radial distribution function (2.98--3.48~\AA{}). For \ce{Mg^2+}, \ce{Zn^2+}, and \ce{Ca^2+} the RDF vanishes between the first and second shells, so the result does not depend on the cutoff.

\begin{table*}[h]
\centering
\caption{\textbf{Single-ion hydration structure from SO3LR-T-M at $300$~K and $1$~atm.} First-shell ion--oxygen peak position $r_\mathrm{max}$ and coordination number CN, compared with experiment. Each simulation contains one cation with one or two charge-balancing \ce{Cl-} anions in $897$--$898$ water molecules ($\approx$$0.06$~M) and was run for $2$~ns in the NPT ensemble with PME electrostatics; the final $75\%$ of each trajectory ($1509$ frames, 1.5 ns) is analyzed. Coordination numbers are direct per-frame atom counts, given as mean $\pm$ standard deviation. Experimental values are quoted as ranges where independent determinations disagree; note that they derive from solutions at $1$--$4$~M, whereas these simulations are at the dilute limit}
\label{tab:ion_hydration}
\begin{tabular}{lcccc}
\toprule
Ion & \multicolumn{2}{c}{$r_\mathrm{max}$ (\AA)} & \multicolumn{2}{c}{CN} \\
\cmidrule(lr){2-3} \cmidrule(lr){4-5}
 & SO3LR-T-M & Exp. & SO3LR-T-M & Exp. \\
\midrule
\ce{Na+}   & $2.375$ & $2.37$--$2.38$\,\footnotemark[1] & $5.53 \pm 0.58$ & $5.5 \pm 0.3$\,\footnotemark[1] \\
\ce{K+}    & $2.775$ & $2.73$--$2.81$\,\footnotemark[2] & $5.93 \pm 1.03$ & $5.6$--$8.3$\,\footnotemark[2] \\
\ce{Ca^2+} & $2.425$ & $2.39$--$2.46$\,\footnotemark[3] & $6.62 \pm 0.66$ & $6$--$8$\,\footnotemark[3] \\
\ce{Mg^2+} & $2.075$ & $2.09$--$2.10$\,\footnotemark[4] & $6.00 \pm 0.00$ & $6$\,\footnotemark[4] \\
\ce{Zn^2+} & $2.125$ & $2.08$--$2.09$\,\footnotemark[4] & $5.99 \pm 0.07$ & $6$\,\footnotemark[4] \\
\ce{Cl-}   & $3.125$--$3.175$ & $3.10$--$3.20$\,\footnotemark[4] & $6.06$--$6.43$ & $5$--$7$\,\footnotemark[4] \\
\bottomrule
\end{tabular}
\footnotetext[1]{Ref.~\citenum{galib2017}, X-ray diffraction and \ce{Na} K-edge EXAFS.}
\footnotetext[2]{Ref.~\citenum{mahler2012}, large-angle X-ray scattering; the upper end of the CN range comes from distance-correlation analyses.}
\footnotetext[3]{Ref.~\citenum{jalilehvand2001}, EXAFS and large-angle X-ray scattering; direct integration of diffraction data gives the lower end of the range.}
\footnotetext[4]{Ref.~\citenum{ohtaki1993}, compilation of X-ray, neutron and EXAFS determinations.}
\end{table*}

\clearpage
\section{Proton transfer in aqueous solution}\label{sec:si-pka} 

\subsection{Preparation of simulation structures}
Initial structures of substituted penta-alanine peptides, Ac-Ala-Ala-X-Ala-Ala-NHMe, with X = Asp, Glu, Cys, Tyr, His, Lys or Arg,  and of the terminally titratable peptides Ac-(Ala)$_5$-COOH and H$_3$N$^+$-(Ala)$_5$-NHMe, were dissolved in a periodic water box of $35\,\text{\AA}$ edge length and 1410 water molecules via Packmol, generating four independent replicas. The substituted peptides were capped by an acetyl group at the N-terminus, and N-methyl amide at the C-terminus. To directly obtain the p$K_\mathrm{a}$ value from the simulation (and not the p$K_\mathrm{b}$), the transferable proton was in all but one case attached to the initial peptide structure, thus sampling the acidic reaction

\begin{equation*}
  \ce{HA + H2O <=> A- + H3O+}\text{,}
\end{equation*}

where for the acidic residues, Asp, Glu, Tyr, Cys, and C-terminus HA contains a neutral COOH, OH or SH group, while for the basic residues His, Lys, and the N-terminus HA is the positively charged imidazolium (ImH$^+$) or NH$_3^+$ group. To neutralize the protonated basic residues a Cl$^{-}$ ion was added, rendering the final simulation cell charge-neutral in all cases.

To determine the water autoionization constant, p$K_\mathrm{w}$ = 
p$K_\mathrm{a}$ + p$K_\mathrm{b}$, the Lys peptide was additionally 
simulated as a base, with a neutral NH$_2$ group and no counter-ion, 
sampling \ce{B + H2O <=> BH+ + OH-}. The Arg peptide is the only 
exception to the acidic setup: since its p$K_\mathrm{a}$ of 13.8 is 
close to p$K_\mathrm{w}$, deprotonation of the guanidinium group 
would compete with the autoionization of water in a direct 
simulation. Arg was therefore simulated as a base only, and its 
p$K_\mathrm{a}$ was obtained as p$K_\mathrm{w} -$ p$K_\mathrm{b}$, 
using the computed p$K_\mathrm{w}$.

\subsection{Molecular dynamics and enhanced-sampling protocol}
\textbf{General settings.} All SO3LR-T-M MD simulations were run in single precision (float32) with a $0.5\,\text{fs}$ time step at $300\,\text{K}$ controlled by a Nosé–Hoover chain thermostat (chain length 3, time constant $50\,\text{fs}$). To account for long-range interactions beyond the semi-local SO3krates model, a long-range cutoff of $12\,\text{\AA}$ was used, electrostatics beyond that range were captured by PME with a $2.0\, \text{\AA}$ Gaussian smearing and a $128\times128\times128$ $k$-space grid.\newline
\textbf{Equilibration.} After a geometry optimization of the starting structure of at most 1000 steps with the FIRE algorithm, a three-step NVT$\to$NPT$\to$NVT equilibration protocol was employed, each for $300\,\text{ps}$. The NPT step was performed at ambient pressure ($1.013\,\text{bar}$) and modified the simulation cell lattice parameters by only up to $0.3\,\text{\AA}$. Final lattice parameters were between $34.9\,\text{\AA}$ and $35.3\,\text{\AA}$.\newline
\textbf{Enhanced sampling.} To sample the otherwise rare transfer and diffusion of the proton from the peptide residue into bulk water, the adaptive biasing force (ABF) method as implemented in PySAGES was employed. As the collective variable (CV), an adapted version of the charge-separation CV of Xu et al.~\cite{xu2026quantitative} was used. The CV assigns each non-hydrogen (heavy) atom an excess charge by smoothly counting the number of hydrogen atoms bound to it and comparing this number with that in the starting structure of the simulation. To this end, every hydrogen atom $j\in\mathcal{H}$, where $\mathcal{H}$ is the set of all hydrogen atoms in the system, is distributed over the set $\mathcal{S}$ of all heavy atoms (C, N, O and S) with normalised Gaussian weights
\begin{equation}
  n_{j\to i} =
  \frac{\exp\!\left(-\left|\mathrm{mic}(\mathbf{r}_j-\mathbf{r}_i)\right|^2/2\sigma^2\right)}
       {\sum_{k\in\mathcal{S}}\exp\!\left(-\left|\mathrm{mic}(\mathbf{r}_j-\mathbf{r}_k)\right|^2/2\sigma^2\right)},
  \qquad j\in\mathcal{H},\; i\in\mathcal{S},
  \label{eq:cs_weights}
\end{equation}
where $\mathbf{r}_i$ is the position of atom $i$, $\sigma=0.3\,\text{\AA}$ is the width of the Gaussians, and $\mathrm{mic}(\cdot)$ denotes the minimum-image convention of the periodic simulation cell. Since the weights in Eq.~\eqref{eq:cs_weights} sum to one over $\mathcal{S}$, each hydrogen atom contributes exactly one unit of charge, which is shared smoothly among its nearest heavy atoms. The excess charge of heavy atom $i$ is then
\begin{equation}
  q_i = \sum_{j\in\mathcal{H}} n_{j\to i} \;-\; n_i^{0},
  \label{eq:cs_charge}
\end{equation}
where $n_i^{0}$ is the integer number of hydrogen atoms bound to atom $i$ in the starting structure. Hence, $q_i\approx+1$ for the oxygen atom of a hydronium ion H$_3$O$^+$, $q_i\approx-1$ for a hydroxide ion or a deprotonated site, and $q_i\approx0$ otherwise.

The excess charges are measured relative to a reference point. For this, we use the set $\mathcal{A}$ of the atoms of the titratable group that can carry the transferring proton (e.g.\ the two carboxyl oxygen atoms of Asp) and take their centroid,
\begin{equation}
  \mathbf{r}_\mathrm{ref} = \mathbf{r}_{a_1} + \frac{1}{|\mathcal{A}|}
  \sum_{a\in\mathcal{A}} \mathrm{mic}(\mathbf{r}_a-\mathbf{r}_{a_1}),
  \label{eq:cs_ref}
\end{equation}
where $a_1$ is the first atom in $\mathcal{A}$, so that the centroid is well defined across the periodic boundaries. The value of the CV is the sum, over all heavy atoms except the reference atoms, of the excess charge times the distance to the reference point:
\begin{equation}
  s(\mathbf{R}) = \sum_{i\in\mathcal{S}\setminus\mathcal{A}}
  q_i\,\left|\mathrm{mic}(\mathbf{r}_i-\mathbf{r}_\mathrm{ref})\right|,
  \label{eq:cs_cv}
\end{equation}
where $\mathbf{R}$ denotes the coordinates of all atoms. For an acid, a CV value close to $0$ therefore corresponds to the state in which the proton is still attached to the residue and no proton transfer has taken place yet, while a CV value of $d$ corresponds to a state in which the hydronium ion H$_3$O$^+$ is approximately at a distance $d$ from the titratable group. In the original formulation of the charge-separation CV, the hydrogen atoms are distributed over the oxygen atoms only, and excess charges are counted only on the water molecules. This is appropriate for the small molecules discussed there, such as H$_3$PO$_4$, in which proton transfer can only occur at a single group of the molecule. Since the peptides simulated here are more complex, and proton transfer can also involve the peptide groups R-NH-CO-R$'$, our CV distributes the hydrogen atoms over all heavy atoms (Eq.~\eqref{eq:cs_weights}) and counts the excess charges of all non-reference heavy atoms (Eq.~\eqref{eq:cs_cv}).

For the ABF simulations, the CV was sampled between $-0.25\,\text{\AA}$ and $14.25\,\text{\AA}$ (for base simulations between 
$0.25\,\text{\AA}$ and $-14.25\,\text{\AA}$) on a very fine grid of 1450 bins of width $0.01\,\text{\AA}$, to ensure that the very steep free-energy minimum around $s=0$ is captured accurately. The biasing force in each bin is scaled down linearly with the number of samples until the bin has been visited $150$ times. To restrain the exploration of CV values beyond this range, harmonic walls with a force constant of $0.2\,\text{eV/\AA}^2$ are applied at its borders. The ABF simulations were run for $8\,\text{ns}$ ($12\,\text{ns}$ for LysH$^+$).

\begin{table}[h]
  \centering
  \small
  \setlength{\tabcolsep}{4pt}
  \caption{Reference atoms $\mathcal{A}$ of the charge-separation CV (Eq.~\eqref{eq:cs_ref}) and CV range of the ABF grid (bins of $0.01\,\text{\AA}$) for each simulated reaction}
  \label{tab:cs_reference}
  \begin{tabular}{lllc}
    \toprule
    Peptide & Reaction & Reference atoms $\mathcal{A}$ & CV range (\AA) \\
    \midrule
    Asp & COOH $\rightarrow$ COO$^-$            & both carboxyl O      & $-0.25$ to $14.25$ \\
    Glu & COOH $\rightarrow$ COO$^-$            & both carboxyl O      & $-0.25$ to $14.25$ \\
    Cys & SH $\rightarrow$ S$^-$                & thiol S              & $-0.25$ to $14.25$ \\
        &                                       &                      & $-0.25$ to $6.25$\,$^{a}$ \\
    Tyr & OH $\rightarrow$ O$^-$                & phenol O             & $-0.25$ to $14.25$ \\
    His & ImH$^+$ $\rightarrow$ Im              & both imidazole N     & $-0.25$ to $14.25$ \\
    Lys & NH$_3^+$ $\rightarrow$ NH$_2$         & amino N              & $-0.25$ to $14.25$ \\
    Lys & NH$_2$ $\rightarrow$ NH$_3^+$ (basic) & amino N              & $-14.25$ to $0.25$ \\
    Arg & guanidine $\rightarrow$ guanidinium (basic) & three guanidine N & $-14.25$ to $0.25$ \\
    C-term. & COOH $\rightarrow$ COO$^-$              & both carboxyl O   & $-0.25$ to $14.25$ \\
    N-term. & NH$_3^+$ $\rightarrow$ NH$_2$           & amino N           & $-0.25$ to $14.25$ \\
    \bottomrule
  \end{tabular}

  \raggedright\footnotesize $^{a}$\,Short-range simulations.
\end{table}

\subsection{Post-processing of the ABF simulations and p$K_\mathrm{a}$ estimation}
All quantities below were computed separately for each of the four independent replicas of a system, using the PySAGES restart files, the CV logs and the trajectories as inputs.

\textbf{Free energy along the CV.} The analysis starts from the ABF estimate of the mean force in every bin $k$ of the CV grid (bin centres $s_k$, width $h=0.01\,\text{\AA}$),
\begin{equation}
  f_k=\frac{\Sigma_k}{N_k}\approx-\left.\frac{\mathrm{d}F}{\mathrm{d}s}\right|_{s=s_k},
  \label{eq:pk_meanforce}
\end{equation}
where $N_k$ is the number of samples in bin $k$, $\Sigma_k$ the sum of the instantaneous forces along $s$ in these samples, and $F(s)$ the free energy along the CV of Eq.~\eqref{eq:cs_cv}. To discard the first 2\,ns of every simulation, during which the bias builds up and the system is not yet in equilibrium, $N_k$ and $\Sigma_k$ at 2\,ns were subtracted from those at the end of the simulation, and only this difference (the ``2\,ns--end block'') was used for all reported values. This needs no reweighting, because the ABF bias depends on the coordinates only through $s$ and thus does not change the conditional average in Eq.~\eqref{eq:pk_meanforce}.

The free energy was obtained by integrating the mean force with the trapezoidal rule,
\begin{equation}
  F_{k+1}=F_k-\frac{h}{2}\left(f_k+f_{k+1}\right),
  \label{eq:pk_integration}
\end{equation}
over the bins with $N_k\ge150$, the same threshold as in the ABF simulations. These bins have to form a single connected range that contains both the undissociated basin and the plateau region defined below; otherwise no p$K_\mathrm{a}$ value can be extracted from the simulation. The probability of bin $k$ is then $p_k=\mathrm{e}^{-\beta F_k}/\sum_l\mathrm{e}^{-\beta F_l}$, with $\beta=1/k_\mathrm{B}T$ and $T=300\,\text{K}$.

\textbf{Dissociation constant.} Following Xu et al.~\cite{xu2026quantitative}, the dissociation constant was evaluated with the reversible-work theorem,
\begin{equation}
  K=\frac{1}{c^\circ}\left[\int_0^{r_\mathrm{c}}4\pi r^2\,\mathrm{e}^{-\beta W(r)}\,\mathrm{d}r\right]^{-1},
  \label{eq:pk_rwt}
\end{equation}
where $c^\circ=1\,\text{mol/L}$ is the standard concentration and $r_\mathrm{c}$ is a cutoff distance that separates associated from dissociated states. $r$ is the distance between the two products of the dissociation, i.e.\ between the deprotonated (or, for a base, protonated) residue and the free H$_3$O$^+$ (or OH$^-$) ion. $W(r)$ is the free energy of the two products at distance $r$ relative to infinite separation, i.e.\ the reversible work needed to bring the free ion from far away to the distance $r$ from the residue. For the acidic dissociation HA + H$_2$O $\rightarrow$ A$^-$ + H$_3$O$^+$, which gives the p$K_\mathrm{a}$, we set $r=s$; for the basic dissociation B + H$_2$O $\rightarrow$ BH$^+$ + OH$^-$, which gives the p$K_\mathrm{b}$, we set $r=-s$. In both cases, $r$ is the distance of the free ion from the reference point of Eq.~\eqref{eq:cs_ref}.

The integral in Eq.~\eqref{eq:pk_rwt} can be obtained from the free-energy profile $F(s)$ as follows. Far from the residue, the free ion is distributed uniformly over the box with a probability per unit volume $\rho_\infty$. Closer to the residue, this probability per unit volume is changed by the Boltzmann factor of $W$ and becomes $\rho_\infty\,\mathrm{e}^{-\beta W(r)}$. The probability of finding the ion in a thin spherical shell of radius $r$ and thickness $\mathrm{d}r$ around the residue is therefore
\begin{equation}
  p(r)\,\mathrm{d}r=4\pi r^2\rho_\infty\,\mathrm{e}^{-\beta W(r)}\,\mathrm{d}r.
  \label{eq:pk_shell}
\end{equation}
Xu et al.\ use the same relation in the form $W_\mathrm{sim}(r)=G(r)+k_\mathrm{B}T\ln 4\pi r^2$ (their Eq.~14), where $G$ is the free-energy profile along $s$ and $W_\mathrm{sim}$ is the PMF up to an additive constant $C$ that they determine separately (their Eqs.~5 and 6). In Eq.~\eqref{eq:pk_shell}, the density $\rho_\infty$, determined below, takes the role of this constant. The integrand in Eq.~\eqref{eq:pk_rwt} is thus $p(r)/\rho_\infty$, and the integral equals the probability $P_\mathrm{bound}$ of finding the ion closer than $r_\mathrm{c}$ to the residue, divided by $\rho_\infty$. $P_\mathrm{bound}$ is the sum of the bin probabilities $p_k$ below $r_\mathrm{c}$, so that
\begin{equation}
  \mathrm{p}K=-\log_{10}K=\log_{10}\frac{P_\mathrm{bound}}{\rho_\infty V^\circ},
  \qquad P_\mathrm{bound}=\sum_{k:\,r_k<r_\mathrm{c}}p_k,
  \label{eq:pk_pk}
\end{equation}
where $V^\circ=1/(N_\mathrm{A}c^\circ)=1660.54\,\text{\AA}^3$ is the volume per molecule at the standard concentration. As in Ref.~\citenum{xu2026quantitative}, $r_\mathrm{c}=7\,\text{\AA}$.

$P_\mathrm{bound}$ contains the undissociated state as well as close ion pairs. In the undissociated state, $s$ fluctuates around zero and also takes small negative values in the acidic (small positive values in the basic) dissociation. Consistent with Ref.~\citenum{xu2026quantitative}, these values are included, and the sum in Eq.~\eqref{eq:pk_pk} starts at the edge of the ABF grid, $s=-0.25\,\text{\AA}$ (acidic) or $s=+0.25\,\text{\AA}$ (basic).

\textbf{Density of the free ion.} The free-energy profile gives only relative probabilities. The density $\rho_\infty$ in Eq.~\eqref{eq:pk_pk} provides the missing reference: The $p_k$ add up to one over the whole grid, and $\rho_\infty$ is defined with the same normalisation: $\rho_\infty\,\mathrm{d}V$ is the probability, out of this total of one, of finding the free ion in a small volume $\mathrm{d}V$ far from the residue. It was obtained from the plateau region $8\,\text{\AA}\le r<r_\mathrm{max}$, with $r_\mathrm{max}=13.75\,\text{\AA}$ ($0.5\,\text{\AA}$ inside the outer edge of the grid). In this region the ion is separated from the residue by several water layers, but, if the residue is charged after the reaction, it is still attracted by the residue through the screened Coulomb interaction. Its probability per unit volume at a position $\mathbf{x}$ in the plateau region is therefore not yet $\rho_\infty$ but $\rho_\infty\,\mathrm{e}^{-\beta U(\mathbf{x})}$, where $U(\mathbf{x})$ is the electrostatic interaction between the ion and the residue in the periodic box (Eq.~\eqref{eq:pk_U} below).

The total probability of the plateau region is then $\sum_{k\in\mathrm{plateau}}p_k=\rho_\infty\int_{\mathcal{V}_\mathrm{plateau}}\mathrm{e}^{-\beta U(\mathbf{x})}\,\mathrm{d}^3x$, which gives
\begin{equation}
  \rho_\infty=\frac{\sum_{k\in\mathrm{plateau}}p_k}
  {\int_{\mathcal{V}_\mathrm{plateau}}\mathrm{e}^{-\beta U(\mathbf{x})}\,\mathrm{d}^3x}.
  \label{eq:pk_rho}
\end{equation}
Here $\mathcal{V}_\mathrm{plateau}$ is the spherical shell $8\,\text{\AA}\le|\mathrm{mic}(\mathbf{x}-\mathbf{r}_\mathrm{ref})|<r_\mathrm{max}$ around the reference point. Part of this shell is occupied by the peptide itself, where the ion cannot be. This part is not counted: points closer than $2.6\,\text{\AA}$ to any non-hydrogen atom of the peptide, roughly the closest distance between the oxygen atom of the ion and a heavy atom of the peptide, are removed from $\mathcal{V}_\mathrm{plateau}$. For Asp, this removes about 4\,\% of the shell volume. Since $r_\mathrm{max}<L/2=17.5\,\text{\AA}$, the shell lies entirely inside the box. The integral was evaluated by Monte Carlo integration, with $5\times10^4$ random points in each of the first five saved configurations (the first 2.5\,ps) of the last segment of each replica.

For the interaction we use the energy of two point charges in the periodic box, as Xu et al.~\cite{xu2026quantitative},
\begin{equation}
  U(\mathbf{x})=\frac{e^2}{4\pi\varepsilon_0\varepsilon}\,q_\mathrm{res}\,q_\mathrm{ion}
  \left[\frac{1}{|\mathcal{A}|}\sum_{a\in\mathcal{A}}\phi_\mathrm{E}(\mathbf{x}-\mathbf{r}_a)-\psi\right],
  \label{eq:pk_U}
\end{equation}
with $\varepsilon=80$. $q_\mathrm{ion}=+1$ for H$_3$O$^+$ and $-1$ for OH$^-$. $q_\mathrm{res}$ is the net charge of the residue after the reaction: $-1$ for Asp, Glu, Cys, Tyr, and the C-terminus, $+1$ for the basic dissociation of Lys and Arg, and $0$ for HisH$^+$, LysH$^+$, and the N-terminus for which $U=0$ and $\rho_\infty$ is simply the plateau probability divided by the plateau volume. The residue charge is spread evenly over the reference atoms $\mathcal{A}$ of Eq.~\eqref{eq:cs_ref}. $\phi_\mathrm{E}$ is the Ewald potential of a unit point charge with its periodic images and a neutralising background in the cubic box of edge length $L$, and $\psi=-2.837297/L$ is its self-term~\cite{hummer1996free}. With this term, $U$ is the energy of the two separated charges relative to the associated state, in which both charges sit at the same point. This is the reference interaction of Xu et al., including their removal of the self-energy of the net charge, with three differences: the residue charge is spread over the reference atoms instead of being placed at a single point; the volume occupied by the peptide is excluded, as described above; and their correction for spherical shells that extend beyond $L/2$ is not needed, because the plateau region ends at $r_\mathrm{max}<L/2$.

Xu et al.\ fix the corresponding constant as the offset $C$ of the PMF: the arithmetic mean of the difference between the simulated PMF and the reference interaction (their Eq.~6), taken over a window from $r_\mathrm{c}$ to 4\,\AA{} below the wall, i.e.\ 7--10\,\AA{} in their phosphate simulations. Both approaches give the same result when this difference is constant across the window. We checked this by fitting a straight line to the difference over the plateau region. Using the arithmetic mean of Xu et al.\ over our plateau region changed the p$K$ of each Asp replica by at most 0.04, and using their window ($r_\mathrm{c}$ to 4\,\AA{} below our wall, i.e.\ 7--10.25\,\AA) instead of ours changed the Asp mean by 0.06.

\textbf{Averages and replica selection.} The p$K$ was computed for every replica, and we report the mean over replicas $\pm$ the standard deviation. A replica was left out if (i) its bins with $N_k\ge150$ did not form one connected range through the bound region and the plateau, so that no p$K$ could be computed, or (ii) the integrated range did not cover the whole grid. In the analyses reported here, all four replicas of every system passed both criteria. The p$K_\mathrm{w}$ of the model was obtained as p$K_\mathrm{a}$(LysH$^+$) + p$K_\mathrm{b}$(Lys).

In some configurations, at intermediate distances and in the plateau region, the transferred proton has been passed on through the water to the oxygen atom of a carbonyl group of the peptide backbone. This is a genuine, though minor, protonation state, since the carbonyl oxygen is a weak base. However, the CV of Eq.~\eqref{eq:cs_cv} gives it the same value as a free H$_3$O$^+$ at the same distance, and a move of the proton from the carbonyl group back to a neighbouring water hardly changes $s$, so the bias does not help it leave and this state can be sampled for longer than its equilibrium weight. 
In the actual simulations, these backbone protonation states lower the p$K$ estimates by less than 0.1 units for all systems except LysH$^+$, where they lower the mean by 0.15 units.

\textbf{The Cys peptide.} In the full-range Cys simulations the undissociated state was poorly sampled. We therefore ran four additional ABF simulations in which $s$ was limited to $-0.25$--$6.25\,\text{\AA}$ (650 bins). All other settings and the starting structures were the same as in the full-range simulations. For bins with $s<5.25\,\text{\AA}$, the mean force was taken from these short-range simulations, provided the bin had at least 150 samples. For all other bins, it was taken from the full-range simulations. Because the mean force in each bin is a local average, the two sets can be directly joined in this way before the integration of Eq.~\eqref{eq:pk_integration}. Replica $n$ of one set was paired with replica $n$ of the other, and, as for all simulations, the first 2\,ns were discarded.
Between 2 and 8\,ns, the proton moved from the undissociated basin ($s<1\,\text{\AA}$) beyond $s=3\,\text{\AA}$ and back 529--705 times in each short-range simulation, but only 0--247 times in each full-range simulation, and one full-range simulation barely sampled $s<2.4\,\text{\AA}$, highlighting the necessity of sampling the first half of the full CV range separately. As a result, relative to $s=5.25\,\text{\AA}$, the undissociated basin of the full-range simulations lies 6--7\,kJ/mol higher than in the short-range simulations, with the difference mainly arising between 2.5 and 3.5\,\AA.

Even so, the Cys p$K_\mathrm{a}$ varies more between replicas than the other p$K$ values. The reason is not a lack of transitions: in each short-range simulation, $s$ moves
from the undissociated basin ($s<1\,\text{\AA}$) beyond $s=3\,\text{\AA}$ and back hundreds of times.
The reason is a slow change of the peptide conformation.
 In some conformations, the C-terminus folds into a helical turn and backbone N--H groups form hydrogen bonds with the thiolate sulfur (H$\cdots$S $<2.7\,\text{\AA}$). Once formed, the turn can persist for several nanoseconds, and the bias along $s$ does not speed up its formation or unfolding.
The free energy difference between the undissociated basin and $s=5.25\,\text{\AA}$ is clearly lower when the proton transfer occurs  while the helical turn is present. 
In the short-range simulations, this happens in a single 2\,ns block before any other replica adopts the turn: replica 2 between 2 and 4\,ns. Replica 1 adopts it only later (4--8\,ns).
If this 2--4\,ns block is used for the $s<5.25\,\text{\AA}$ portion of the free energy profile, together with the full 2--8\,ns data of the full-CV-range run for $s\geq5.25\,\text{\AA}$, replica 2 gives a p$K_\mathrm{a}$ of 6.7 and replicas 1, 3 and 4 give 9.7, 9.3 and 9.0.
Because the p$K_\mathrm{a}$ estimated this way only uses a selected subset of the simulation for the free energy profile in the $-0.25<s<5.25\,\text{\AA}$ CV range, it serves only to show the magnitude of the p$K_\mathrm{a}$ sensitivity to the peptide conformation and is not used for the reported p$K_\mathrm{a}$.

\begin{table}[h]
  \centering\small
  \setlength{\tabcolsep}{2.5pt}
  \caption{p$K$ values of the individual replicas, with the mean $\pm$ standard deviation over
    replicas, compared with experiment. Unless noted otherwise, the first 2\,ns of each replica were
    discarded and 2--8\,ns were used. $\Delta$p$K$ = mean $-$ experiment;
    $\Delta\Delta G = RT\ln 10\,\Delta\mathrm{p}K$ in kcal/mol at 300\,K}
  \label{tab:pk_replicas}
  \begin{tabular}{llcccccccc}
    \toprule
    Peptide & & Rep.~1 & Rep.~2 & Rep.~3 & Rep.~4 & Mean & Exp.$^{d}$ & $\Delta$p$K$ & $\Delta\Delta G$ \\
    \midrule
     Asp      & p$K_\mathrm{a}$ & 4.38  & 4.05  & 4.47  & 4.36  & $4.31 \pm 0.18$  & 3.67  & $+0.64$ & $+0.88$ \\
    Glu      & p$K_\mathrm{a}$ & 4.97  & 4.90  & 5.06  & 5.25  & $5.04 \pm 0.15$  & 4.25  & $+0.79$ & $+1.08$ \\
    Cys$^{a}$ & p$K_\mathrm{a}$ & 8.47 & 7.21  & 8.74  & 7.95  & $8.09 \pm 0.68$  & 8.55  & $-0.46$ & $-0.63$ \\
    Tyr      & p$K_\mathrm{a}$ & 10.59 & 11.19 & 10.22 & 10.22 & $10.56 \pm 0.46$ & 9.84  & $+0.72$ & $+0.99$ \\
    HisH$^+$ & p$K_\mathrm{a}$ & 7.02  & 7.21  & 6.79  & 7.32  & $7.09 \pm 0.23$  & 6.54  & $+0.55$ & $+0.75$ \\
    LysH$^+$\,$^{b}$ & p$K_\mathrm{a}$ & 8.01 & 8.53 & 8.52 & 7.87 & $8.23 \pm 0.34$ & 10.40 & $-2.17$ & $-2.98$ \\
    Lys      & p$K_\mathrm{b}$ & 6.32  & 6.59  & 6.51  & 6.13  & $6.39 \pm 0.20$  & 3.60$^{e}$ & $+2.79$ & $+3.83$ \\
    Arg$^{c}$ & p$K_\mathrm{a}$ & 12.90 & 13.32 & 12.79 & 13.12 & $13.03 \pm 0.24$ & 13.8$^{f}$ & $-0.77$ & $-1.06$ \\
    $\alpha$-COOH & p$K_\mathrm{a}$ & 5.03 & 4.64 & 4.65 & 4.90 & $4.81 \pm 0.19$ & 3.67  & $+1.14$ & $+1.56$ \\
    $\alpha$-NH$_2$ & p$K_\mathrm{a}$ & 6.43 & 6.75 & 6.30 & 6.13 & $6.40 \pm 0.26$ & 8.00  & $-1.60$ & $-2.20$ \\
    \bottomrule
  \end{tabular}
  
   \raggedright\footnotesize
  $^{a}$\,Full-range and short-range simulations joined at $s=5.25$\,\AA, both 2--8\,ns.\\
  $^{b}$\,2--12\,ns.\\
  $^{c}$\,p$K_\mathrm{a} = \mathrm{p}K_\mathrm{w} - \mathrm{p}K_\mathrm{b}$, with
  p$K_\mathrm{b}$ from the base-mode simulations (1.72, 1.30, 1.83, 1.49; mean $1.59 \pm 0.24$) and
  $\mathrm{p}K_\mathrm{w} = \mathrm{p}K_\mathrm{a}(\mathrm{LysH^+}) + \mathrm{p}K_\mathrm{b}(\mathrm{Lys}) = 14.62$.
  The spread reflects only the Arg replicas; the uncertainty of p$K_\mathrm{w}$ is not included.\\
  $^{d}$\,Alanine pentapeptides, Thurlkill et al.~\cite{thurlkill2006pk}.\\
  $^{e}$\,$14.00 - 10.40$, from the experimental p$K_\mathrm{a}$ of LysH$^+$ and p$K_\mathrm{w} = 14.00$.\\
  $^{f}$\,Fitch et al.~\cite{Fitch2015}.
\end{table}

\begin{figure*}
    \centering
    \includegraphics[width=0.6\linewidth]{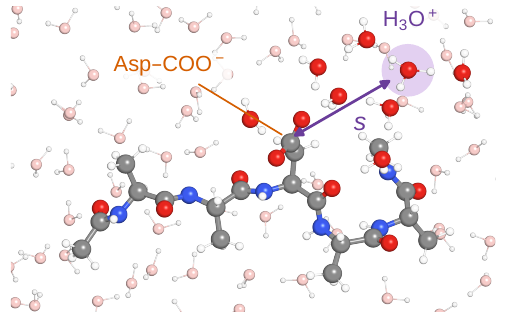}
    \caption{Representative snapshot of the simulation of Asp pentapeptide in water. The carboxylic proton has been transferred to a water molecule (purple area), which is tracked by the charge separation CV $s$ (purple arrow)}
    \label{fig:charge-separation-illustration}
\end{figure*}

\begin{figure*}
\centering
\includegraphics[width=1.0\linewidth]{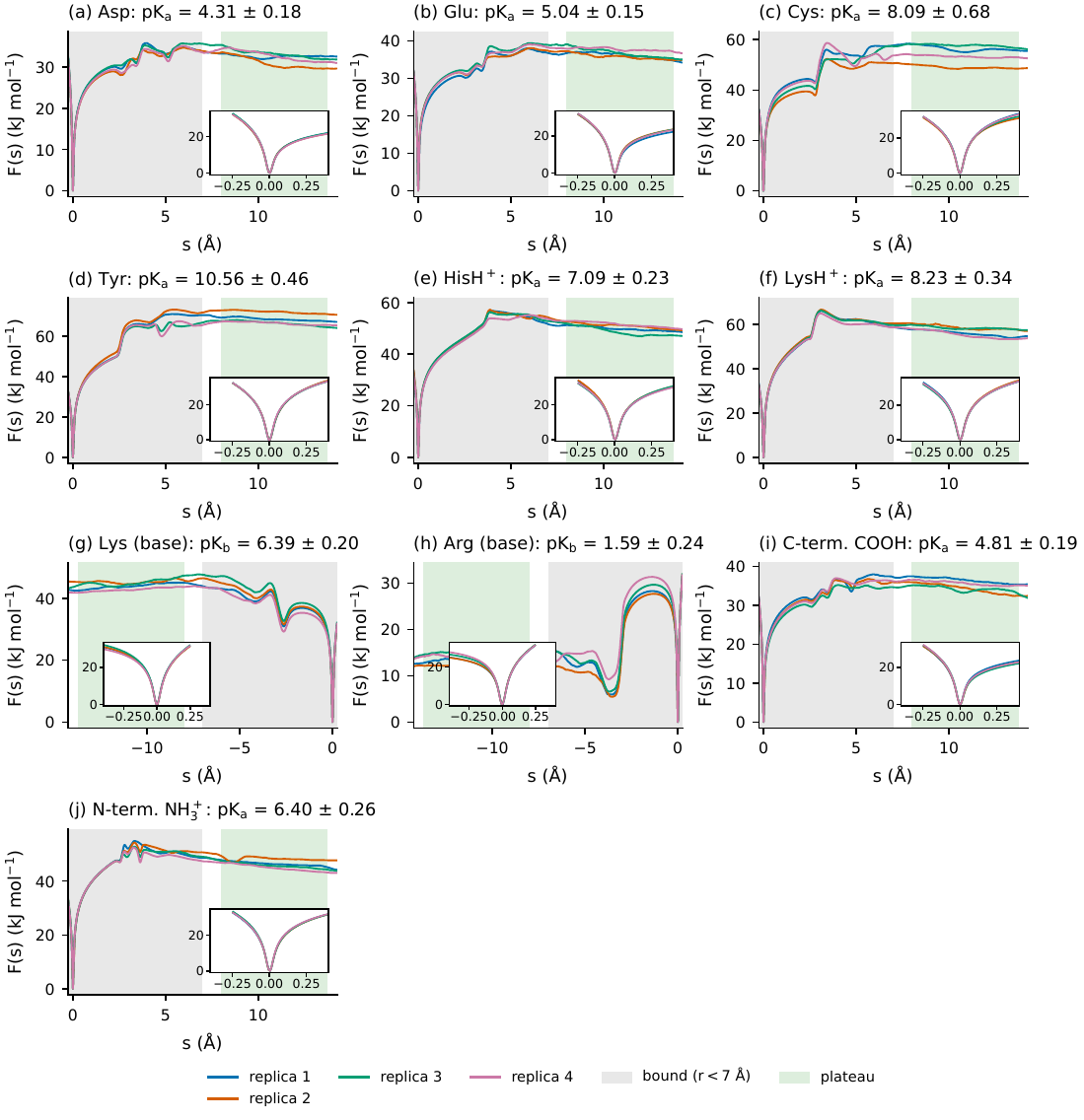}
\caption{\textbf{Free energy profiles of proton transfer between peptides and water.} Each plot shows the free energy as a function of the charge separation CV s. The profile is split into two regions: $-0.25\,\text{\AA}<s<7\,\text{\AA}$ (shaded grey) in which the proton is still considered bound to the peptide and $8\,\text{\AA}<s<13.75\,\text{\AA}$ (shaded green) in which the proton is considered to be freely diffusing in bulk water. Insets show the free energy profile around $s=0$, where the proton is still non-dissociated from the peptide residue}
\label{fig:peptide_pmf}
\end{figure*}

\clearpage

\section{Biomolecular dynamics}\label{sec:md-biomol}

We first revisited the biomolecular systems used to characterize SO3LR to check that the larger corpus and the refined interaction modules maintain their behavior in production simulations. We then simulated a metal-containing DNA crosslink and an unmodified B-DNA duplex.

\vspace{-2mm}
\subsection{Polyalanine}
\begin{figure*}[h]
  \centering
  \includegraphics[width=0.75\textwidth]{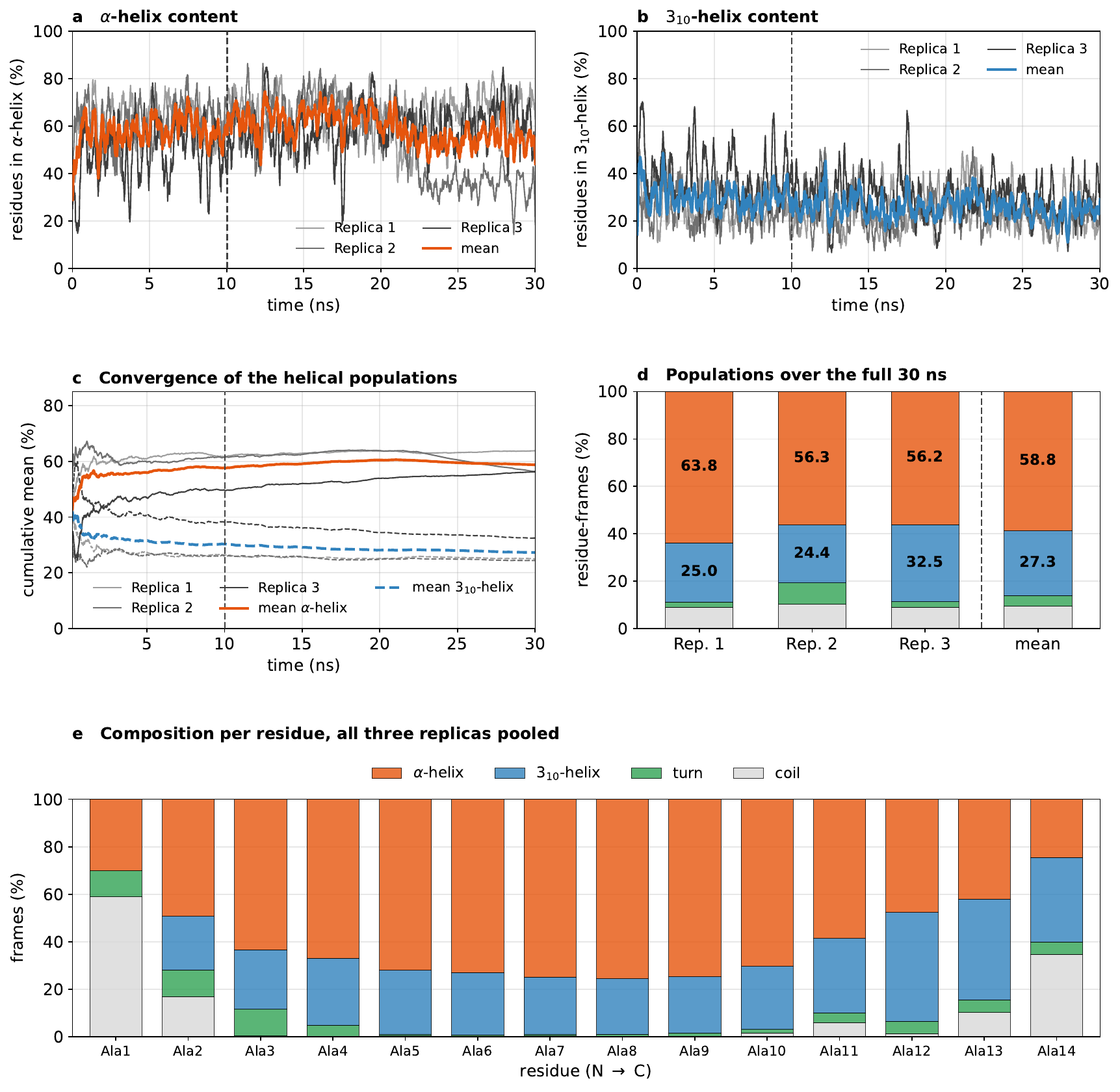}
  \caption{\textbf{Secondary structure of the Ala$_{15}$ helix under SO3LR-T-M.} Three independent \qty{30}{\nano\second} gas-phase trajectories of Ace-(Ala)$_{15}$-NMe at \qty{300}{\kelvin} (Nos\'e--Hoover NVT, \qty{0.5}{\femto\second} step), started from the folded $\alpha$-helix; secondary structure is assigned per frame with STRIDE~\cite{frishman1995knowledge}. \textbf{(a,b)}~$\alpha$- and $3_{10}$-helix fractions as a \qty{200}{\pico\second} rolling mean (grey, individual replicas; colour, their mean); the dashed line marks the \qty{10}{\nano\second} restart. \textbf{(c)}~Cumulative means. \textbf{(d)}~Populations averaged over each trajectory and over all three. \textbf{(e)}~Per-residue composition. Quantities are evaluated over residues Ala1--Ala14; STRIDE discards the caps and the $\psi$-less C-terminal alanine}
  \label{fig:ala15-ss}
\end{figure*}

\vspace{-2mm}
A gas-phase Ace-(Ala)$_{15}$-NMe helix was propagated in three independent \qty{30}{\nano\second} NVT trajectories at \qty{300}{\kelvin}, starting from the folded $\alpha$-helix (Fig.~\ref{fig:ala15-ss}). The chain adopts a dynamic mixture of $\alpha$- and $3_{10}$-helix, $58.8\%$ and $27.3\%$ of residues on average, a $68{:}32$ ratio among helical residues, with per-replica values of $72{:}28$, $70{:}30$ and $63{:}37$; both populations are stationary beyond ${\sim}\qty{5}{\nano\second}$. Recovering an appreciable $3_{10}$ population is notable, since additive force fields tend to over-stabilize the $\alpha$-helix~\cite{unke2024biomolecular}, and the exact $\alpha/3_{10}$ balance is governed by small free-energy differences that are sensitive to sampling. Unlike SO3LR, whose training deliberately excluded such fragments as a transferability test, SO3LR-T includes the top-down GEMS fragments (which contain Ala$_{15}$-type peptides) in its training corpus, since the aim here is to obtain the most accurate model.

\clearpage

\subsection{Crambin} 
\begin{figure*}[h]
  \centering
  \includegraphics[width=\textwidth]{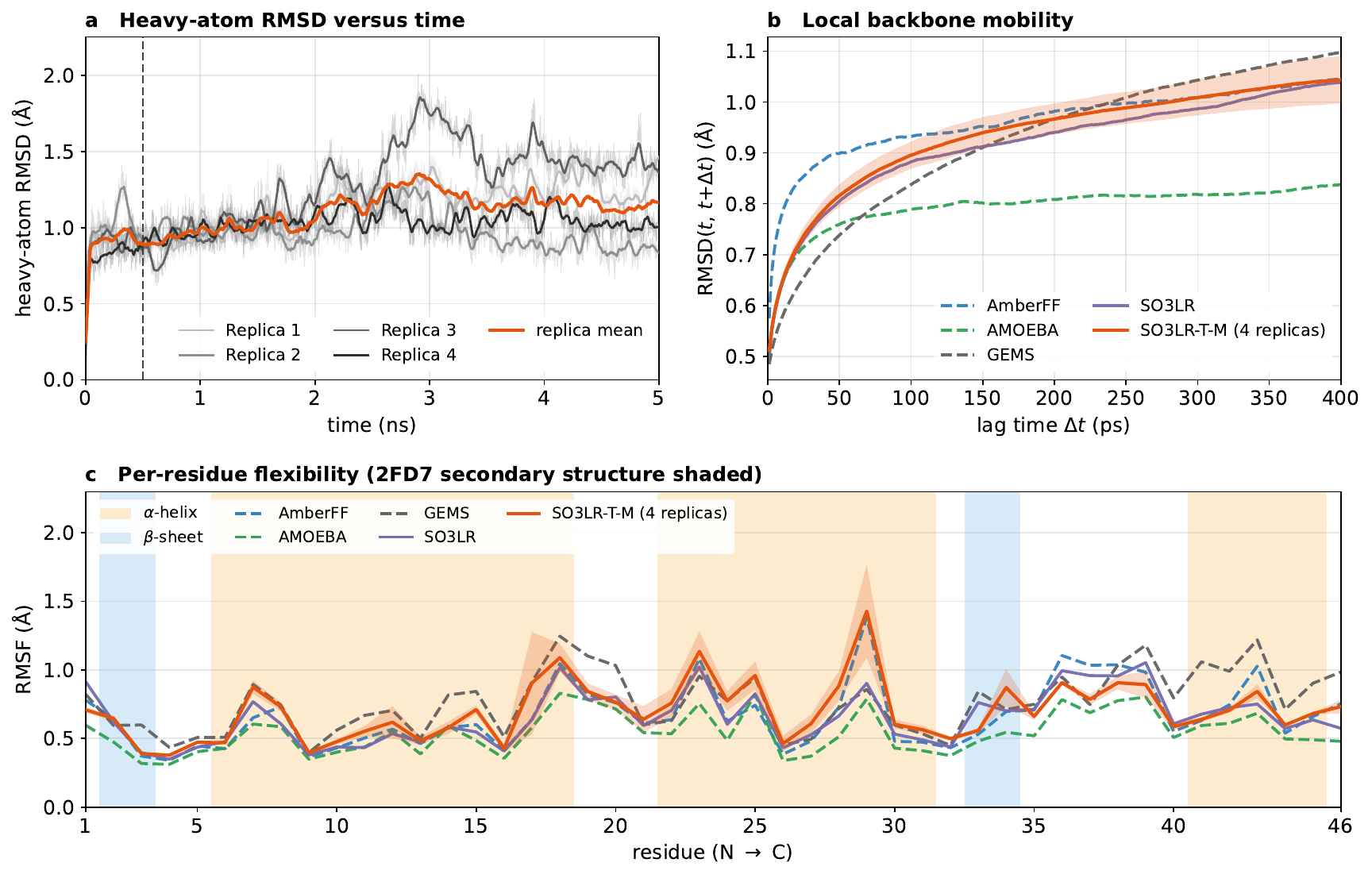}
  \caption{\textbf{Solvated crambin.} Crambin (PDB 2FD7, 46 residues) in explicit water, \num{25257} atoms, NPT at \qty{300}{\kelvin} and \qty{1}{\bar} with PME, \qty{0.5}{\femto\second} step; four velocity-seed replicas analysed over their first \qty{5}{\nano\second}. References are the classical AmberFF, the polarizable AMOEBA, and the machine-learned GEMS~\cite{unke2024biomolecular} and SO3LR~\cite{kabylda2025molecular} (AmberFF, AMOEBA and GEMS dashed; SO3LR solid). \textbf{(a)}~Heavy-atom RMSD against each replica's first frame; the native fold is maintained (pooled $\num{1.10}\pm\num{0.22}$~\AA, radius of gyration stable to \qty{0.9}{\percent}). \textbf{(b)}~Lag-time RMSD, a measure of local backbone mobility insensitive to slow drift; SO3LR-T-M is statistically indistinguishable from SO3LR, AmberFF and GEMS, while AMOEBA is markedly more rigid. \textbf{(c)}~Per-residue RMSF; the two inter-helix loops (residues 18--21 and 35--40) are among the most mobile segments in all five models}
  \label{fig:si-crambin}
\end{figure*}

Solvated crambin (PDB 2FD7, \num{25257} atoms) was run in explicit water under NPT at \qty{300}{\kelvin} in four replicas (Fig.~\ref{fig:si-crambin}). The native fold is maintained throughout, with a pooled heavy-atom RMSD of $\num{1.10}\pm\num{0.22}$~\AA{} and a radius of gyration stable to \qty{0.9}{\percent}. Local backbone mobility, measured by the lag-time RMSD, is statistically indistinguishable from SO3LR and from the classical AmberFF and machine-learned GEMS references (about $\num{0.90}$ and $\num{1.04}$~\AA{} at lag times of \qty{100}{\pico\second} and \qty{400}{\pico\second}), while the polarizable AMOEBA is markedly more rigid~\cite{unke2024biomolecular}. The two inter-helix loops (residues 18--21 and 35--40) are the most mobile part of the chain in every model, a structurally sensible pattern, with SO3LR-T-M showing more mobile conformational exploration around the second helix. The step from SO3LR to SO3LR-T-M thus leaves crambin flexibility essentially unchanged, consistent with the high conformational variability seen by NMR~\cite{ahn2006three}.

\newpage

\subsection{Glycoprotein} 
\begin{figure*}[h]
  \centering
  \includegraphics[width=0.6\textwidth]{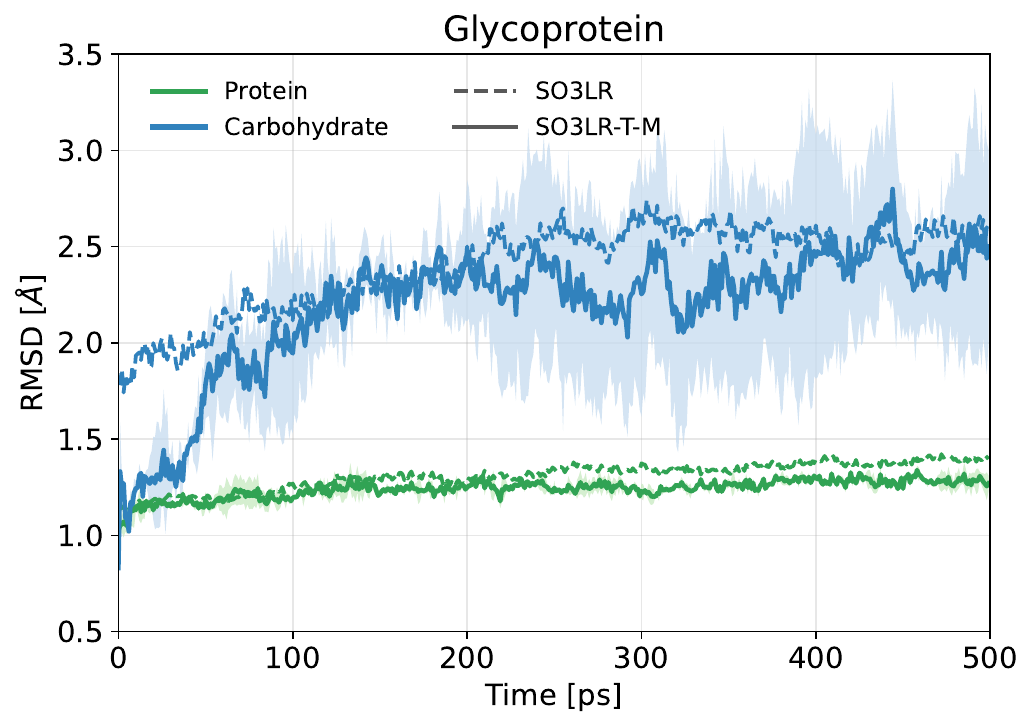}
  \caption{\textbf{The 1K7C glycoprotein.} Root-mean-square deviation of the protein (segment \texttt{PROA}, \num{3377} atoms) and the carbohydrate (segments \texttt{CARA}/\texttt{CARB}, \num{167} atoms) from the starting structure, each segment superposed on itself so the carbohydrate trace reports internal glycan rearrangement rather than motion relative to the protein. Lines are replica means (4 for SO3LR, 2 for SO3LR-T-M); the band is the SO3LR-T-M standard deviation. Over the final \qty{50}{\pico\second} the protein reaches \qty[separate-uncertainty]{1.40(9)}{\angstrom} (SO3LR) and \qty[separate-uncertainty]{1.29(3)}{\angstrom} (SO3LR-T-M), the carbohydrate \qty[separate-uncertainty]{2.56(50)}{\angstrom} and \qty[separate-uncertainty]{2.41(42)}{\angstrom}: the protein stays folded while the glycans remain mobile, and the two models agree within the replica scatter}
  \label{fig:glyco-so3lrm}
\end{figure*}

The 1K7C glycoprotein (\num{47863} atoms; the CHARMM-GUI~\cite{jo2008charmm} box as built, neutralized by 11 protonated carboxylates without added ions) was simulated with both SO3LR and SO3LR-T-M (Fig.~\ref{fig:glyco-so3lrm}). With both models the protein keeps its fold, with an RMSD of 1.3--1.4~\AA{} over the final \qty{50}{\pico\second}, while the glycans are markedly more mobile (2.4--2.6~\AA{} within the carbohydrate segments). This greater flexibility of the glycans relative to the protein agrees with dedicated carbohydrate force fields~\cite{guvench2011charmm}, and the two SO3LR generations agree within the replica scatter.

\newpage

\subsection{POPC bilayer} 
\begin{figure*}[h]
  \centering
  \includegraphics[width=\textwidth]{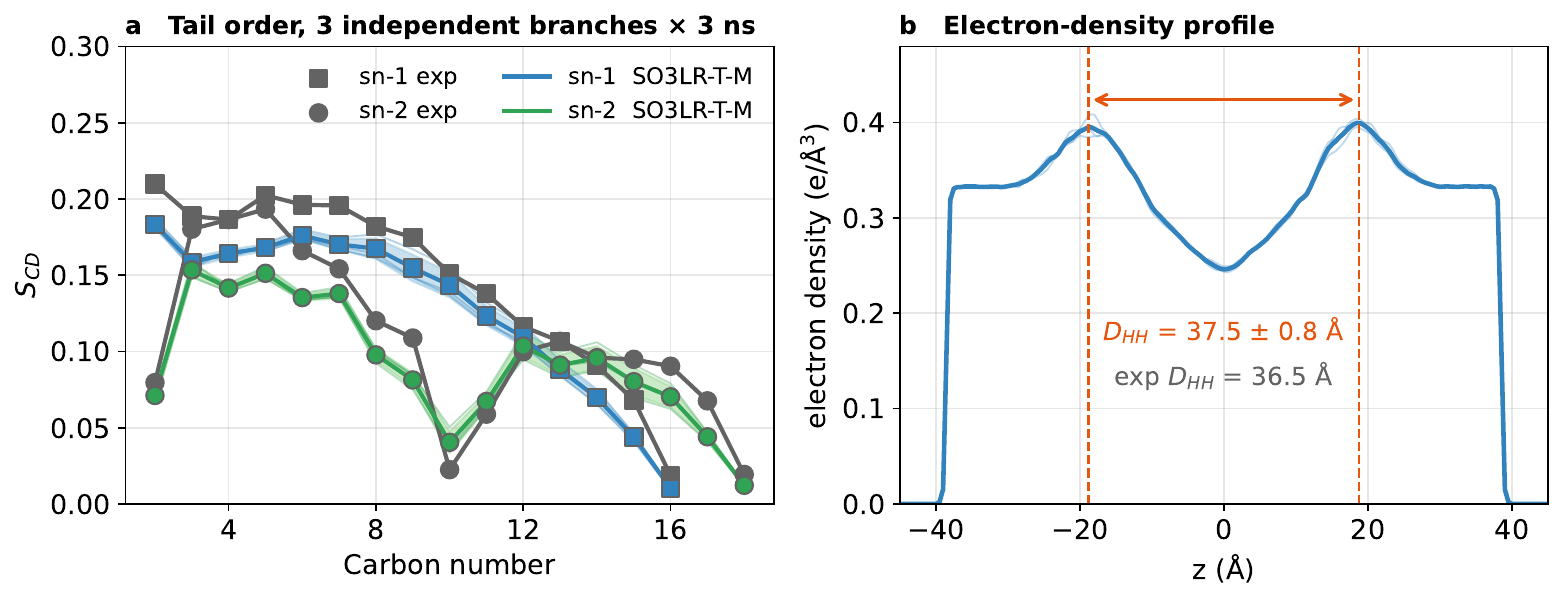}
  \caption{\textbf{POPC bilayer at \qty{303}{\kelvin} and \qty{1}{\bar}.} A hydrated bilayer of 128 POPC and 5120 water (\num{32512} atoms), isotropic NPT, \qty{1}{\femto\second} step, PME for electrostatics and dispersion, averaged over three \qty{3}{\nano\second} trajectories. \textbf{(a)}~Deuterium order parameters $S_{\mathrm{CD}}$ along the \textit{sn}-1 (blue) and \textit{sn}-2 (green) chains against experiment~\cite{ferreira2013cholesterol}; the dip at the \textit{sn}-2 double bond is reproduced (RMSD from experiment $0.021$ and $0.023$, combined $0.022$). \textbf{(b)}~Electron-density profile; the head-to-head thickness $D_{\mathrm{HH}} = \qty[separate-uncertainty]{37.5(8)}{\angstrom}$ agrees with the experimental \qty{36.5}{\angstrom}~\cite{kuvcerka2011fluid}. In the isotropic ensemble the z/xy ratio is set by the starting box rather than predicted}
  \label{fig:popc-so3lrm}
\end{figure*}

A hydrated POPC bilayer (128 lipids, 5120 water, \num{32512} atoms) was simulated in the isotropic NPT ensemble at \qty{303}{\kelvin}, with PME for both electrostatics and dispersion, over three \qty{3}{\nano\second} trajectories (Fig.~\ref{fig:popc-so3lrm}). The head-to-head thickness, $D_{\mathrm{HH}} = \qty[separate-uncertainty]{37.5(8)}{\angstrom}$, agrees with the experimental \qty{36.5}{\angstrom}~\cite{kuvcerka2011fluid}, and the deuterium order parameters follow the experimental profile along both acyl chains (combined RMSD $0.022$)~\cite{ferreira2013cholesterol}, reproducing the characteristic dip at the \textit{sn}-2 double bond. In the isotropic ensemble the box shape is fixed, so the area per lipid stays close to its starting value rather than being predicted; determining the tensionless area in a semi-isotropic ensemble remains a challenge for the current models.

\newpage

\subsection{Cisplatin--DNA crosslink} 
\begin{figure*}[h]
\centering
\includegraphics[width=0.99\linewidth]{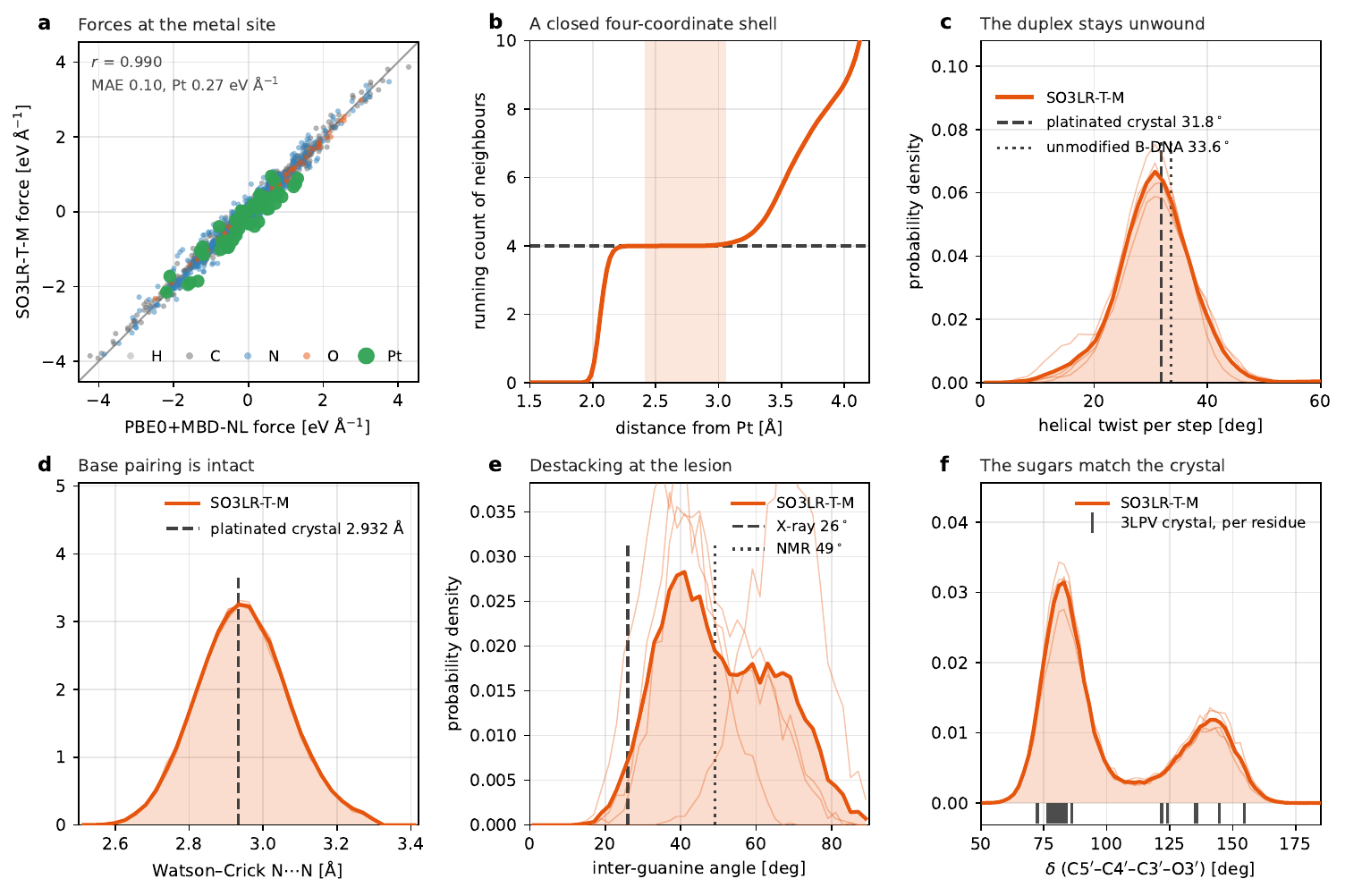}
\caption{\textbf{Cisplatin--DNA crosslink.} Four \qty{2.6}{\nano\second} SO3LR-T-M replicas of the 1,2-d(GpG) crosslink (PDB 3LPV). Thick lines are distributions pooled over the four replicas and thin lines the individual replicas. \textbf{(a)}~SO3LR-T-M forces compared to all-electron PBE0+MBD-NL on sixteen snapshots of cis-[Pt(NH$_3$)$_2$(9H-guanine)$_2$]$^{2+}$, colored by element with platinum enlarged. \textbf{(b)}~Running neighbor count around the platinum for one replica, with a dashed line at four neighbors. The shaded band marks the 2.41--3.06~\AA{} window discussed in the text. \textbf{(c)}~Helical twist per base-pair step, with the platinated crystal and unmodified B-DNA values as vertical lines. \textbf{(d)}~Watson--Crick N$\cdots$N distance, with the crystal value as a vertical line. \textbf{(e)}~Angle between the two platinated guanine planes, with the X-ray and NMR values as vertical lines (not converged, see text). \textbf{(f)}~Sugar pucker through the backbone torsion $\delta$. Ticks mark the per-residue values of the 3LPV crystal}
\label{fig:si_cis}
\end{figure*}

The cisplatin 1,2-d(GpG) crosslink (PDB 3LPV), a Pt(II) centre bridging two guanines in a solvated DNA duplex, was run with SO3LR-T-M in four \qty{2.6}{\nano\second} replicas (Fig.~\ref{fig:si_cis}). The system contains \num{33217} atoms, comprising the duplex with its crosslink, the two crystallographic Mg$^{2+}$ ions (each given an explicit octahedral hexahydrate, since the deposited structure resolves at most one first-shell water), \num{10775} added water molecules, 26 Na$^+$, and 10 Cl$^-$. It was run in the NPT ensemble at 300~K and 1~bar, with a 1~fs time step, a 10~\AA{} long-range cutoff, and PME for electrostatics and dispersion.

The platinum stays four-coordinate, and no fifth ligand enters over \qty{10.5}{\nano\second} of pooled trajectory. The running neighbor count reaches four at \qty{2.41}{\angstrom} and increases by less than $0.07$ up to \qty{3.06}{\angstrom} (Fig.~\ref{fig:si_cis}b). This coordination is maintained by the potential alone, without any bonded restraint, and the N--Pt--N angles (Fig.~\ref{fig:main3}A of the main text) confirm the square-planar geometry. To validate the forces at the metal site, we compared SO3LR-T-M with all-electron PBE0+MBD-NL, the level the model was trained on, for sixteen snapshots of a cis-[Pt(NH$_3$)$_2$(9H-guanine)$_2$]$^{2+}$ model of the site (Fig.~\ref{fig:si_cis}a). The forces correlate at $r = 0.990$, with a mean absolute error of $0.105$~eV~\AA$^{-1}$ ($0.267$~eV~\AA$^{-1}$ on platinum).

The lesion leaves the duplex underwound, with a mean helical twist of $30.6^\circ$, compared to $33.6^\circ$ for unmodified B-DNA and $31.8^\circ$ in the platinated crystal (Fig.~\ref{fig:si_cis}c). Watson--Crick pairing stays intact (mean N$\cdots$N distance \qty{2.946}{\angstrom} vs \qty{2.932}{\angstrom} in the crystal; Fig.~\ref{fig:si_cis}d). The sugar pucker, followed through the backbone torsion $\delta$, adopts the bimodal distribution the lesion imposes, with $66\%$ of sugars in the low-$\delta$ basin compared to $73\%$ in the starting crystal (Fig.~\ref{fig:si_cis}f). However, the angle between the two platinated guanine planes is not converged at this trajectory length (Fig.~\ref{fig:si_cis}e). Its mean is $50.3^\circ$ (NMR $49^\circ$, crystal $26^\circ$), but the four replica means span $39.6$--$65.8^\circ$. For this observable, the relevant uncertainty is the spread between the replica means.

\subsection{Drew--Dickerson dodecamer} 
\begin{figure*}[h]
\centering
\includegraphics[width=0.8\linewidth]{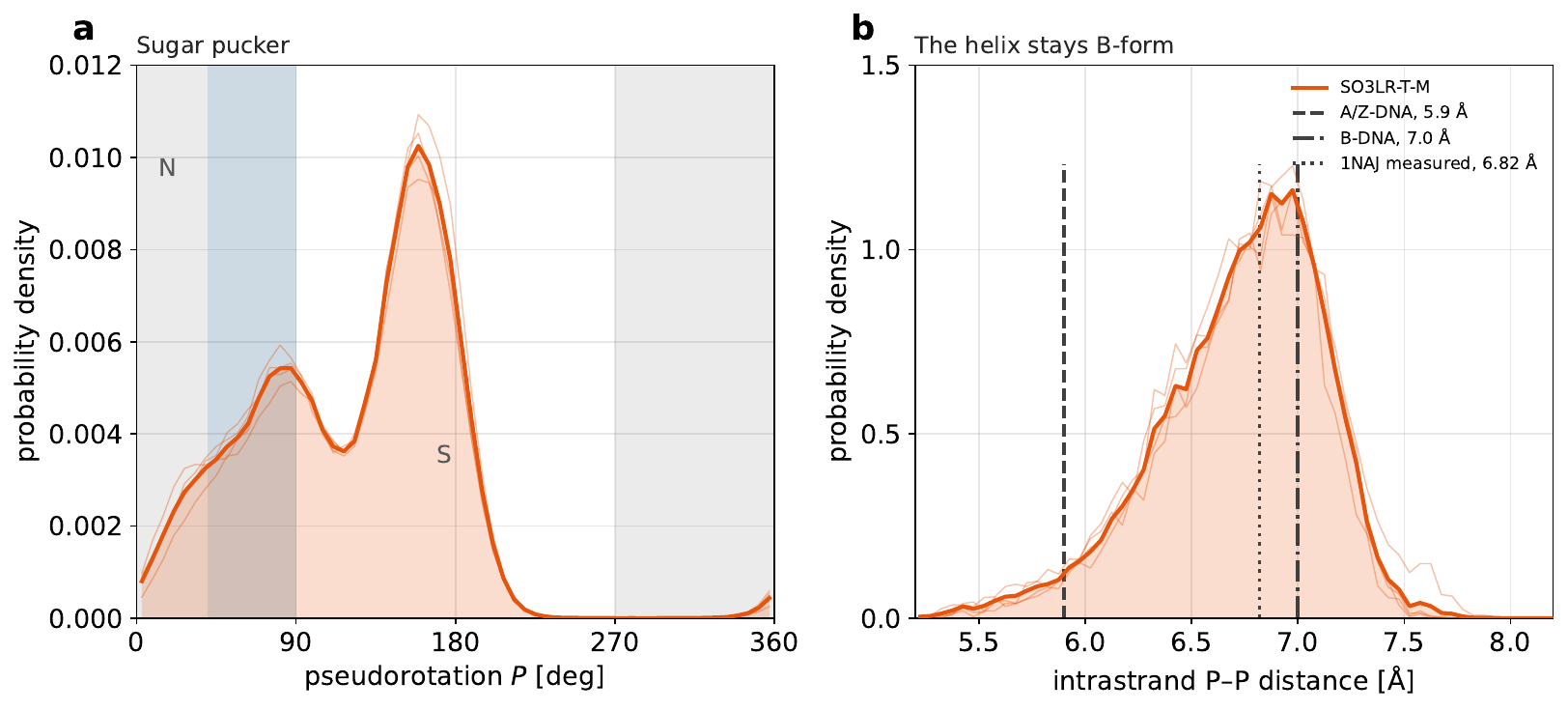}
\caption{\textbf{Drew--Dickerson dodecamer under SO3LR-T-M.} Four \qty{3.8}{\nano\second} replicas. Thick lines are distributions pooled over all replicas and thin lines the individual replicas. \textbf{(a)}~Pseudorotation phase $P$, with the northern (N) hemisphere shaded gray and the O4$'$-endo region ($P = 40$--$90^\circ$) shaded blue. \textbf{(b)}~Intrastrand P$\cdots$P distance, with the canonical A/Z- and B-DNA values and the 1NAJ solution structure as vertical lines}
\label{fig:si_ddd}
\end{figure*}

We simulated the unmodified Drew--Dickerson dodecamer with SO3LR-T-M in four \qty{3.8}{\nano\second} replicas (Fig.~\ref{fig:si_ddd} and Table~\ref{tab:ddd_table3}). The system contains \num{30956} atoms (the duplex, \num{10052} water molecules, 32 Na$^+$, and 10 Cl$^-$) and was run under the same protocol as the crosslink above.

Table~\ref{tab:ddd_table3} compares backbone and helical parameters with the X-ray and NMR structures and with the DNA-specific force fields OL21 and OL24, following the averaging protocol of Ref.~\citenum{zgarbova2025}. The mean helical twist is $33.4^\circ$, close to the X-ray value of $33.6^\circ$. The backbone torsions $\alpha$, $\beta$, $\gamma$, $\varepsilon$ and $\zeta$ differ from the X-ray values by at most $7.6^\circ$ ($\beta$). For the same torsions, OL21 deviates by up to $4.1^\circ$, while OL24 and the NMR structure deviate by up to $9.5^\circ$ and $9.9^\circ$, respectively. The $\alpha$ and $\beta$ values of SO3LR-T-M ($293.2^\circ$ and $169.0^\circ$) closely match OL24 ($293.3^\circ$ and $169.7^\circ$). The glycosidic torsion $\chi$, tilt and propeller twist agree with the X-ray structure within $3^\circ$. The mean $\delta$ ($120.5^\circ$) and $P$ ($117.9^\circ$) require care, because both distributions are multimodal and their circular means fall between basins. We therefore discuss the sugar pucker through the full $P$ distribution (Fig.~\ref{fig:si_ddd}a).

The sugar pucker is dominated by the southern (C2$'$-endo) basin, which peaks at $P = 160^\circ$, as expected for B-DNA. However, the northern hemisphere is over-populated, with $31.4\%$ of sugars compared to $12$--$15\%$ from NMR. This excess does not indicate a drift toward A-form. The true C3$'$-endo population is $9.0\%$, close to experiment, and the remaining $22.4\%$ lie in an O4$'$-endo shoulder at $P = 40$--$90^\circ$, where reference free-energy profiles place a barrier. The excess is concentrated in pyrimidines ($53.7\%$ vs $23\%$ in experiment; purines $12.8\%$ vs $7\%$). It does not propagate into the backbone. The intrastrand P$\cdots$P distance has a median of \qty{6.79}{\angstrom} and a mode of \qty{6.97}{\angstrom} (Fig.~\ref{fig:si_ddd}b). For comparison, the 1NAJ solution structure gives \qty{6.82}{\angstrom}, and the canonical values are \qty{7.0}{\angstrom} (B) and \qty{5.9}{\angstrom} (A/Z). The helix therefore stays B-form throughout. This suggests a local deficiency in the sugar torsional barriers that leaves the description of the duplex intact.

\begin{table*}[h]
\centering
\caption{\textbf{Drew--Dickerson dodecamer: backbone and helical parameters.} Means over four replicas of \qty{3.8}{\nano\second} ($t \geq \qty{200}{\pico\second}$), following the protocol of Table~3 of Ref.~\citenum{zgarbova2025} (two base pairs trimmed at each end, canonical BI backbone only). Deviations are from the X-ray reference; the two DNA-specific force fields OL21 and OL24 and the solution NMR structure are listed on the same footing, so the spread among them is the scale on which a deviation is meaningful. All values in degrees. $^{\dagger}$~Both $\delta$ and $P$ are multimodal here, so a circular mean describes no actual conformer; the $P$ distribution is $55.7\,\%$ south, $35.2\,\%$ east and $9.0\,\%$ north, and the mean falls between basins (Fig.~\ref{fig:si_ddd}a). The last row gives the mean absolute deviation from X-ray over all eleven parameters. $^{\ddagger}$~Not trimmed}
\label{tab:ddd_table3}
\begin{tabular}{lrrrrrr}
\toprule
 & X-ray & NMR & OL21 & OL24 & SO3LR-T-M & $\Delta$(SO3LR-T-M$-$X-ray) \\
\midrule
$\alpha$ & 298.7 & 298.4 & 294.9 & 293.3 & 293.2 & $-5.5$ \\
$\beta$ & 176.6 & 173.5 & 180.7 & 169.7 & 169.0 & $-7.6$ \\
$\gamma$ & 52.6 & 50.3 & 50.2 & 51.6 & 57.3 & $+4.7$ \\
$\delta$\,$^{\dagger}$ & 122.4 & 126.7 & 133.5 & 142.5 & 120.5 & $-1.9$ \\
$\varepsilon$ & 180.1 & 188.4 & 181.3 & 189.6 & 187.6 & $+7.5$ \\
$\zeta$ & 268.6 & 258.7 & 266.7 & 268.5 & 269.9 & $+1.3$ \\
$\chi$\,$^{\ddagger}$ & 247.6 & 249.2 & 253.3 & 246.4 & 244.8 & $-2.8$ \\
twist & 33.6 & 35.7 & 35.2 & 35.0 & 33.4 & $-0.2$ \\
tilt & $-0.2$ & 0.0 & 0.1 & 0.0 & $-0.0$ & $+0.2$ \\
propeller & $-11.8$ & $-17.5$ & $-11.7$ & $-10.9$ & $-10.5$ & $+1.3$ \\
$P$\,$^{\dagger}$ & 130.1 & 137.2 & 151.7 & 150.6 & 117.9 & $-12.2$ \\
\midrule
MAD vs.\ X-ray & -- & 4.1 & 4.9 & 6.1 & 4.1 & -- \\
\bottomrule
\end{tabular}
\end{table*}

\clearpage
\section{Computational performance}\label{sec:si-perf}

\begin{figure*}[h]
  \centering
  \includegraphics[width=\textwidth]{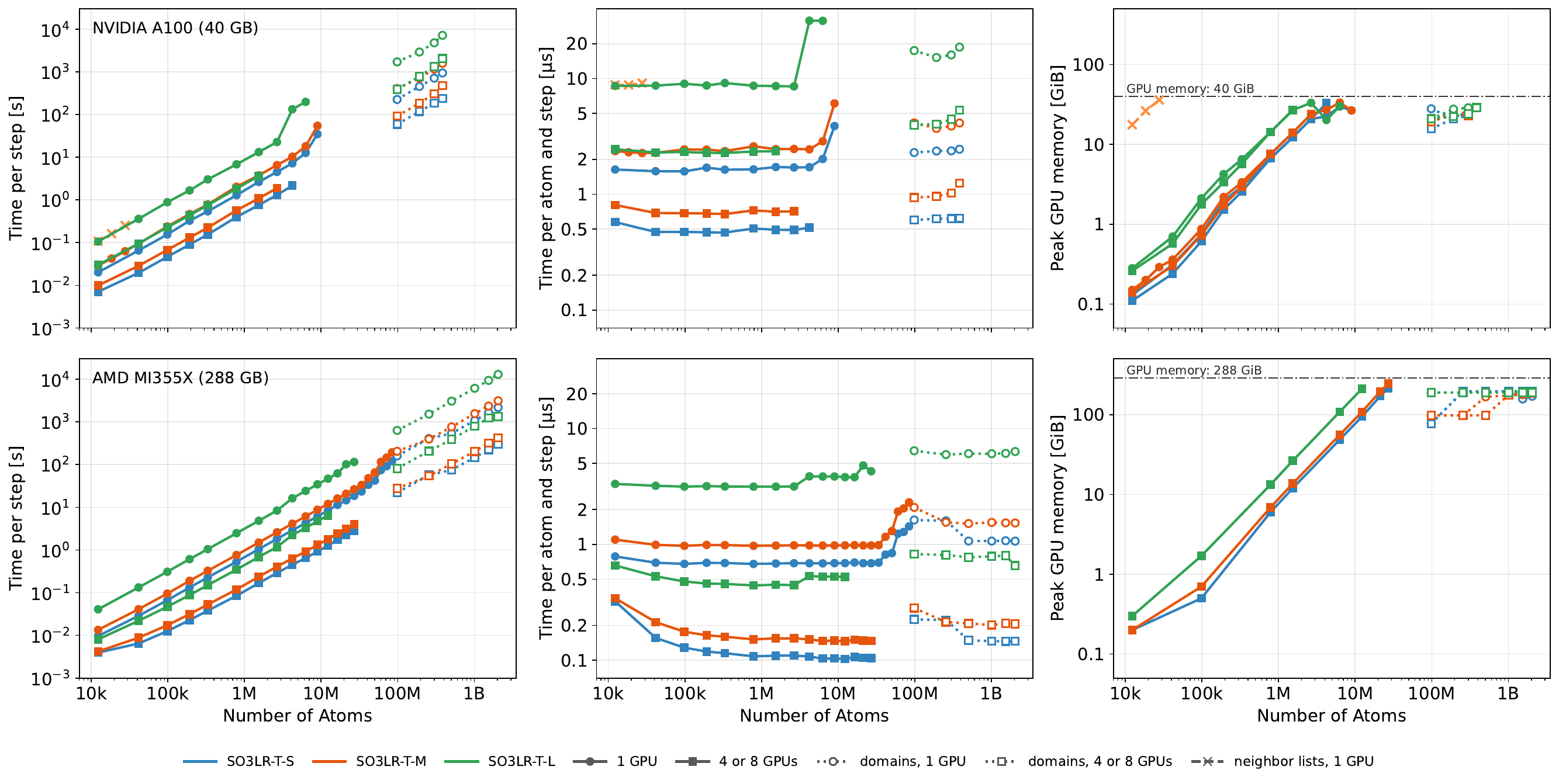}
  \caption{\textbf{Cost and memory on A100 and MI355X GPUs.} Time per MD step (left), time per atom and step (middle) and peak memory per GPU (right) as a function of the number of atoms, for SO3LR-T-S, -M and -L on periodic liquid water, measured on NVIDIA A100 (top) and AMD MI355X (bottom) GPUs following the protocol of Sec.~\ref{sec:si-perf}. Filled circles show flash on one GPU, and filled squares show flash on all GPUs of one node (four A100 or eight MI355X). Open circles and squares show the corresponding domain-wise evaluation, with one force call per point and real-space long-range interactions. Crosses and dashed lines show the neighbor-list implementation of SO3LR-T-M on one A100. The steps in the middle panels mark the transitions from the fastest memory mode to modes that recompute intermediate arrays. Dash-dotted lines indicate the memory of one GPU. No peak memory was recorded for whole systems on one MI355X}
  \label{fig:si-perf-all}
\end{figure*}

\textbf{Overview.} All timings in this section were measured with flash, the GPU engine of the \texttt{so3lr} package, using the settings in Table~\ref{tab:si-flash-protocol}. It can evaluate a system as a whole on one GPU, as a whole on all GPUs of one node, or one domain at a time (Fig.~\ref{fig:so3lr-single-gpu-scaling} of the main text and Fig.~\ref{fig:si-perf-all}). Its main features are the following.

\begin{itemize}
\item Fused GPU kernels, written in JAX Pallas on Triton, compute all interactions of each atom in a single pass over its neighbors, without storing any per-edge quantities.
\item Neighbors are found on a cell grid that is rebuilt at every call, so that no neighbor list needs to be buffered or padded. The long-range pairs are streamed from the same grid.
\item The radial filters and the electrostatic pair function are tabulated as cubic Hermite splines in the interatomic distance.
\item Forces are computed with dedicated backward kernels instead of automatic differentiation through a per-edge graph.
\item Per-atom operations (projections, MLPs and output heads) are processed in tiles of a fixed number of atoms.
\item Five memory modes (\texttt{fast}, \texttt{default}, \texttt{lean}, \texttt{leanest} and \texttt{minimal}) store progressively less data between the forward and backward passes and recompute more instead. By default, flash selects the fastest mode that is expected to fit in memory.
\end{itemize}
The tabulated radial functions~\cite{lysogorskiy2021performant}, the edge streaming and the tiled execution follow Symmetrix-XL~\cite{fu2026train}.

\begin{table}[h]
    \centering
    \caption{\textbf{Settings of the timing runs.} Settings are identical
    for both GPU types unless stated otherwise}
    \label{tab:si-flash-protocol}
    \begin{tabular}{ll}
        \toprule
        \textbf{Setting} & \textbf{Value} \\
        \midrule
        System & \parbox[t]{0.62\linewidth}{\raggedright periodic liquid water, $\rho \approx \qty{1.00}{g/cm^3}$, \num{12288} to \num{84283392} atoms, replicated from a \qty{50}{ps} NVT equilibration of a 512-molecule box (\num{1536} atoms, \qty{24.8}{\angstrom}) with each model} \\
        \addlinespace
        Dynamics & \parbox[t]{0.62\linewidth}{\raggedright NVT, \qty{300}{K}, Nos\'e--Hoover chain, \qty{0.5}{fs} time step, no prior relaxation} \\
        \addlinespace
        Precision & \parbox[t]{0.62\linewidth}{\raggedright model in \texttt{float32}; MD state in \texttt{float64} (flash) or \texttt{float32} (neighbor lists); \texttt{matmul\_precision} set to \texttt{highest}} \\
        \addlinespace
        Long range & \parbox[t]{0.62\linewidth}{\raggedright \qty{10}{\angstrom} real-space cutoff; PME for electrostatics and dispersion with a smearing of \qty{2.0}{\angstrom} and a grid spacing of \qty{1.55}{\angstrom} (A100) or \qty{2.0}{\angstrom} (MI355X)} \\
        \addlinespace
        Neighbor lists & \parbox[t]{0.62\linewidth}{\raggedright buffer factors of $1.15$ (short range) and $1.10$ (long range); neighbor-list runs only} \\
        \addlinespace
        Steps & \parbox[t]{0.62\linewidth}{\raggedright four cycles (three above $10^{6}$ atoms) of $\max(2, \min(100, 2\times10^{6}/N))$ steps for $N$ atoms} \\
        \addlinespace
        Reported time & \parbox[t]{0.62\linewidth}{\raggedright median over all cycles except the first} \\
        \addlinespace
        Memory & \parbox[t]{0.62\linewidth}{\raggedright A100 with \qty{95}{\percent} of the GPU memory (\qty{38}{GB}) allocated to XLA, peak memory taken as the peak XLA allocation; MI355X with the XLA platform allocator from \num{6.0e7} atoms onward, no peak memory recorded for whole systems on one GPU} \\
        \bottomrule
    \end{tabular}
\end{table}

\textbf{Accuracy.} In \texttt{float64}, on a water box of \num{12288} atoms, flash reproduces the energies of the neighbor-list implementation to within \num{8e-13}~eV/atom and the forces to within \num{1.2e-8}~eV/\AA{}. This holds for SO3LR-T-S, -M and -L in the \texttt{fast}, \texttt{leanest} and \texttt{minimal} modes. The strain derivative used in NPT simulations agrees to a relative deviation of \num{5e-10}. On an AMD MI355X, the same comparison yields deviations of \num{8.6e-13}~eV/atom and \num{1.2e-8}~eV/\AA{}.

\textbf{Timing protocol.} Each system size was run once, as a small number of MD cycles (Table~\ref{tab:si-flash-protocol}). The reported time per step is the median over all cycles except the first, which includes the just-in-time compilation.

\textbf{Cost and maximum system size.} Table~\ref{tab:si-flash-cost} summarizes the cost per atom and the largest systems that could be run, and Fig.~\ref{fig:si-perf-all} shows the complete scans. On one GPU, the cost per atom remains nearly constant as long as the fastest memory mode fits in memory. The steps in the middle panels of Fig.~\ref{fig:si-perf-all} mark the system sizes at which flash switches to more memory-efficient modes. Per GPU, the MI355X is $2.5$ (S, M) and $2.7$ (L) times faster than the A100, and accommodates $9$ (S, M) and $4$ (L) times more atoms.

\begin{table}[h]
    \centering
    \caption{\textbf{Cost per atom and the largest systems.} Whole systems: median over the \texttt{fast} memory mode from $100$k atoms on, MD with PME. Domains: median over $98$M--$384$M atoms (A100) and $98$M--$2.0$B atoms (MI355X), one force call with real-space long range; the largest domain runs are the same for all three models}
    \label{tab:si-flash-cost}
    \begin{tabular}{lccc}
        \toprule
        & \textbf{SO3LR-T-S} & \textbf{SO3LR-T-M} & \textbf{SO3LR-T-L} \\
        \midrule
        \multicolumn{4}{l}{\textit{Whole system, \si{\micro\second} per atom and step}} \\
        1 A100 & $1.7$ & $2.5$ & $8.7$ \\
        4 A100 & $0.49$ & $0.71$ & $2.3$ \\
        1 MI355X & $0.69$ & $0.98$ & $3.2$ \\
        8 MI355X & $0.11$ & $0.15$ & $0.46$ \\
        \addlinespace
        \multicolumn{4}{l}{\textit{One domain at a time, \si{\micro\second} per atom and force call}} \\
        1 A100 & $2.4$ & $4.0$ & $17$ \\
        4 A100 & $0.62$ & $0.99$ & $4.2$ \\
        1 MI355X & $1.1$ & $1.5$ & $6.1$ \\
        8 MI355X & $0.15$ & $0.21$ & $0.79$ \\
        \addlinespace
        \multicolumn{4}{l}{\textit{Whole system of \num{12288} atoms, ms per step}} \\
        1 A100 & $20.1$ & $29.0$ & $106$ \\
        4 A100 & $7.07$ & $9.93$ & $30.1$ \\
        1 MI355X & $9.67$ & $13.5$ & $40.9$ \\
        8 MI355X & $3.96$ & $4.21$ & $8.08$ \\
        \addlinespace
        \multicolumn{4}{l}{\textit{Largest whole system (atoms)}} \\
        1 A100 & $9.0$M & $9.0$M & $6.3$M \\
        4 A100 & $4.2$M & $2.7$M & $1.5$M \\
        1 MI355X & $84$M & $84$M & $27$M \\
        8 MI355X & $27$M & $27$M & $12$M \\
        \bottomrule
    \end{tabular}
\end{table}

\textbf{Comparison with the neighbor-list implementation.} On one A100, flash runs SO3LR-T-M $3.7$ to $4.0$ times faster than the neighbor-list implementation for the system sizes that both can handle, which range from \num{12288} to \num{27648} atoms (\num{29.0} compared with \qty{108}{ms} per step at \num{12288} atoms; crosses in Fig.~\ref{fig:si-perf-all}). The neighbor-list implementation stores per-edge arrays requiring \qtyrange{1.3}{1.5}{MB} per atom and runs out of memory at the next system size tested, \num{41472} atoms. In comparison, flash requires \qty{9.4}{KB} per atom in its fastest mode (about \num{8} and \qty{18}{KB} for SO3LR-T-S and -L) and \qty{3.1}{KB} per atom in the \texttt{minimal} mode. In this mode, systems of up to 9M atoms fit on the same GPU, more than $300$ times the largest system accessible to the neighbor-list implementation.

\textbf{Memory modes.} Table~\ref{tab:si-flash-modes} lists the cost of each memory mode relative to \texttt{fast}. The \texttt{default} mode has the same cost as \texttt{fast}, and \texttt{lean} and \texttt{leanest} cost $1.1$ to $1.4$ times as much. The \texttt{minimal} mode costs $1.9$ to $2.1$ times as much on the MI355X (S, M) and $2.3$ to $3.7$ times as much on the A100.

\begin{table}[h]
    \centering
    \caption{\textbf{Memory modes on one GPU.} Cost per atom and MD step of each memory mode relative to \texttt{fast} at the largest system that \texttt{fast} ran (median over the runs in each mode from $100$k atoms on). On the MI355X, the mode of each size is the first that fits, by the largest system measured in each mode. A dash: not run}
    \label{tab:si-flash-modes}
    \begin{tabular}{llcccc}
        \toprule
        \textbf{GPU} & \textbf{Model} & \texttt{default} & \texttt{lean} & \texttt{leanest} & \texttt{minimal} \\
        \midrule
        A100 & SO3LR-T-S & $1.0$ & $1.2$ & $1.3$ & $2.3$ \\
         & SO3LR-T-M & $1.0$ & $1.2$ & $1.4$ & $2.5$ \\
         & SO3LR-T-L & $1.0$ & $1.2$ & -- & $3.7$ \\
        \midrule
        MI355X & SO3LR-T-S & $1.0$ & $1.2$ & $1.2$ & $1.9$ \\
         & SO3LR-T-M & $1.0$ & $1.2$ & $1.3$ & $2.1$ \\
         & SO3LR-T-L & $1.0$ & $1.3$ & $1.1$ & -- \\
        \bottomrule
    \end{tabular}
\end{table}

\textbf{Several GPUs.} With \texttt{--flash-devices}~$N$, the $N$ GPUs of one node share each force call. Every GPU holds the arrays of all atoms, computes the rows of $1/N$ of them, and exchanges rows with the others after each interaction layer and for the forces; the MD state is replicated, so the integrator is unchanged. In \texttt{float64}, the forces on four A100s and on eight MI355Xs equal those on one GPU to \num{7e-15} and \num{1.5e-14}~eV/\AA{} (\num{41472} atoms). From $10^{5}$ atoms on, eight MI355Xs are $5.8$--$6.7$ (S, M) and $6.9$--$7.3$ (L) times faster than one, and four A100s $3.2$--$3.6$ (S, M) and $3.6$--$4.0$ (L) times; smaller systems gain less, because a step is too short to hide the exchanges. Since every GPU holds all atoms, several GPUs fit what one GPU fits in the \texttt{fast} mode, the only mode they run.

\textbf{A domain at a time.} Systems larger than the memory of one GPU can be evaluated a domain at a time. The box is cut into $k\times k\times k$ domains, and each force call makes three sweeps over them, one domain after another:
\begin{enumerate}
\item each domain, with a halo of $T r_\mathrm{c} + 0.1$~\AA{} (\qty{9.1}{\angstrom}, \qty{13.6}{\angstrom}, and \qty{15.1}{\angstrom} for SO3LR-T-S, -M, and -L), gives the local energies, charges, and $a_0$ and $C_6$ ratios of its own atoms;
\item with the charges equilibrated over the whole system, each domain, with a \qty{10.1}{\angstrom} halo, gives the long-range energy of its own atoms and its derivatives with respect to the charges and ratios;
\item each domain backpropagates the terms of its own atoms; the gradients with respect to its halo atoms are partial forces, summed over the domains.
\end{enumerate}
All domains are padded to the same size, so that a single compiled engine can be used for all of them. In \texttt{float64}, the forces agree with those of a single whole-system evaluation to within \num{6.4e-13}~eV/\AA{} and the energy to within \num{1e-14}~eV/atom. Owing to the halos and the repeated forward pass, a force call costs $1.4$ to $1.9$ times as much per atom as the fastest whole-system mode (Table~\ref{tab:si-flash-cost}). For a system of \num{2.04e9} atoms, a force call takes \qtylist{2181;3129;12950}{s} on one MI355X with SO3LR-T-S, -M and -L, and \qtylist{298;421;1339}{s} on eight. On A100 GPUs, four devices are $3.3$ to $4.4$ times faster than one. %
The domain-wise evaluation includes only real-space long-range interactions, since a PME mesh requires the whole box, and it is not yet coupled to the MD integrator.

\textbf{Numerical precision.} Both implementations evaluate the model in \texttt{float32}. The neighbor-list implementation performs selected, numerically sensitive JAX-MD operations in \texttt{float64}. flash keeps the positions, momenta and energy sums in \texttt{float64}, and computes the interatomic displacements in \texttt{float64} within its kernels. This is necessary for large systems, since in \texttt{float32} the coordinates in a \qty{500}{\angstrom} box can only be represented to about \qty{3e-5}{\angstrom}.

The \texttt{so3lr} MD commands set the JAX option \texttt{matmul\_precision} to \texttt{highest}, so that all matrix products are computed in full \texttt{float32} precision. This setting is used for all timings in this section and for all production simulations. With the \texttt{default} setting, XLA may use the TF32 format on the tensor cores of NVIDIA A100 GPUs, which is not sufficiently accurate. The \texttt{float32} energy then depends on how the computation is compiled and differs by \numrange{0.01}{0.07}~meV/atom between a whole-system evaluation and an evaluation one domain at a time. Its deviation from the \texttt{float64} energy also increases to \qty{1.3}{meV/atom} for SO3LR-T-S and \qtyrange{0.24}{0.32}{meV/atom} for SO3LR-T-M, compared with \qty{1e-4}{meV/atom} with \texttt{highest}. Matrix multiplication in bf16 also failed our accuracy tests, and neither TF32 nor bf16 was used in any simulation. The attention kernels of flash control their precision independently. The forward pass computes all products in IEEE \texttt{float32}. In the backward pass, the products with the filter weights are computed as three TF32 products on NVIDIA tensor cores, which recovers \texttt{float32} accuracy and leaves the force errors relative to \texttt{float64} unchanged. On AMD GPUs, these products are computed in IEEE \texttt{float32}.